\documentclass[aip,jcp,amsmath,amssymb,reprint]{revtex4-2}
\usepackage{enumitem}
\usepackage{graphicx}
\usepackage{multirow}
\usepackage{dcolumn}
\usepackage{bm}
\usepackage[utf8]{inputenc}
\usepackage[T1]{fontenc}
\usepackage{mathptmx}
\usepackage{etoolbox}
\usepackage{soul}
\usepackage{color}
\usepackage{float}
\usepackage{xr}
\makeatletter
\newcommand*{\addFileDependency}[1]{
\typeout{(#1)}
\@addtofilelist{#1}
\IfFileExists{#1}{}{\typeout{No file #1.}}
}\makeatother

\newcommand*{\myexternaldocument}[1]{%
\externaldocument{#1}%
\addFileDependency{#1.tex}%
\addFileDependency{#1.aux}%
}

\myexternaldocument{SM/sm}

\begin{document}

\title{Comprehensive molecular dynamics study of the dynamical properties of a dense binary hard-sphere mixture}

\author{Sabry G. Moustafa}
\email[Electronic mail: ]{smoustafa@una.edu}
\affiliation{Department of Engineering and Industrial Professions, University of North Alabama, Florence, Alabama 35632, USA}
\author{Andrew J. Schultz}
\affiliation{Department of Chemical and Biological Engineering, University at Buffalo, The State University of New York, Buffalo, New York 14260, USA}

\begin{abstract}
We present an extensive molecular dynamics (MD) study of the dynamical properties of a binary hard-sphere fluid over a wide range of packing fractions, $\phi \approx 0.357-0.582$. The self-diffusivity, $D$, and shear viscosity, $\eta$, are computed using an efficient implementation of the Einstein--Helfand method. The finite-size effects in $D$ scale as $1/N^\alpha$, with $\alpha$ increasing from approximately $1/3$ to $3/4$ with increasing $\phi$, whereas those in $\eta$ are negligible except for $\phi \gtrsim 0.554$, where they scale as $1/N$. The data are then extrapolated to the thermodynamic limit to obtain $D_{\infty}$ and $\eta_{\infty}$. Both coefficients show a super-Arrhenius dependence on $\phi$ for dense states, accompanied by a breakdown of the Stokes--Einstein relation. Although both $D_{\infty}(\phi)$ and $\eta_{\infty}(\phi)$ data are well described by an exponential form, we demonstrate that these fits do not provide reliable estimates of the critical packing fraction, $\phi_0$, owing to the substantial extrapolation required beyond the accessible equilibrium range. We find the commonly assumed proportionality between $\eta$ and the structural relaxation time, $\tau_\alpha$ , to not hold for this system. For $\phi\gtrsim 0.570$, the van Hove self-correlation function $G_s(r, \tau)$ exhibits a spatial exponential decay at intermediate times, $\tau$, signaling dynamic heterogeneity. The characteristic decay length scales as $\lambda \sim \tau^\nu$, with $\nu \approx 1/3$, in contrast to the conventional square-root scaling. We also investigate temporal heterogeneity through the four-point dynamic susceptibility, $\chi_4(\tau)$, and its peak time, $\tau_4$. These findings provide rigorous benchmark MD data for computational studies of glassy dynamics and establish a reference for testing theoretical models in dense disordered systems.
\end{abstract}

\maketitle

\section{INTRODUCTION}
\label{sec:intro}
Glasses are ubiquitous in nature and play a central role in a wide range of technological applications due to their unique structural, mechanical, and optical properties.~\cite{angell1995,debenedetti2001rev} Examples of glass-forming liquids include colloidal suspensions, such as polymethyl methacrylate (PMMA) particles; metallic glasses, such as zirconium-based alloys; and molecular liquids, such as glycerol.~\cite{hunter2012physics,weeks2017intro,berthier2023mod} As glass-forming fluids approach the glass transition—either through decreasing temperature or increasing density—their dynamics slow dramatically by orders of magnitude, while their structural properties change comparatively little.~\cite{allen2017book} This decoupling between dynamical and structural behavior is one of the defining characteristics of glass-forming systems and has motivated extensive theoretical, experimental, and computational efforts to elucidate the microscopic mechanisms underlying the dramatic dynamical slowdown near the glass transition.

Among the model systems commonly employed to investigate glassy dynamics is the binary hard-sphere (BHS) mixture, which provides a minimal yet versatile framework for studying slow dynamics.~\cite{hunter2012physics,weeks2017intro,zhang2023effects} The model isolates the effects of excluded-volume interactions, making it one of the simplest and most fundamental architectures for exploring glassy phenomena. Moreover, by appropriately choosing the composition and particle-size ratio, crystallization can be suppressed, enabling the study of dense states at equilibrium without the rapid compression required for monodisperse fluids.~\cite{royall2024colloidal}

Despite their geometric simplicity, dense BHS fluids exhibit a rich range of dynamical behavior, including particle caging, non-Gaussian displacement statistics, hopping, and the breakdown of the hydrodynamic Stokes--Einstein relation. Because BHS mixtures capture many essential features of concentrated disordered systems—such as colloidal suspensions,\cite{royall2013} granular materials,\cite{srivastava2021} and metallic glasses\cite{angell2000relaxation,zhang2014connection}—they have been extensively investigated using both experimental techniques and event-driven MD simulations.~\cite{weeks2011s,weeks2017intro} The slowing dynamics in these systems is characterized by a pronounced non-Arrhenius increase in shear viscosity, $\eta$,\cite{puertas2007visco,mittal2009u,charbonneau2013dim} and a corresponding decrease in self-diffusivity, $D$,\cite{foffi2003mixing,mittal2009u,flenner2011analysis,charbonneau2013dim,charbonneau2013decor} as the fluid packing fraction, $\phi$, increases toward the glassy regime. This behavior is accompanied by a breakdown of the Stokes--Einstein relation, which predicts a constant $D\eta$ product for homogeneous dynamics, reflecting a heterogeneous nature of the dynamics.~\cite{puertas2007visco,charbonneau2013dim,royall2024colloidal}

Transport coefficients from equilibrium simulations can be obtained using the Green--Kubo\cite{green1954,kubo1957} or Einstein--Helfand\cite{Helfand1960} formulations. Although mathematically equivalent,\cite{hess2002d,puertas2007visco,charbonneau2013dim,nieszporek2016c,jamali2019,mangaud2020sampling,celebi2021finite,li2023md2d,malaspina2023,mabillard2023poles} the Green--Kubo method is inapplicable to systems with discontinuous potentials, such as hard spheres, because the stress autocorrelation function diverges singularly at zero time.\cite{garcia2006t,bannerman2009t,allen2017book,heyes2022bulk,pieprzyk2024r} The Einstein--Helfand approach, by contrast, relies on the mean square Helfand moment (MSH) and does not suffer from this limitation, providing a rigorous and appropriate framework for evaluating viscosity in hard-sphere architectures.\cite{pieprzyk2019t,puertas2007visco,charbonneau2013dim,nieszporek2016c,moustafa2024}

Despite extensive studies of transport in hard-sphere systems, a comprehensive study that systematically unifies macroscopic transport coefficients ($\eta$, $D$) with microscopic two-point and four-point density correlations within a common simulation framework remains scarce. Much of the existing literature focuses on monodisperse fluids\cite{yeh2004system,krekelberg2007short,pieprzyk2019t} or limits its scope to self-diffusivity,\cite{flenner2010dynamic,flenner2011analysis,charbonneau2013dim} leaving systematic assessments of shear viscosity in dense BHS mixtures severely limited.\cite{mittal2009u,bannerman2009t,puertas2007visco,charbonneau2013dim} 

Finite-size effects represent a critical but frequently overlooked source of uncertainty in glassy systems. While hydrodynamic arguments traditionally dictate an $N^{-1/3}$ scaling for diffusivity corrections in ordinary fluids,\cite{yeh2004system} whether and to what extent these relations hold in dense, non-continuous BHS mixtures approaching the glassy regime has not been systematically established.\cite{charbonneau2013dim} Furthermore, system-size dependencies in the shear viscosity and structural relaxation times of slowly relaxing BHS fluids remain poorly quantified,\cite{pieprzyk2019t} making the extraction of reliable thermodynamic-limit ($N\to\infty$) values a major challenge.

Moreover, there is a critical need to rigorously test the validity of commonly assumed dynamical scalings in hard-sphere fluids. In computational studies, it is common practice to substitute the structurally demanding calculation of shear viscosity with the structural relaxation time, $\tau_\alpha$, under the assumed proportionality $\eta \propto \tau_\alpha$.\cite{bannerman2009t,schober2016h} While this relationship has been shown to hold in specific continuous-potential models,\cite{yamamoto1998,sengupta2013,kawasaki2017,kawasaki2014d} a systematic, thermodynamic-limit verification of this scaling—alongside the fractional Stokes--Einstein relation ($D \eta^c = \text{const.}$)—remains absent for dense BHS mixtures, especially since evidence exists that such couplings can fail.\cite{shi2013r,flenner2019v} 

Additionally, at high densities, the emergence of dynamic heterogeneity manifests spatially as a distinct exponential tail in the van Hove self-correlation function, $G_{\rm s}(r, \tau)$, which is widely attributed to particle hopping events.\cite{chaudhuri2007u} Although this tail length, $\lambda$, is conventionally assumed to follow a standard square-root time scaling ($\lambda \sim \tau^{1/2}$), the exact nature of this spatial decay length near the glass transition requires high-precision verification.

To resolve these questions, we present an extensive equilibrium event-driven MD study of a 50:50 BHS fluid over a wide range of packing fractions ($0.357\leq\phi\leq0.582$) approaching the glassy regime. Using an efficient on-the-fly, binary-based implementation of the Einstein--Helfand formulation combined with Generalized Least Squares (GLS) fitting,\cite{moustafa2024} we systematically characterize finite-size effects in diffusivity and viscosity and extrapolate these properties, together with the structural relaxation time, to the thermodynamic limit ($N\to\infty$). We then examine the density dependence of $D_\infty$ and $\eta_\infty$, assess the reliability of empirical models for the critical packing fraction $\phi_0$, and test the conventional and fractional Stokes--Einstein relations and the assumed proportionality $\eta\propto\tau_\alpha$. Finally, we investigate dynamic heterogeneity using the non-Gaussian parameter $\alpha_2(\tau)$,\cite{kob1997d} the four-point susceptibility $\chi_4(\tau)$,\cite{glotzer2000time} and the spatial decay of the van Hove self-correlation function, $G_{\rm s}(r,\tau)$. We find that the characteristic exponential-tail length follows a sub-diffusive scaling, $\lambda\sim\tau^{1/3}$, rather than the conventional $\tau^{1/2}$ dependence. Collectively, these results provide a unified thermodynamic-limit dataset for assessing transport, relaxation, and dynamic heterogeneity in dense BHS fluids.

The structure of this paper is organized as follows. In Section~\ref{sec:methods}, we describe the technical details of the Einstein--Helfand transport formulation, the definitions of the two-point and four-point correlation functions, and the parameters governing our event-driven MD simulations. Section~\ref{sec:results} presents our primary results and discussion, which is partitioned into calculations of macroscopic transport coefficients, their finite-size scaling behavior, the evaluation of empirical glass transition models, the breakdown of the Stokes--Einstein and $\eta \propto \tau_\alpha$ relationships, and an in-depth assessment of spatiotemporal dynamic heterogeneity. Finally, Section~\ref{sec:conclusions} summarizes our core conclusions and suggests avenues for future research.

\section{METHODS}
\label{sec:methods}
\subsection{Transport coefficients}
\subsubsection*{Einstein–Helfand formulation}
According to the Einstein–Helfand (also known as the generalized Einstein) formulation, the time-dependent self-diffusivity, $D_s$, and the shear viscosity, $\eta$, of a fluid occupying a volume $V$ in $d$ dimensions at temperature $T$ are given by:\cite{Helfand1960,Alder1970,hess2002d}
\begin{subequations} 
\label{eq:DV_def}
\begin{align}
\label{eq:D_t}
D_s\left(\tau\right) = \frac{{\rm MSD}_s\left(\tau\right)}{2d \tau} 
&= 
 \frac{1}{2 d \tau N_s} \sum_{i=1}^{N_s} \left< \left[ {\bf r}_i\left(\tau+t_0\right) - {\bf r}_i\left(t_0\right) \right]^2 \right>_{t_0}, \\
 \label{eq:V_t}
\eta\left(\tau\right) = \frac{{\rm MSH}\left(\tau\right)}{2 n \tau}  &= 
\frac{V}{2 n \tau k_{\rm B}T} \sum_{\alpha < \beta}  \left<   \left[ \int_{t_0}^{t_0+\tau}  P_{\alpha\beta}\left(t\right) {\rm d}t  \right]^2  \right>_{t_0},
\end{align}
\end{subequations}
where $\tau$ is the time interval, MSD\textsubscript{$s$} is the mean square displacement of species $s$, MSH is the mean square Helfand moment, ${\bf r}_i(t)$ is the position of particle $i$ at time $t$, $N_s$ is the number of particles of species $s$, and $k_{\rm B}$ is the Boltzmann constant. The notation $\langle \dots \rangle_{t_0}$ denotes an average over the sampled time origins $t_0$. For simplicity, hereafter, we omit the subscript $t_0$ and set the time origin $t_0=0$. Here, $n=d(d-1)/2$ is the number of independent off-diagonal components of the instantaneous pressure tensor $\mathbf{P}$, where $\alpha,\beta=1,2,\dots,d$. Only the independent components are included because the pressure tensor is symmetric, $P_{\alpha\beta}=P_{\beta\alpha}$. 

In the long-time diffusive regime, both MSD$_s(\tau)$ and MSH$(\tau)$ increase linearly with $\tau$. Consequently, $D_s(\tau)$ and $\eta(\tau)$ converge to time-independent values corresponding to the actual transport coefficients $D_s$ and $\eta$, respectively.\cite{haile_book} Note that, for binary mixtures, the total self-diffusivity is given by $D = x_A D_A + x_B D_B$, where $x_s$ denote the mole fraction of species $s$. This average diffusivity corresponds to Darken's ideal-mixture approximation for the mutual diffusivity, $D_{AB}$.

For the shear viscosity calculations, we employ the pressure tensor integration formulation of the Helfand moment because it is compatible with periodic boundary conditions. \cite{Alder1970,allen2017book} For the hard-sphere model, the pressure tensor is evaluated as a time average over the event-driven MD time step, $\Delta t$, as\cite{alder1960s,haile_book}
\begin{equation}
\label{eq:Pavg}
\bar {\mathbf P}(t; \Delta t) = \frac{1}{V\Delta t} \sum_{\rm collisions} \left[ \tau_c \sum_i m_i{\bf v}_i\otimes{\bf v}_i - \sum_{i<j} 2\mu_{ij}  \frac{b_{ij}}{r_{ij}^{2}} {\bf r}_{ij}\otimes{\bf r}_{ij} \right].
\end{equation}
The integral in Eq.~\ref{eq:V_t} is then evaluated by accumulating these time-averaged pressure-tensor contributions over successive event-driven MD time steps \cite{moustafa2024}. Here, $m_i$ and ${\bf v}_i$ are the mass and velocity of particle $i$, respectively, and $\tau_{\rm c}$ is the time elapsed since the preceding collision event. The reduced mass and separation vector between particles $i$ and $j$ are defined as $\mu_{ij}=m_im_j/(m_i+m_j)$ and ${\bf r}_{ij}={\bf r}_i-{\bf r}_j$, respectively, where $r_{ij}=|{\bf r}_{ij}|$. In our case, $\mu_{ij}=m/2$ since all masses are equal $m_i=m_j=m$. The quantity $b_{ij}$ is the radial relative-velocity projection at collision, $b_{ij}={\bf v}_{ij}\cdot{\bf r}_{ij}$, with ${\bf v}_{ij}={\bf v}_i-{\bf v}_j$. Note that $b_{ij}<0$ because particles involved in a collision are approaching each other immediately before contact. The outer summation is taken over all collision events occurring during the interval $\Delta t$. This expression preserves the symmetry of the pressure tensor, $P_{\alpha\beta}=P_{\beta\alpha}$, as expected.

\label{sec:fit_gamma_t}
The transport coefficients are extracted from the time-dependent data (Eq.~\ref{eq:DV_def}) using a recently proposed fitting procedure based on a power-law function,\cite{moustafa2024} 
\begin{eqnarray}
    \label{eq:gamma_fit}
    \gamma(\tau) = a \left(1 -  \frac{b}{\tau^c}\right),
\end{eqnarray}
where $\gamma(\tau)$ represents either $D_s(\tau)$ or $\eta(\tau)$, and $a$, $b$, and $c$ are fitting parameters. Here, $a$ corresponds to the infinite-time limiting value (plateau) of the respective transport coefficient ($D_s$ or $\eta$). The parameter $b$ is always positive for viscosity, whereas for diffusivity it depends on the density: it is positive at low densities and negative at high densities, as discussed below. The exponent $c$ depends on both the density and the specific transport property, but is typically close to unity.

The fitting is performed using the GLS method, which explicitly accounts for correlations between data points at different time intervals. This is important because the statistical fluctuations of $\gamma(\tau)$ at different $\tau$ are not independent,\cite{moustafa2024} and ignoring these correlations can lead to biased estimates of the infinite-time limit $a$ and associated uncertainty. Moreover, the long-time fitting domain is determined automatically by the algorithm to maximize the agreement between the data and the model function in Eq.~\eqref{eq:gamma_fit}, ensuring a robust estimation of the limiting values. Further technical details and validation of the procedure can be found in our on-the-fly implementation,\cite{moustafa2024} and in the recently introduced postprocessing implementation in the \texttt{MDTransport} Python package.\cite{verma2026md}

\subsection{Two-point density correlations}  
\label{sec:2pt_corr}
The van Hove self-correlation function, $G_{\rm s}({\bf r},\tau)$, quantifies correlations in the microscopic self-density $\rho_{\rm s}({\bf r},\tau) = \delta({\bf r} - {\bf r}_{\rm s}(\tau))$ at two points separated by a displacement ${\bf r}$ and a time interval $\tau$ \cite{HansenBook2013}:
\begin{eqnarray}
\label{eq:Gs_def}
G_{\rm s}\left({\bf r},\tau\right) 
&\equiv&  \frac{\left\langle \rho_{s}({\bf r},\tau)\, \rho_{s}({\bf 0},0) \right\rangle}{\rho_{\rm s}} \nonumber \\
&=& \frac{1}{\rho_{\rm s} }  \frac{1}{N_{\rm s}}\sum_{i=1}^{N_{\rm s}} \frac{1}{V}\int \left< \rho_{i}\left({\bf r}+{\bf r}',\tau\right) \rho_{i}\left({\bf r}',0\right) \right> {\rm d} {\bf r}'  \nonumber \\
&=& \frac{1}{N_{\rm s}}  \sum_{i=1}^{N_{\rm s}} \left< \delta\left({\bf r} - \left[ {\bf r}_i\left(\tau\right) - {\bf r}_i\left(0\right) \right] \right)  \right>,
\end{eqnarray}
where $\delta(x)$ is the Dirac delta function and $\rho_{\rm s} \equiv \langle \rho_{\rm s}({\bf r},t) \rangle = 1/V$ is a normalization constant ensuring that $
\int G_{\rm s}({\bf r},\tau)\, {\rm d}{\bf r} = 1 $. Here, the sum averages over all particles of the same species, and the spatial integral averages over the volume $V$. In a homogeneous fluid, $G_{\rm s}({\bf r},\tau)$ depends only on the magnitude of the displacement, $r$, reducing the vector expression to its radial form $G_{\rm s}(r,\tau)$. In this case, the radial probability density function of finding a particle at a distance $r$ from its initial position after a time interval $\tau$ is given as
\begin{eqnarray}
\label{eq:Ps}
P_s(r, \tau) \equiv 4\pi r^2 G_{\rm s}\left(r,\tau\right) 
= \frac{1}{N_{\rm s}}  \sum_{i=1}^{N_{\rm s}} \left< \delta\left(r - \left| {\bf r}_i\left(\tau\right) - {\bf r}_i\left(0\right) \right| \right)  \right>.
\end{eqnarray}
such that $\int_0^\infty P_s(r, \tau) {\rm d}r = 1$. This function was determined numerically using the standard histogram technique, where the Dirac delta function is replaced by a rectangular function.

The spatial-dependence of $G_{\rm s}(r,\tau)$ is Gaussian at both short and long time limits, reflecting the ballistic (Maxwellian) and diffusive (Brownian) nature of particle motion, respectively,
\begin{align}
\label{eq:Gs_gauss}
    G^{\rm g}_{\rm s}(r,\tau) = \left(\frac{3}{2 \pi \langle r^2(\tau)\rangle}\right)^{3/2} \exp\left(-\frac{3 r^2}{2 \langle r^2(\tau)\rangle}\right),
\end{align}
where $\langle r^2(\tau)\rangle$ is the MSD as given by Eq.~\ref{eq:D_t}. The long-time Gaussian behavior can also be interpreted as a consequence of the central limit theorem. The deviation from this behavior is often quantified using the non-Gaussian parameter $\alpha_{2,{\rm s}}(\tau)$, which is defined as
\begin{eqnarray}
\label{eq:alpha2}
\alpha_{2,{\rm s}}\left(\tau\right) \equiv \frac{d}{d+2} \frac{\left<\frac{1}{N_s}\sum_{i=1}^{N_{\rm s}} \left|{\bf r}_i\left(\tau\right) - {\bf r}_i\left(0\right)\right|^4 \right>}{\left< \frac{1}{N_s}\sum_{i=1}^{N_{\rm s}} \left|{\bf r}_i\left(\tau\right) - {\bf r}_i\left(0\right)\right|^2 \right>^2} - 1,
\end{eqnarray}
where $d$ is the dimensionality. In fact, the excess kurtosis of the displacement distribution is given by  $\left(d+2\right) \alpha_{2,{\rm s}}\left(\tau\right)$; hence, the non-Gaussian parameter provides a measure of the distribution’s “tailedness.”  It is important to point out that, since the sum of two different Gaussian distributions is not itself Gaussian, the non-Gaussian parameter should be evaluated separately for each particle species rather than for the system as a whole as sometimes is done. Since particle motion is Gaussian at both short and long times, $\alpha_{2,{\rm s}}(\tau)$ vanishes in these limits, reaching a maximum at an intermediate timescale, denoted $\tau_{\rm s}^*$. This timescale is widely regarded as a characteristic timescale of dynamic heterogeneity, at which the coexistence of relatively slow and fast particles is most pronounced. We estimate $\tau_{\rm s}^*$ by performing a quadratic fit to five data points: the maximum and the two points immediately before and after it. Hence, $\tau_{\rm s}^*$ is then taken as the time corresponding to the maximum of the fitted quadratic function.

It is worth noting that this quantification is limited to the fourth-order moments of $G_{\rm s}$. Alternative measures, such as the statistical distance $D_{\rm KL}$ between the true and Gaussian distributions, do not suffer from this limitation. Nevertheless, in practice, the resulting characterization is often found to differ only marginally from that provided by the conventional non-Gaussian parameter.\cite{dandekar2020n}

The single-particle dynamic relaxation is often determined, both experimentally and computationally, through the incoherent intermediate scattering function, $F_{\rm s}\left({\bf q},\tau\right)$, which is defined as the spatial Fourier transform of $G_{s}({\bf r},\tau)$, Eq.~\ref{eq:Gs_def}:
\begin{align}
F_{\rm s}\left({\bf q},\tau\right) &= \frac{1}{N_{\rm s}} \sum_{i=1}^{N_{\rm s}} 
 \left<\rho_{i}\left({\bf q},\tau\right) \rho_{i}\left(-{\bf q},0\right)\right> \nonumber \\
&= \frac{1}{N_{\rm s}}  \sum_{i=1}^{N_{\rm s}} \left< e^{i {\bf q}\cdot \left( {\bf r}_i \left(\tau\right) - {\bf r}_i\left(0\right) \right) } \right>
\end{align}
where $\bf q$ is the wave vector and $\rho_{\rm s}\left({\bf q}, \tau\right) \equiv \exp\left(-i {\bf q} \cdot {\bf r}_{\rm s}\left(\tau\right)\right)$ is the Fourier transform of the self particle density.~\cite{HansenBook2013} For homogeneous fluids, the mean displacement is statistically zero, such that the imaginary contribution vanishes, $\left\langle \sin\!\left({\bf q}\cdot[{\bf r}_i(\tau) - {\bf r}_i(0)]\right)\right\rangle = 0$, and only the real (cosine) component remains. The magnitude of the wave vector is chosen to correspond to the first peak of the static structure factor ($q = 7.0$), which roughly matches the wavelength associated with the first-nearest-neighbor separation. At this length scale, $F_{\rm s}(q,\tau)$ directly probes structural relaxation of particles. We note, however, that nearby choices of $q$ yield qualitatively similar results. We define the structural relaxation time, $\tau_\alpha$, as the time at which $F_{\rm s}(q,\tau)$ decays to $1/e$. We estimate $\tau_\alpha$ by locating the two sampled times at which $F_{\rm s}(q,\tau)$ lies immediately above and below $1/e$ and linearly interpolating between them to determine the value of $\tau$ satisfying $F_{\rm s}(q,\tau_\alpha) = 1/e$.

For purely diffusive (Brownian) dynamics, $F_{\rm s}(q,\tau)$ decays exponentially as $\exp(-q^{2} D_{\rm s} \tau)$, with a characteristic decay time of $1/(q^{2} D_{\rm s})$. In real fluids, however, the decay is slower than a simple exponential, and the coupling between the structural relaxation time $\tau_\alpha$ and the self-diffusivity breaks down. According to mode-coupling theory, for example, the decay follows a stretched-exponential form, $\exp[-(\tau/\tau_c)^{\beta}]$, where $\beta$ and $\tau_c$ are constants.

\subsection{Four-point density correlation}
While two-point density correlation functions capture average relaxation, they cannot quantify the spatially correlated motion that underlies dynamic heterogeneity. To probe these correlated fluctuations directly, higher-order measures are required, most notably four-point density correlations.

The four-point correlation function $G_4(\mathbf{r},\tau)$ quantifies spatiotemporal correlations of local immobility fluctuations and is defined as
\begin{align}
G_4(\mathbf{r}, \tau) &\equiv \left\langle \delta q(\mathbf{0}, \tau)\, \delta q(\mathbf{r}, \tau) \right\rangle \nonumber \\
&= \frac{1}{V} \int \mathrm{d}\mathbf{r}' \, \left\langle \delta q(\mathbf{r}', \tau)\, \delta q(\mathbf{r}'+\mathbf{r}, \tau) \right\rangle ,
\end{align}
where $q(\mathbf{r},\tau) = \sum_{i} \rho_{i}(\mathbf{r},0)\,\rho_{i}(\mathbf{r},\tau)$ represents the \emph{local overlap} field of immobile particles and $
\rho_i({\bf r},\tau) = \delta({\bf r} - {\bf r}_i(\tau))$ is the usual microscopic density. The corresponding fluctuation field is defined as $\delta q(\mathbf{r},\tau) = q(\mathbf{r},\tau) - \left\langle q(\mathbf{r}, \tau)\right\rangle_V$, where the spatial average $\left\langle q(\mathbf{r}, \tau)\right\rangle$ is
\begin{align}
\label{eq:Q_def}
\left\langle q(\mathbf{r}, \tau) \right\rangle_V &\equiv \frac{1}{V} \int \mathrm{d}\mathbf{r}'\, \left\langle q(\mathbf{r}', \tau) \right\rangle \nonumber \\
&= \frac{1}{V} \left\langle \sum_{i=1}^N \delta\!\left(\mathbf{r}_i(\tau) - \mathbf{r}_i(0) \right) \right\rangle
\equiv \frac{\left\langle Q(\tau)  \right\rangle}{V} .
\end{align}
and $Q(\tau)$ is defined as the \textit{global} overlap function. As mentioned earlier, the Dirac delta function is replaced by a rectangular function $w\left(x\right)$ of width $l$,
\begin{eqnarray}
\label{eq:Qt}
Q\left(\tau\right) =  \sum_{i=1}^{N} w\left( \left|{\bf r}_i(\tau) - {\bf r}_i(0)\right|\right).
\end{eqnarray}
Accordingly, $Q$ represents the number of immobile particles, which are then defined as the particles which stay in a volume $V_a=\frac{4}{3}l^3$. The zero-time limit of $\left<Q\left(\tau\right)\right>/N$ is $1$, while the infinite-time limit is nearly zero as there is always a small random probability for a particle to visit its initial location. Moreover, for a given time interval $\tau$, $Q(\tau)$ fluctuates among different time origins due to the spatial heterogeneity of particle dynamics.

A common measure of dynamic heterogeneity in supercooled fluids is the volume integral of the four-point correlation function, which defines the four-point dynamical susceptibility,
\begin{align}
\label{eq:chi4}
\chi_4(\tau) &\equiv \beta \int \mathrm{d}\mathbf{r}  g_4(\mathbf{r},\tau) \nonumber \\
&= \frac{\beta}{V\rho^2}\left\langle \delta Q(\tau)^2 \right\rangle,
\end{align}
where $\delta Q(\tau)=Q(\tau)-\langle Q(\tau)\rangle$ and $g_4(\mathbf{r},\tau)\equiv G_4(\mathbf{r},\tau)/\rho^2$. At $\tau=0$, this normalization ensures $g_4(\mathbf{r},0)=g(r)$, the static radial distribution function. The spatial integral probes the \textit{range} of correlated immobility, while the variance form captures temporal fluctuations in the global dynamical order parameter. Together, these representations establish $\chi_4(\tau)$ as a quantitative measure of the spatiotemporal heterogeneity of the dynamics. 

In the short-time limit, all particles remain caged, resulting in zero fluctuations in immobility, $\chi_4 = 0$. At long times, particles have escaped their cages, leading to minimal immobility and, consequently, $\chi_4 \approx 0$. At intermediate times, $\chi_4(\tau)$ exhibits a peak at a characteristic timescale $\tau_4$. Similar to the non-Gaussian parameter case, $\tau_4$ is determined through a quadratic fit (see Sec.~\ref{sec:2pt_corr}). In this work, we examine the relationship between $\tau_4$ and the structural relaxation time, $\tau_\alpha$, across different densities.

We also note that, the four-point dynamical susceptibility is also related to the four-point structure factor through $\chi_4(\tau)=\lim_{q\to0} S_4(q,\tau)$, where $S_4(q,\tau)$ is the Fourier transform of $G_4(\mathbf{r},\tau)$. This relation links temporal fluctuations in the dynamics to the spatial correlations of particle mobility. Moreover, analysis of the low-$q$ behavior of $S_4(q,\tau)$ provides a dynamical correlation length, $\xi_4(\tau)$, which characterizes the spatial extent of dynamically correlated regions. Together, $\chi_4(\tau)$ and $\xi_4(\tau)$ quantify the strength and range of dynamic heterogeneity, respectively.

\subsection{Simulation details}
The simulated three-dimensional system consists of a 50:50 BHS mixture with diameters $\sigma_{\rm A}=1$ and $\sigma_{\rm B}=1/1.4$ (or, $5/7$), where $\sigma_{\rm A}$ is taken as the unit of length. This model has been widely used in previous computational studies because it efficiently suppresses crystallization and promotes glass formation.\cite{hern2002random,brambilla2009probing,flenner2014u,callaham2017p} All particles have identical mass, $m=1$. The number density, $\rho\equiv N/V$, is varied from $1.00$ to $1.63$, corresponding to packing fractions $\phi = \frac{\pi}{6} \rho \left( x_A \sigma_{\rm A}^3 + x_B \sigma_{\rm B}^3 \right)$, where $x_i = 1/2$ is the mole fraction of species $i$. For these parameters, the relation simplifies to $\phi \approx 0.357 \,\rho$, so the explored densities correspond to $0.357 \le \phi \le 0.582$. 

Unless stated otherwise, results are reported for systems of size $N = 3200$. Simulations with $N = 200,\ 400,\ 800,$ and $1600$ were additionally performed to assess finite-size effects. All simulations were carried out using event-driven MD, which is particularly suitable for hard-sphere systems because particle trajectories consist of ballistic motion interrupted by instantaneous binary collisions. All simulations were performed at $k_{\rm B}T=1$.

Simulations started from an FCC configuration with A and B spheres randomly assigned to sites.  If this configuration included overlaps, $\sigma_{\rm A}$ started at the maximum value that would avoid overlaps with a strong square shoulder added that extended to 1.0 (so that non-overlapping spheres would stay apart).  The event-driven MD simulation then ran with the configurations checked periodically and $\sigma_{\rm A}$ increased whenever possible until it reached 1.0.  During this part, Monte Carlo moves to swap the identities of A and B spheres were also attempted to accelerate the process of resolving overlaps. Once this initialization finished, $10^7$ steps of equilibration and $10^8$ steps of data collection were run, all with a time step of $\Delta t = 0.01$. 

Nonlinear on-the-fly time intervals are employed using our recently introduced binary-based sampling method \cite{moustafa2024}, in which the intervals are defined as
\begin{align}
\tau_i = 2^i \times \Delta t,
\end{align}
with $i = 0, 1, 2, \dots$ denoting the interval index. Unlike linear or order-$n$ schemes, this approach yields a uniform distribution of data points on a logarithmic time scale, which is the natural scale for describing the long-time behavior of $D(\tau)$ and $\eta(\tau)$. We note that a postprocessing implementation of this approach was recently introduced in the \texttt{MDTransport} Python package to compute transport coefficients.\cite{verma2026md}

The time-dependent transport coefficients in Eqs.~\ref{eq:DV_def} are computed by averaging over consecutive, non-overlapping, intervals. As an example, for an index of $i=1$, the intervals correspond to the following pair of start-end simulation times: $(0,2\Delta t)$, $(2\Delta t, 4\Delta t)$, $(4\Delta t, 6\Delta t)$, $\dots$. The  statistical uncertainty associated with the average is computed using the usual standard deviation of the mean, $\sigma_i/\sqrt{M_i}$, where $\sigma_i$ is the standard deviation of the $M_i$ intervals. This estimate assumes statistically independent samples. As shown previously,\cite{moustafa2024} this approximation is valid for the long-time intervals relevant to extracting the transport coefficients using the Einstein--Helfand formulation.

Both the event-driven MD simulations and the on-the-fly implementation of the binary-based sampling method were carried out within the \texttt{Etomica} simulation package. \cite{schultz2015etomica} The binary-based sampling was integrated directly into the simulation workflow, allowing time-dependent transport coefficients to be accumulated during runtime without the need to store full particle trajectories. This approach significantly reduces memory requirements while ensuring consistent sampling across the wide range of time intervals required for the Einstein--Helfand analysis. The \texttt{Etomica} package is publicly available at \url{https://github.com/etomica/etomica/tree/master/etomica-modules/src/main/java/etomica/modules/glass}.

\section{RESULTS AND DISCUSSION}
\label{sec:results}
\subsection{Transport coefficients}

\subsubsection{Time-dependent transport coefficients}
\label{sec:y_t}
\begin{figure*}
\centering
\includegraphics[width=\textwidth]{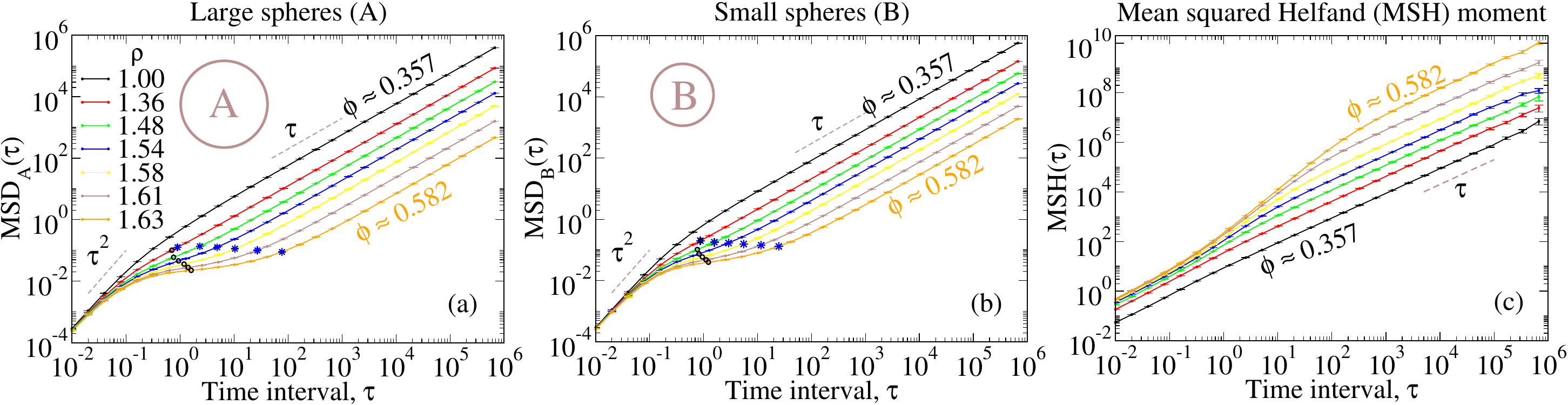}
\caption{Time dependence of the mean square displacement (MSD) for the large (a) and small (b) particles, and mean square Helfand (MSH) moment (c) at seven densities, for a system size of $N=3200$. The ballistic, $\sim\tau^{2}$, and diffusive, $\sim\tau$, regimes are indicated by dashed lines. Error bars correspond to 68\% confidence limits. Lines connect the data points. In panels (a) and (b), black open circles denote the inflection points (see text) and blue stars correspond to the peak of the $\alpha_2(\tau)$ function. (see Fig.~\ref{fig:alpha2}).}
\label{fig:msd_t}
\end{figure*}

We begin by presenting in Fig.~\ref{fig:msd_t} the time-dependent MSD and MSH data on a log–log scale, which are used to evaluate the transport coefficients (Eq.~\ref{eq:DV_def}). The data are reported at seven packing fractions for the largest system-size considered ($N=3200$ particles). In both the large ($\sigma_{\rm A}=1$) and small ($\sigma_{\rm B}=1/1.4$) particles cases, MSD$(\tau)$ exhibits the normal ballistic ($\sim\tau^{2}$) and diffusive ($\sim\tau$) scalings at short and long time intervals, respectively. Because of their smaller size, B particles reach the diffusive regime earlier than particles A and display a larger MSD at any given time interval; hence, $D_{\rm B} > D_{\rm A}$. Moreover, the relative statistical uncertainty for all densities (not shown) scales approximately as $\sqrt{\tau}$, consistent with previous observations.~\cite{moustafa2024}

As the density increases, an inflection point in MSD$(\tau)$, defined by the condition ${\rm d}^2\log{\rm MSD}(\tau)/{\rm d}\log\tau^2=0$, begins to emerge at approximately $\rho \approx 1.36$ ($\phi \approx 0.486$). These points are marked by open black circles in panels (a) and (b) of Fig.~\ref{fig:msd_t}. This plateau-like behavior is characteristic of glassy dynamics and arises from caging effects, in which particles are temporarily trapped by cages formed by their nearest neighbors. We therefore consider this density to represent a rough onset of glassy dynamics. 

The MSD value at the inflection point can be interpreted as the ``cage size,''\cite{weeks2011s,Swol2014m} such that $\sqrt{\rm MSD}$ representing the corresponding cage radius. For both species, the cage size decreases systematically with increasing density, with the cage size of B particles consistently larger than that of A particles. Although the cavity that a B particle occupies is smaller, the cage here confines the particle center; having fewer neighbors around B particles may allow it slightly more freedom to rattle.

Panel (c) shows the time dependence of the MSH associated with shear viscosity. In the long-time limit, the MSH exhibits the same linear (diffusive) scaling with $\tau$ as the MSD, with larger magnitudes at higher densities, reflecting the increase in viscosity. This behavior arises because the Helfand moment associated with viscosity, $\int_{t_0}^{t_0+\tau} P_{\alpha\beta}(s)\,{\rm d}s$ in Eq.~\ref{eq:V_t}, is a stochastic variable that follows Brownian (Wiener) statistics.\cite{moustafa2024} In contrast, at short times, the MSH does not exhibit ballistic $\tau^2$ scaling due to the ill-defined nature of the stress at a point of hard-sphere models.\cite{haile_book} A related distinction is the absence of a peak in $\eta(\tau)$ at all densities, in contrast to $D(\tau)$, which displays a peak at high densities as shown in Fig. S1 of the supplementary material.

\begin{figure}
\centering
\includegraphics[width=0.5\textwidth]{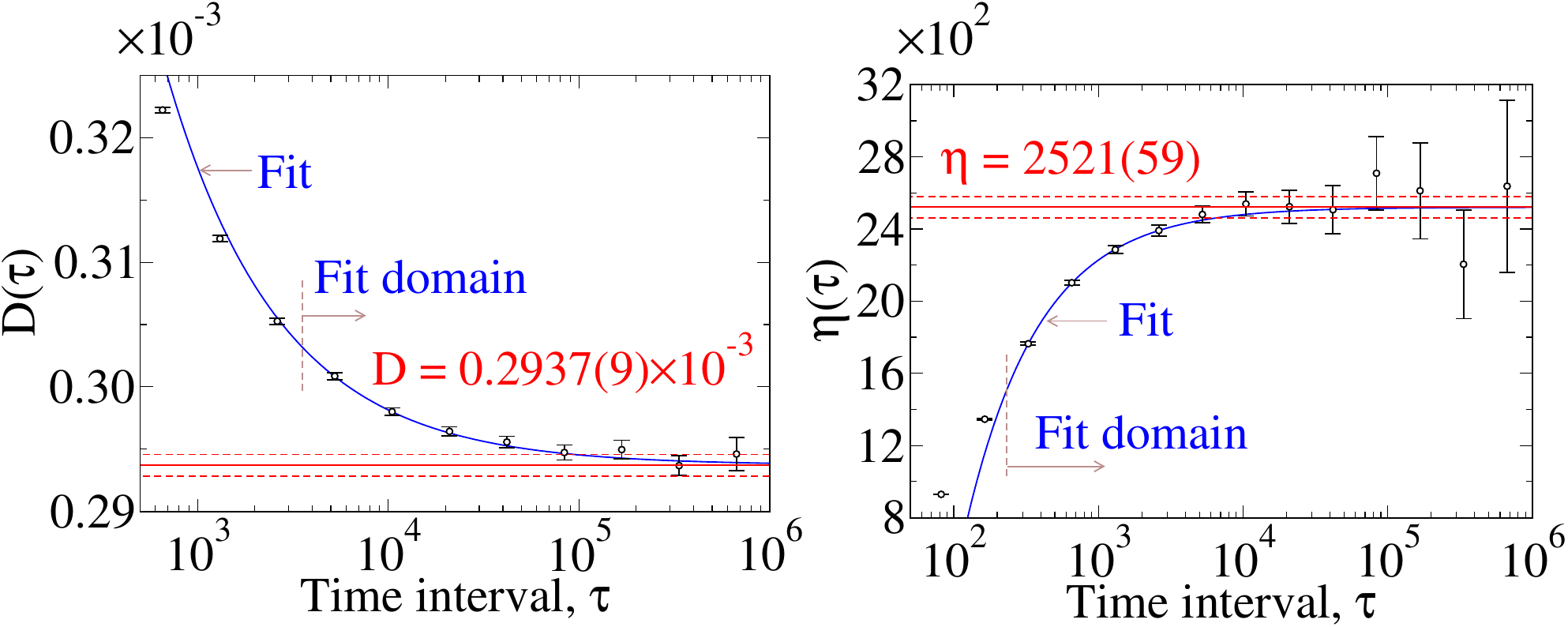}
\caption{Time dependence of the total self-diffusivity (left) and shear viscosity (right) at density $\rho = 1.63$ ($\phi \approx 0.582$) for a system size of $N = 3200$. The blue curves represent fits using Eq.~\ref{eq:gamma_fit}, both yielding reduced $\chi^2$ value less than unity. The red solid lines  indicate the estimated infinite-time limits transport coefficients, with the dashed lines representing the uncertainty bounds obtained from the fits. Numbers in parentheses indicate the uncertainty in the last digit.}
\label{fig:fit_t}
\end{figure}

Given the raw data of MSD$(\tau)$ and MSH$(\tau)$, we compute the corresponding time-dependent transport coefficients, $D(\tau)$ and $\eta(\tau)$, using Eq.~\ref{eq:DV_def}. The infinite-time limit transport coefficients are then estimated by fitting the long-time data using Eq.~\ref{eq:gamma_fit}, as described earlier (Sec.~\ref{sec:fit_gamma_t}). Figure~\ref{fig:fit_t} shows a representative example of this procedure, applied to the total self-diffusivity and shear viscosity at $\rho = 1.63$ ($\phi \approx 0.582$). As noted above, the fitting window is selected automatically to optimally capture the regime in which the data are well described by the power-law fitting form.\cite{moustafa2024} For both coefficients, the fits yield reduced $\chi^2$ values lower than unity, indicating high-quality fits, as also supported by visual inspection. The same fitting protocol is applied consistently across all densities and system sizes considered, yielding similarly good performance. The complete fitting results obtained using the Python script \texttt{fit\_transport\_phi.py}\cite{moustafa2024} are reported in the supplementary material.

\subsubsection{Finite-size effects and thermodynamic-limit}
\label{sec:fse} 

\begin{figure*}
\centering
\includegraphics[width=\textwidth]{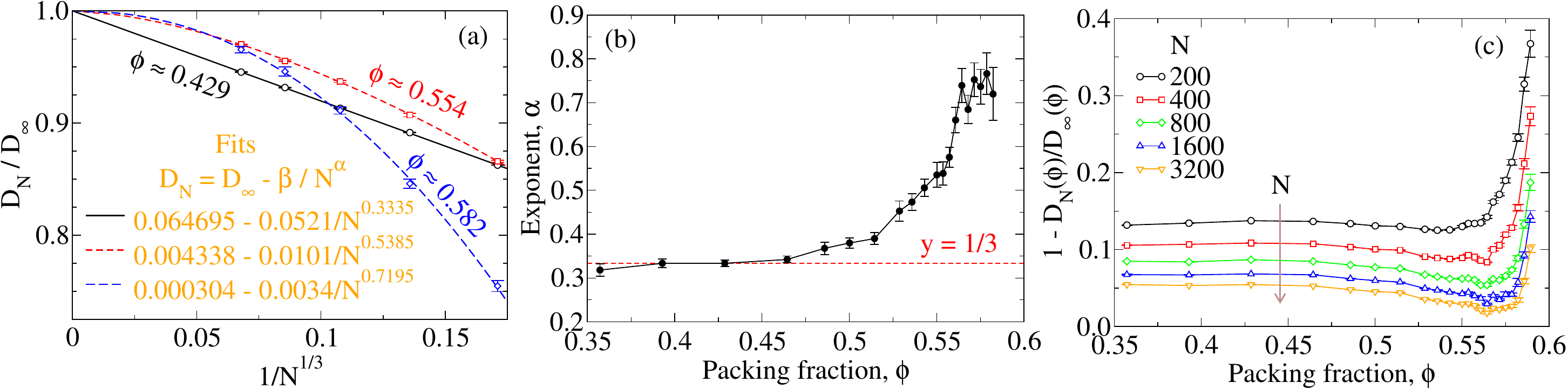}
\caption{(a) Finite-size effects on the total self-diffusivity for system sizes $N = 200, 400, 800, 1600, 3200$ at densities $\rho = 1.20, 1.55, 1.63$. Data are normalized by the thermodynamic-limit, $D_\infty$, for each density. Solid lines are WLS fits to Eq~\ref{eq:D_N_fit}, with reduced $\chi^2$ values of $1.76, 0.40, 0.36$ for the three densities, respectively. Density dependence of the fitted exponent $\alpha$ (b) and the normalized self-diffusivity $D_N/D_\infty$ at different system sizes (c), respectively. Lines connect the data points.}
\label{fig:fse_D}
\end{figure*}
Having extracted the transport coefficients for all system sizes, we next investigate the finite-size effects in order to estimate the thermodynamic-limit ($N \to \infty$) values. 

\subsubsection*{Self-diffusivity}
Figure~\ref{fig:fse_D}(a) shows the system-size dependence of the self-diffusivity, normalized by its thermodynamic-limit value (see Eq.~\ref{eq:D_N_fit}). The data are plotted as a function of $N^{1/3}$, equivalently the box length $L$, which is the natural scaling variable based on hydrodynamic arguments.\cite{yeh2004system} We find that the data are well-described by a power-law form:\cite{heyes2007self}
\begin{align}
\label{eq:D_N_fit}
D(N) = D_{\infty} - \frac{\beta}{N^{\alpha}},
\end{align}
where $D_{\infty}$ is the thermodynamic-limit diffusivity, and $\alpha$ and $\beta$ are positive, density-dependent fitting parameters. The negative sign reflects the increase in diffusivity with system size, resulting from the reduced suppression of long-wavelength fluctuations by the periodic boundary conditions. To estimate $D_{\infty}$, we perform weighted least squares (WLS) fitting, with weights given by the inverse squared statistical uncertainties. The resulting fits along with the fitted parameters are presented in Figure~\ref{fig:fse_D}(a). The power-law form provides an excellent description of the data, as evidenced by the low reduced $\chi^2$ values reported in the figure caption.

Figure~\ref{fig:fse_D}(b) depicts the variation of the exponent $\alpha$ with packing fraction. At low packing fractions, $\alpha$ value is statistically close to $1/3$ (dashed red line), consistent with the theoretical prediction of Yeh and Hummer\cite{yeh2004system} based on hydrodynamic arguments. However, for packing fractions greater than about $0.464$, the exponent increases and then plateaus at around $3/4$ in the high-density region. A qualitatively similar behavior was observed previously by Heyes \textit{et al.}\cite{heyes2007self} for the pure hard-sphere fluid. This agreement is not surprising given the close relationship between the two models.

We next investigate the density dependence of the finite-size effects for a given system size in terms of the relative deviation, $\left[D_{\infty}(\phi)-D_N(\phi)\right]/D_{\infty}(\phi)$. For all five system sizes considered, the relative finite-size effects remain nearly independent of density in the low-density regime. At intermediate densities, a slight decrease is observed, resulting in a shallow minimum, followed by a pronounced increase as the packing fraction enters the high-density regime. This increase indicates a substantial enhancement of the relative finite-size effects at high densities. Importantly, despite this nonmonotonic density dependence, the flexibility afforded by the density-dependent fitting exponent $\alpha$ in Eq.~\ref{eq:D_N_fit} enables an accurate extrapolation of the finite-system diffusivities to their thermodynamic-limit values. 

\begin{figure}
\centering
\includegraphics[width=0.5\textwidth]{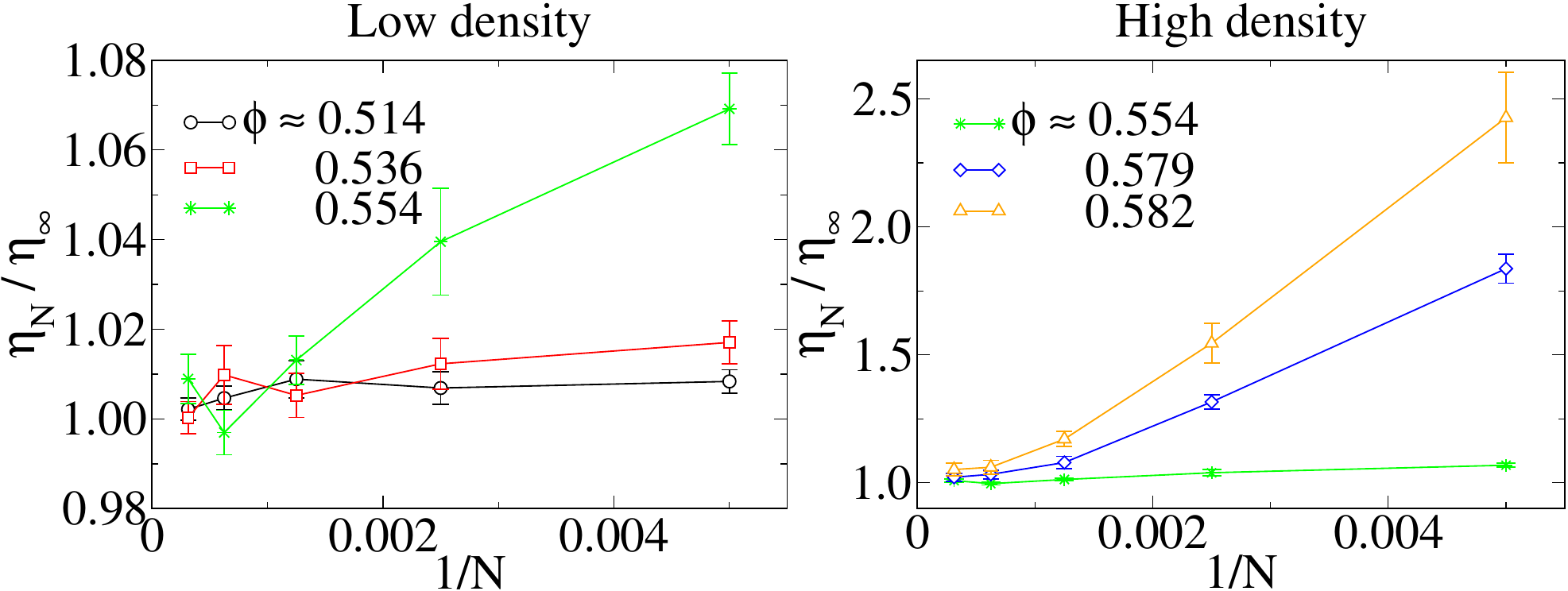}
\caption{Finite-size effects of the shear viscosity for system sizes $N = 200, 400, 800, 1600, 3200$, at relatively low (left; $\rho=1.44, 1.50, 1.55$) and high (right; $\rho=1.55,  1.62, 1.63$)  densities. Data for each density are normalized by the corresponding thermodynamic-limit $\eta_\infty$ obtained from the fits (see text). Lines connect the data points.}
\label{fig:fse_visc}
\end{figure}

\subsubsection*{Shear viscosity}
Figure~\ref{fig:fse_visc} depicts the system-size dependence of the shear viscosity, both at relatively low (left) and high (right) densities. For packing fractions, up to approximately $\phi \approx 0.514$, no statistically significant finite-size effects are detected, which is consistent with previous studies on various systems and thermodynamic states.~\cite{daivis1995transport,yeh2004system,meier2004transport,viscardy2007t,moultos2016system,jamali2018finite} At higher densities, however, finite-size effects become increasingly pronounced—an observation that, to our knowledge, has not been previously reported. In this regime, the shear viscosity systematically decreases with increasing system size, which may be associated with the suppression of long-wavelength fluctuations discussed above for self-diffusivity.

However, the finite-size dependence is more complex than that observed for diffusion, and no single functional form accurately describes the data across the full density range. Although the data appear to converge by $N=3200$, we perform a linear fit in $1/N$ to the three largest system sizes (not shown) in order to obtain a more accurate estimate of the thermodynamic-limit viscosity, $\eta_{\infty}$.

\begin{figure*}
\centering
\includegraphics[width=\textwidth]{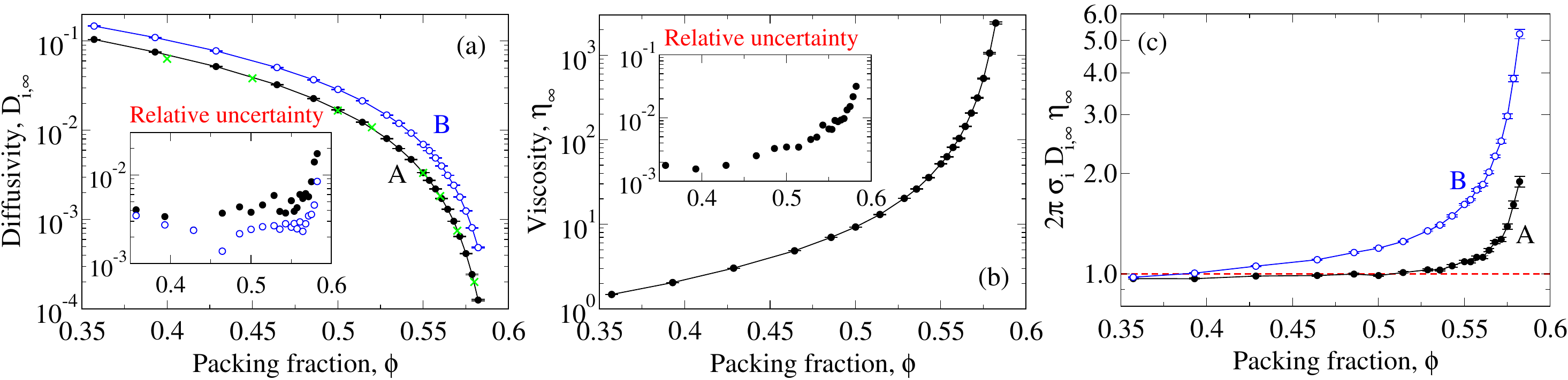}
\caption{Density dependence of the thermodynamic-limit self-diffusivity (a) and shear viscosity (b), with the insets showing the corresponding relative statistical uncertainties. Panel (c): product of the data shown in panels (a) and (b). Lines connect the data points. The green cross markers in panel (a) denote the $D_{\rm A}$ data reported by Charbonneau and Tarjus\cite{charbonneau2013decor} using finite system size.}
\label{fig:DV_phi}
\end{figure*}

\subsubsection{Density dependence data: $D_{\infty}(\phi)$ and $\eta_{\infty}(\phi)$} 
\label{sec:se_relation}
Using the thermodynamic-limit estimates of the self-diffusivity and shear viscosity, their dependence on packing fraction, $D_{\infty}(\phi)$ and $\eta_{\infty}(\phi)$, is shown in Fig.~\ref{fig:DV_phi}(a) and (b), respectively, on a semi-logarithmic scale. At low densities, both transport coefficients exhibit nearly Arrhenius-like behavior, whereas at high densities ($\phi\gtrsim 0.5$) their dependence becomes super-Arrhenius-like, characteristic of the onset of glassy dynamics. Over the density range considered, both quantities span nearly three orders of magnitude. In both cases, the statistical uncertainties, $\sigma$, obtained from the fitting procedure (Sec.~\ref{sec:fse}) are smaller than the symbol sizes. The corresponding relative uncertainties, $\sigma(D_{\infty})/D_{\infty}$ and $\sigma(\eta_{\infty})/\eta_{\infty}$, are shown in the insets on a semi-logarithmic scale. Both quantities  exhibit only a weak $\phi$-dependence at low densities but increase more noticeably at higher densities, reflecting the growing difficulty of obtaining precise estimates of transport coefficients in dense fluids.

To the best of our knowledge, viscosity data for the same model are not available in the literature for direct comparison. However, Charbonneau and Tarjus\cite{charbonneau2012g} reported self-diffusivity data for the large particles (A) in the same model, which are included in Fig.~\ref{fig:DV_phi}(a). Their results are generally consistent with our thermodynamic-limit estimates, with small deviations that may arise from their use of finite-sized systems rather than thermodynamic-limit extrapolations.

A common practice at this point would be to fit the transport coefficient data to an analytical form to describe their high-density dependence and, more importantly, to estimate the critical packing fraction, $\phi_0$. We find that WLS fits using the popular exponential form, $A \exp\left[c_1/(\phi_0-\phi)^{c_2}\right]$,\cite{brambilla2009probing,berthier2009glass,flenner2011analysis} provide good fits to the high-density data. The fitting range was selected such that the resulting reduced $\chi^2$ values are less than unity. Despite the high quality of these empirical fits, we do not consider the resulting estimates of $\phi_0$ to be reliable for two reasons. First, the $\phi_0$ values obtained from fitting $D_\infty(\phi)$ and $\eta_\infty(\phi)$ are not statistically consistent with each other, $0.6045(15)$ and $0.5944(7)$, respectively. This discrepancy suggests that neither value represents the true critical packing fraction, which should be independent of the dynamic property considered. Second, and more importantly, both estimates of $\phi_0$ are substantially smaller than the highest packing fractions accessible in other molecular simulation studies, for example $0.63$\cite{callaham2017p} and $0.65$.\cite{odriozola2011} Moreover, they are also well below the jamming packing fractions, which have been reported to be approximately $0.657$\cite{ozawa2012j}--$0.662$.\cite{berthier2009glass,chaudhuri2010j}

These high-density, non-jammed configurations are generally considered to be out of equilibrium or metastable,\cite{odriozola2011,callaham2017p} because the system cannot readily access new configurations and therefore becomes kinetically ``trapped''. We believe, however, that this apparent limitation is a consequence of the constraints imposed by finite-sized simulations in the NVE/NVT ensembles, which can severely hinder particle rearrangements in highly packed fluids. In principle, increasing the system size should provide access to additional configurations and facilitate equilibration, but this approach quickly becomes computationally prohibitive because both the system size and the simulation time must increase substantially. An alternative is to perform simulations in the NPT ensemble, which allows volume fluctuations and can therefore facilitate transitions between otherwise trapped configurations, even for finite-sized systems. However, measuring dynamical properties in the NPT ensemble introduces additional complications associated with thermostat effects, which warrant a separate investigation in future work.

Despite these difficulties, the maximum packing fraction considered here ($0.582$), as well as those reached in other studies,\cite{odriozola2011,callaham2017p} for which thermalized configurations are obtained, is substantially lower than the expected value of the true $\phi_0$, which should be at least $0.662$. Consequently, estimating $\phi_0$ from the present data necessarily involves substantial extrapolation beyond the range of packing fractions over which equilibrium transport coefficients can be reliably obtained. This extensive extrapolation may, therefore, contribute both to the systematically low values of $\phi_0$ obtained from the two transport coefficients and to their internal inconsistency. A similar extrapolation issue is evident in other studies\cite{berthier2009glass,flenner2010dynamic,flenner2011analysis,odriozola2011} that reached somewhat higher packing fractions than considered here but nevertheless obtained $\phi_0=0.635$, which remains relatively small.

\subsubsection{Stokes--Einstein relation}
Although arising from distinct transport mechanisms, the self-diffusivity and shear viscosity are often related through the Stokes--Einstein relation,
\begin{equation} 
\label{eq:se} 
D_i \, \eta = \frac{k_{\rm B} T}{c \pi \sigma^{\rm h}_i},
\end{equation}
where $\sigma^{\rm h}_i$ denotes the particle hydrodynamic diameter of species $i$ and $c$ is a constant that depends on the assumed boundary condition, taking values of $2$ and $3$ for slip and stick limits, respectively. This relation holds for large Brownian particles in a solvent, such as colloidal suspensions, and provides a good approximation for homogeneous liquids away from the deeply supercooled regime. However, in dense mixtures, particle motion becomes increasingly collective as density increases, effectively reducing the hydrodynamic diameter $\sigma^{\rm h}_i$.~\cite{allen2017book} Consequently, the shear viscosity is expected to increase more rapidly than the diffusivity decreases, leading to a breakdown of the Stokes--Einstein relation—a phenomenon commonly regarded as a hallmark of emerging dynamical heterogeneity.

Figure~\ref{fig:DV_phi}(c) presents the density dependence of the product $2\pi\sigma_i D_i \eta$, which should equal unity if the Stokes--Einstein hydrodynamic relation holds, given the slip boundary condition ($c=2$). The product for the small particles (B) exhibits a systematic increase with density, while that for the large particles (A) remains close to unity at $\phi\lesssim 0.51$ before increasing at higher densities. The validity of the Stokes--Einstein relation for the large particles in the low-density domain can be attributed to their relatively large size compared with the surrounding small particles. In this regime, the small particles can more readily rearrange within the interstitial spaces between the large particles, facilitating their relative motion and effectively acting as a lubricant.\cite{weeks2011s} This behavior is analogous to that of a Brownian particle immersed in a ``sea'' of smaller molecules.

At higher densities, however, the Stokes--Einstein relation breaks down for both particle sizes as the system enters the glassy regime, where particle motion becomes increasingly constrained by the surrounding particles and the dynamics depart from the hydrodynamic regime. This breakdown is qualitatively consistent with previous studies for pure\cite{heyes2007sys,charbonneau2013dim,pieprzyk2019t}, binary\cite{puertas2007visco,charbonneau2013dim}, and polydisperse\cite{kumar2006nature} hard-sphere fluids. 

\begin{figure}
\centering
\includegraphics[width=0.5\textwidth]{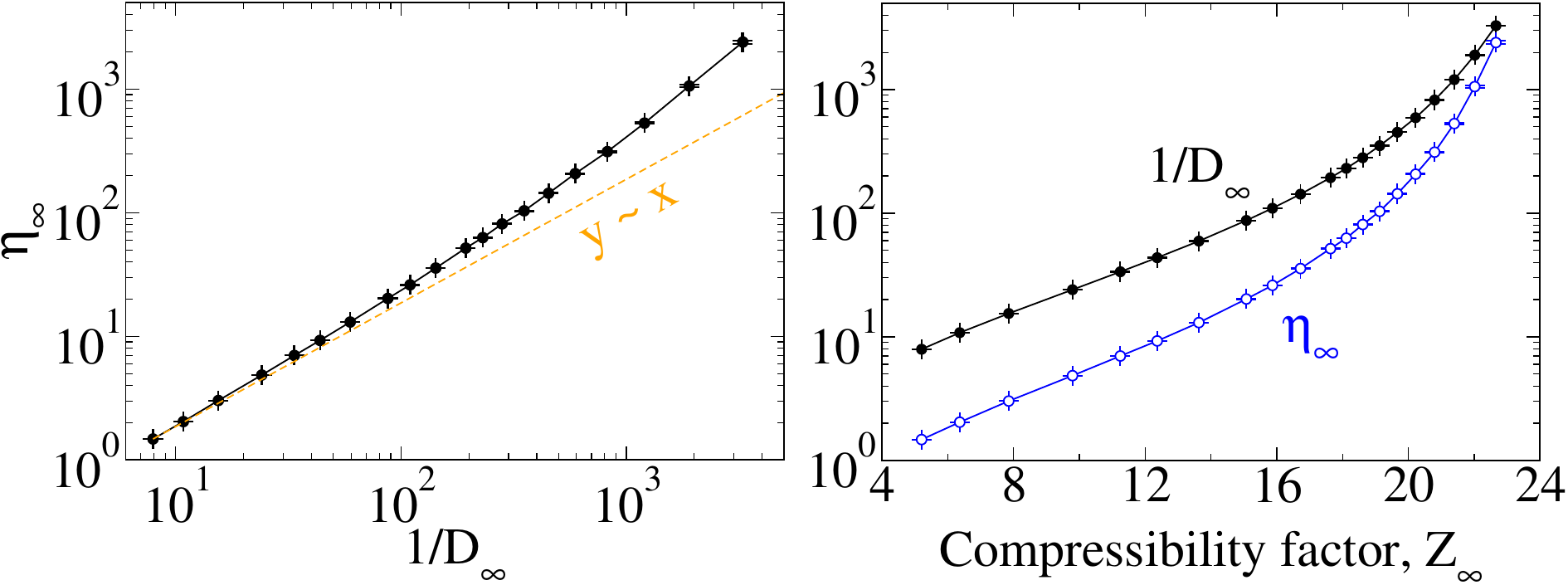}
\caption{Left: scaling of the shear viscosity with the total self-diffusivity on a log-log plot. Right: scaling of both transport coefficients with the compressibility factor on a semi-log plot. All properties are extrapolated to the thermodynamic limit. Lines connect the data points.}
\label{fig:ZDV}
\end{figure}

To investigate the direct relation between the transport coefficients, we plot $\eta_\infty$ versus $1/D_\infty$ (total) in Fig.~\ref{fig:ZDV} (left) on a logarithmic scale. Only in the low-density limit do the results approximately follow the Stokes--Einstein hydrodynamic scaling, $\eta_\infty \sim 1/D_\infty$, as indicated by the orange dashed guideline. However, over the full density range, the data exhibit systematic curvature, with the viscosity increasing more rapidly than the self-diffusivity decreases. Thus, even the ``fractional'' Stokes--Einstein relation, $D_\infty\eta_\infty^{c}=\mathrm{const.}$, often invoked to characterize deviations from the conventional relation, is not supported by our data. This breakdown is consistent with the behavior shown in Fig.~\ref{fig:DV_phi}(c).

\subsubsection{Dynamic--static scaling}
From a statistical thermodynamic point of view, both static (e.g., pressure) and dynamic properties should exhibit singular behavior at the same critical packing fraction, $\phi_0$, although their rates of divergence need not be the same. In fact, one of the fascinating features of glassy dynamics is that the system slows down much more rapidly than the corresponding static properties change as the glass transition is approached.~\cite{janssen2018m} To quantify this contrast, Fig.~\ref{fig:ZDV}(b) shows $1/D_\infty$ and $\eta_\infty$ data on a logarithmic scale as functions of the compressibility factor, $Z_{\infty}=P_{\infty}/\left(\rho k_{\rm B}T\right)$, shown on a linear scale. Here, the thermodynamic-limit pressure, $P_{\infty}$, is obtained by extrapolating the finite-size data using a linear fit to the three largest system sizes; see Sec. Sec. S1 E of the supplementary material. Over the density range considered, $Z_\infty$ increases by a factor of approximately $5$, whereas both $1/D_\infty$ and $\eta_\infty$ increase by nearly two orders of magnitude. 

A qualitatively similar behavior was also reported by Berthier and Witten\cite{berthier2009glass} and by Charbonneau and Tarjus,\cite{charbonneau2013decor} although those studies considered the relaxation time rather than the viscosity. A similar trend was also observed by Mittal,\cite{mittal2009u} albeit for a different diameter ratio, $\sigma_A/\sigma_B=4/3$. This pronounced disparity demonstrates that the dynamic transport properties are substantially more sensitive to increasing density than the static thermodynamic response. The absence of an observable divergence in $Z_\infty$ should not be interpreted as evidence that no singularity exists; rather, it reflects the limited density range accessible in equilibrium simulations and the substantial distance between $\phi_{\max}$ considered here and the expected $\phi_0$.

\subsection{Two-point density correlations}
\subsubsection{Non-Gaussianity}
\label{sec:alpha2}
\begin{figure} 
\centering
\includegraphics[width=0.5\textwidth]{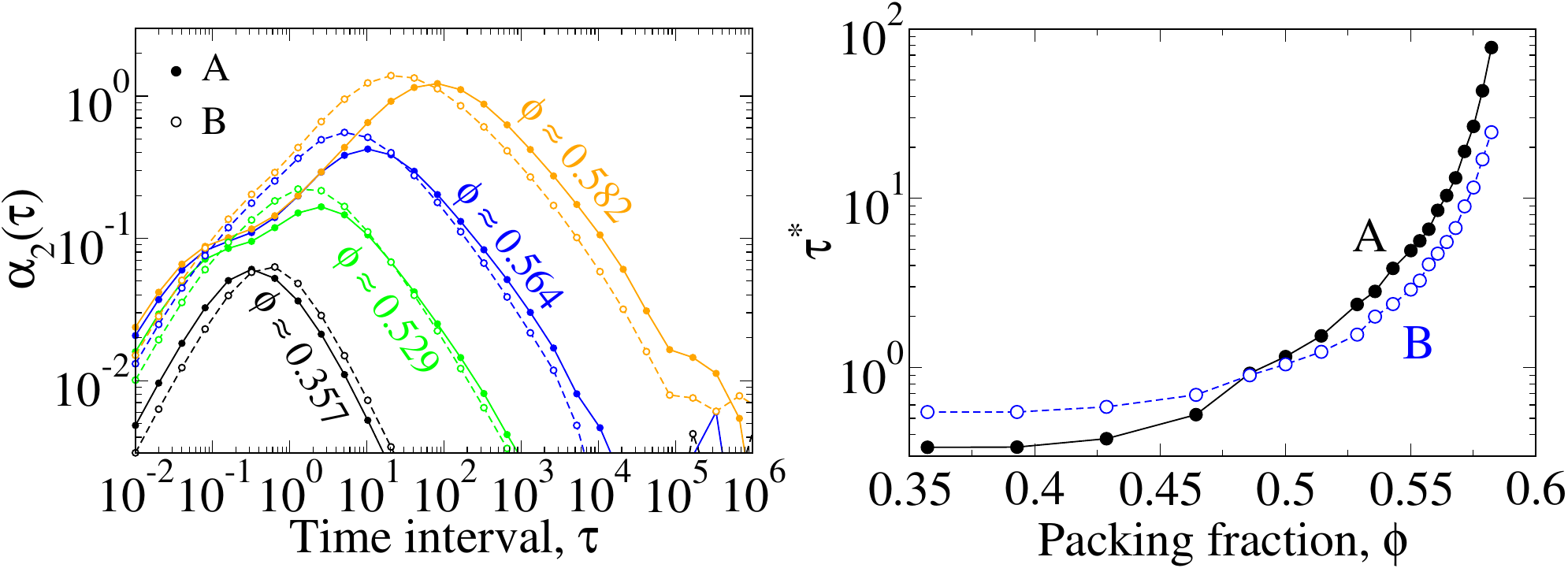}
\caption{Left: time evolution of the non-Gaussian parameter $\alpha_{2}(\tau)$ for the large (solid lines) and small (dashed lines) particles for different densities. Right: density dependence of the characteristic time $\tau^*$, defined as the peak time of $\alpha_{2}(\tau)$. Lines connect the data points. Data are shown for the largest system size considered, $N=3200$.}
\label{fig:alpha2}
\end{figure}

We present in Fig.~\ref{fig:alpha2} (left) the time dependence of the species-dependent non-Gaussian parameter (Eq.~\ref{eq:alpha2}) for several packing fractions. The results are presented for the largest system considered ($N=3200$), for which finite-size effects are negligible (see Sec. S1 B of the supplementary material). For a given packing fraction and species, the parameter vanishes in both the short- and long-time limits due to the Gaussian nature of the displacement distribution in these limits. At intermediate times, it exhibits a pronounced peak at $\tau^*$, where the displacement distribution exhibits enhanced tailing and a substantial deviation from Gaussianity. 

As the fluid becomes denser, $\tau^*$ increases along with the peak height, reflecting increasingly persistent particle cages. One can interpret $\tau^*$ as a characteristic ``cage-escape'' time, when cage rearrangements are most prominent.~\cite{weeks2011s,Swol2014m} The values of $\tau^*$ are also indicated on the MSD$(\tau)$ curves for both species in Figs.~\ref{fig:msd_t}(a) and (b). These points lie between the inflection point and the onset of the diffusive regime.

The packing fraction dependence of $\tau^*$ for both species is shown in Fig.~\ref{fig:alpha2}(right). For packing fractions above the onset of glassy dynamics ($\phi\gtrsim 0.5$), $\tau^*$ increases substantially, spanning nearly two orders of magnitude. In this regime, $\tau^*_B$ is systematically lower than $\tau^*_A$, indicating that the smaller particles (B) escape their cages earlier than A particles, consistent with their greater mobility. At lower packing fractions, however, the opposite trend is observed. In this regime, well-defined cages are not formed, as indicated by the absence of an MSD inflection point (see Sec.~\ref{sec:y_t}). Consequently, $\tau^*$ should not be interpreted as a cage-escape time, and there is therefore no physical basis for expecting the same species-dependent trend as at high densities.

\subsubsection{Displacement distribution}
\label{sec:Gs}
\begin{figure}
\centering
\includegraphics[width=0.5\textwidth]{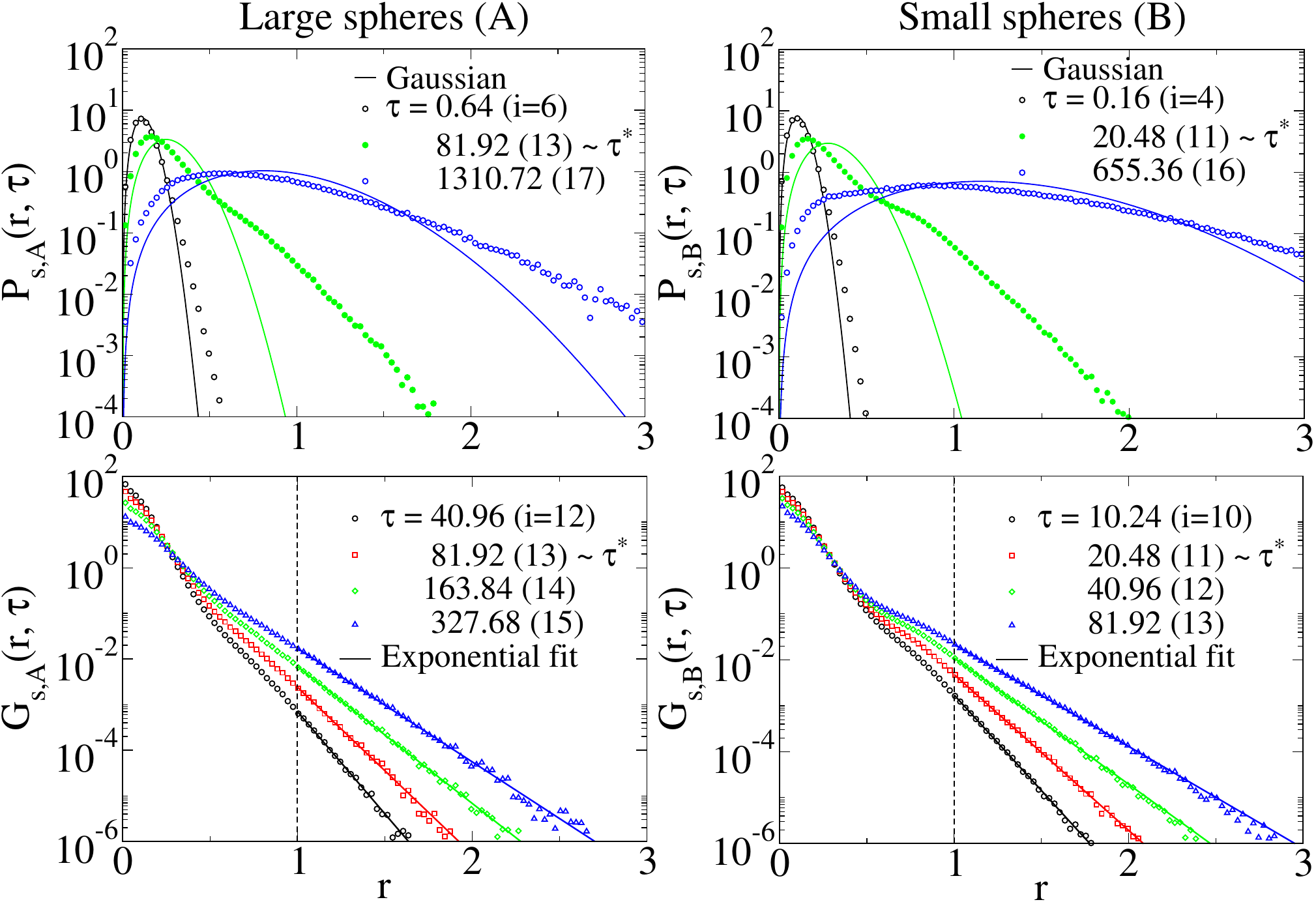}
\caption{Dependence of the radial probability density distribution of the displacement (top row) and the van Hove self-correlation function (bottom row) on the separation distance $r$ for the densest fluid considered ($\phi \approx 0.582$). Results are shown for the large (left) and small (right) spheres at different time intervals, including $\tau^*$. The corresponding Gaussian distributions (Eq.~\ref{eq:Gs_gauss}) and exponential fits (Eq.~\ref{eq:Gs_exp}), obtained over the range $r \ge 1$, are shown as solid lines in the top and bottom rows, respectively. The data correspond to a system size of $N=3200$ particles.}
\label{fig:Gs} 
\end{figure}

Figure~\ref{fig:Gs} (top row) shows the spatial dependence of the radial probability density distributions of the displacement (Eq.~\ref{eq:Ps}) for the large (left) and small (right) particles, using the densest mixture considered, $\phi \approx 0.582$. The results are presented for the largest system considered ($N=3200$), for which finite-size effects are negligible (see Sec. S1 D of the supplementary material). We present the data of three time intervals: one corresponding approximately to the peak time $\tau^{*}$ of $\alpha_{2}(\tau)$ (see Fig.~\ref{fig:alpha2}), and two representing relatively short and long times. At both the short and long times, the Gaussian behavior (solid lines) is approximately recovered. In contrast, significant deviations from Gaussianity are observed at intermediate times, as indicated by the right-hand shoulder and the pronounced long ``tail'' of the distribution. 

Although the shoulder develops for both species, it is more pronounced for the smaller B particles and evolves into a distinct second peak (not shown) at later times, centered at a separation of approximately $\sigma_{\rm B}$. The behavior is indicative of hopping motion, in which a subset of B particles undergoes rare, large displacements between neighboring cages. The hopping events are likely facilitated by transient free volume generated by the motion of the larger A particles, which temporarily opens pathways for cage-to-cage migration.

To investigate the nature of the emergence of the long-displacement tail at intermediate times, we present in the bottom row of Fig.~\ref{fig:Gs} the van Hove self-correlation function (Eq.~\ref{eq:Ps}) for the same density at four intermediate time intervals, including $\tau^*$. The distribution can be divided into two distinct regimes: a Gaussian-like core at short displacements, followed by an exponential decay at large displacements. The coexistence of a localized and a long tail reflects the presence of distinct immobile and mobile particle populations, respectively, and is a hallmark of dynamic heterogeneity in glass-forming systems. 

The long displacement exponential distribution can be described by
\begin{align}
\label{eq:Gs_exp}
 G^{\rm exp}_{\rm s}(r, \tau) \sim  \exp(-r/\lambda(\tau))
\end{align}
where $\lambda(\tau)$ is the characteristic decay length, which increases with $\tau$ as we show below. The emergence of the exponential tail is observed only in the high-density regime ($\phi \gtrsim 0.570$) and is widely attributed to hopping dynamics, in which particles undergo intermittent cage-to-cage displacements after remaining trapped within a cage for a waiting time. Such behavior can be described within the continuous-time random walk (CTRW) framework, where exponential tails naturally arise from intermittent hopping events. The figure also includes exponential fits performed over the range $r \geq 1$, where the data are well described by the exponential form. The fitting parameters are obtained using a WLS procedure with weights proportional to $1/G_{\rm s}(r,\tau)$. This choice is motivated by the fact that the data are obtained from histograms, for which the bin counts follow Poisson statistics, implying that the variance of each data point is proportional to the corresponding value of $G_{\rm s}(r,\tau)$. Consequently, inverse-variance weighting is equivalent to using weights proportional to $1/G_{\rm s}(r,\tau)$.

\begin{figure}
\centering
\includegraphics[width=0.45\textwidth]{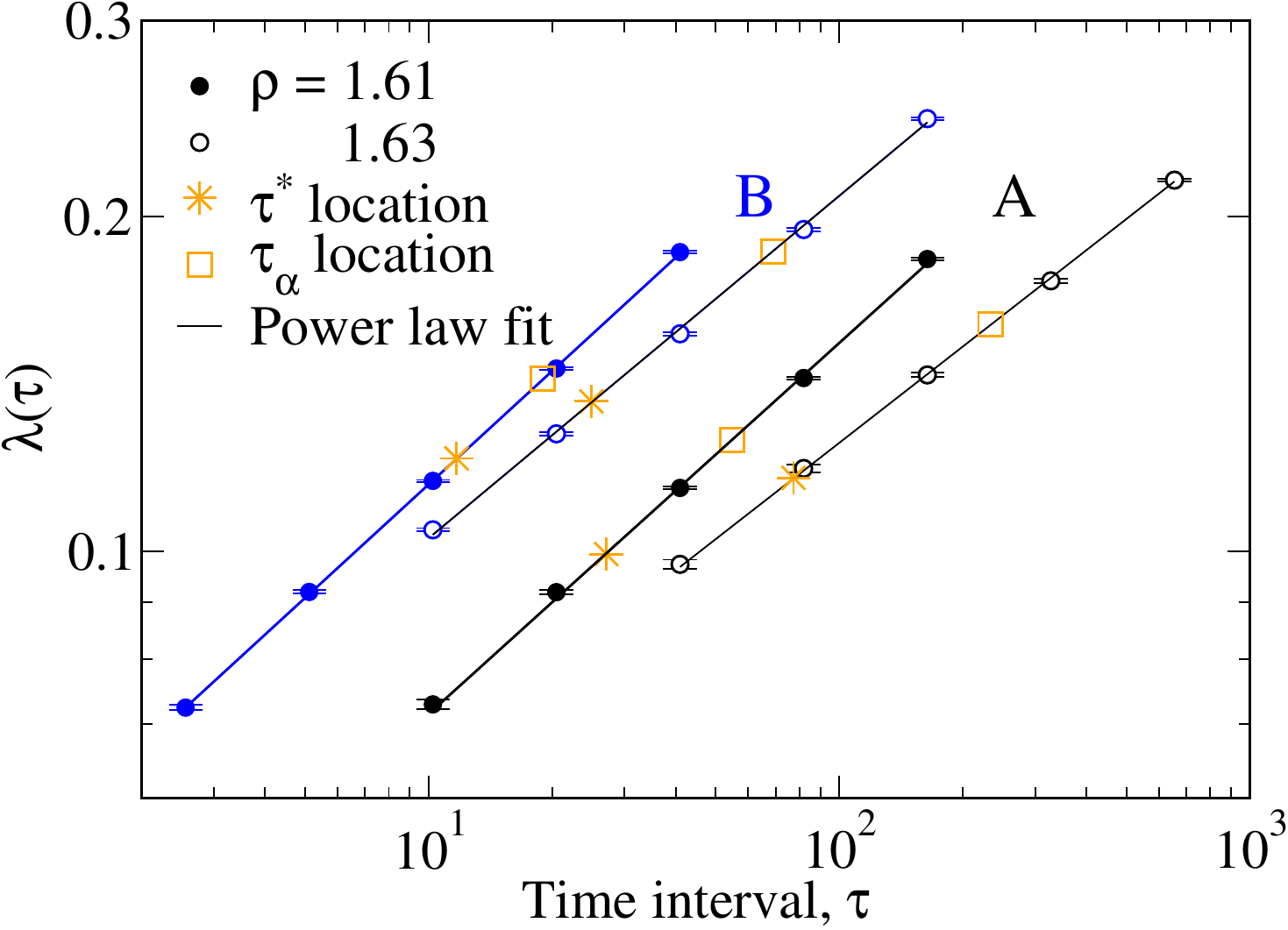}
\caption{Dependence of the characteristic decay length (see Eq.~\ref{eq:Gs_exp}) on the intermediate time interval for A (black) and B (blue) particles at densities $\rho=1.61$ ($\phi \approx 0.575$) and $1.63$ ($0.582$) on logarithmic axes. The solid lines represent power-law fits of the form $\lambda(\tau)=a\tau^{b}$. The orange star and open-square symbols denote the values at $\tau^*$ and $\tau_\alpha$, respectively. Data are shown for the largest system size considered, $N=3200$.}
\label{fig:lambda} 
\end{figure}

Figure~\ref{fig:lambda} shows, on logarithmic axes, the dependence of the fitted $\lambda(\tau)$ on intermediate times for two densities, $\rho=1.61$ and $1.63$. The error bars represent the standard errors of the fitted values obtained from the WLS analysis, scaled such that the reduced $\chi^2$ is unity. For a given density and time interval, we find that $\lambda_B>\lambda_A$, reflecting the greater mobility of the smaller particles. Moreover, on this logarithmic scale, the variation appears to be linear, suggesting a power-law dependence,
\begin{align}
\lambda(\tau)= \theta \,\tau^\nu,
\end{align}
where $\theta$ and $\nu$ are fitting parameters. The resulting fits, shown as solid lines, provide an excellent description of the data. The fitted exponent $\nu$ decreases with increasing density, from $0.334$ to $0.289$ for particle A and from $0.339$ to $0.308$ for particle B as the packing fraction increases from $\phi \approx 0.575$ to $0.582$. These values are lower than the square-root dependence, $\lambda(\tau)\sim\tau^{1/2}$, that is widely reported.

\subsubsection{Structural relaxation} 
\begin{figure}
\centering
\includegraphics[width=0.5\textwidth]{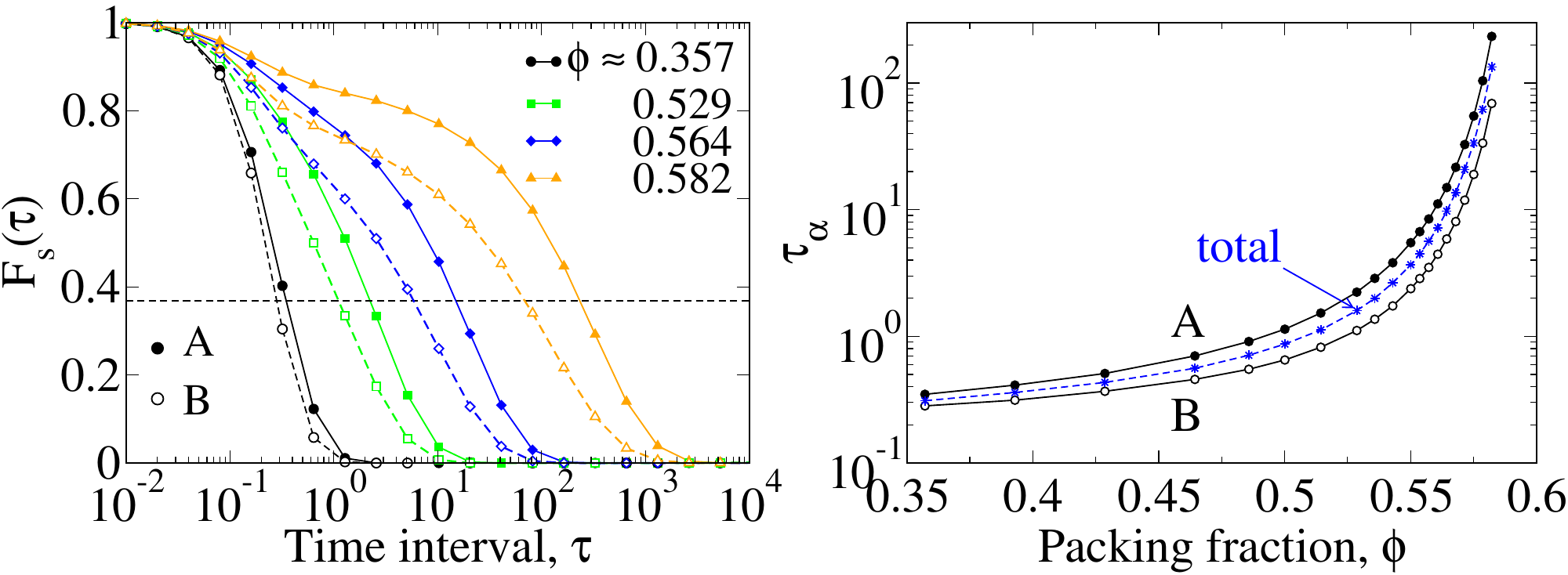}
\caption{Left: time dependence of the self-intermediate scattering function $F_{\rm s}(\tau)$ at wave number $q=7.0$ for both the large (solid lines) and small (dashed lines) particles, at densities $\rho=1.0, 1.48, 1.58, 1.63$, from left to right. Right: density dependence of the structural relaxation times, defined by $F_{\rm s}(\tau_{\alpha})=1/e$ (dashed horizontal line in the left panel). Lines connect the data points. Data are shown for the largest system size considered, $N=3200$.}
\label{fig:fs}
\end{figure}
Figure~\ref{fig:fs}(left) presents the time dependence of the self-intermediate scattering function $F_{\rm s}(\tau, q=7.0)$ for different densities. In the short-time limit, the data show a quadratic dependence associated with the ballistic motion. At intermediate times, however, increasing density gives rise to a $\beta$-relaxation regime, characterized by an extended plateau associated with particle caging effects. This behavior is more pronounced for the larger particles (solid symbols), reflecting stronger caging effects. As a result, the onset of the $\alpha$-relaxation regime is progressively delayed as the system approaches the glassy state.

Figure~\ref{fig:fs}(right) shows the density dependence of the structural relaxation time $\tau_{\alpha}$ defined by the condition $F_{\rm s}(\tau_{\alpha})=1/e$ (see Sec.~\ref{sec:2pt_corr}). As expected, the larger particles consistently exhibit larger $\alpha$ relaxation time than the smaller ones, with the ratio $\tau_{\alpha,A}/\tau_{\alpha,B}$ increasing with density (not shown). At low densities, $\tau_{\alpha}$ exhibits only a weak, approximately Arrhenius-like, density dependence. As the system approaches the glassy regime, however, the relaxation time increases much more rapidly, displaying a super-Arrhenius-like dependence. Over the density range investigated, $\tau_{\alpha}$ spans nearly three orders of magnitude. This behavior closely parallels that of the peak time of the non-Gaussian parameter, indicating that the maximum non-Gaussianity occurs on a similar timescale as the structural relaxation.

The data in Fig.~\ref{fig:fs} are shown for the largest system considered, $N=3200$. However, the structural relaxation time exhibits finite-size effects, as discussed in Sec. S1 C of the supplementary material. Specifically, $\tau_{\alpha}$ is found to vary approximately linearly with $1/N$. We therefore estimate the thermodynamic-limit relaxation time, $\tau_{\alpha,\infty}$, by linearly extrapolating the results for the two largest system sizes to $1/N\rightarrow 0$. Nevertheless, the finite-size results reported here are qualitatively similar to those in the thermodynamic-limit.

\begin{figure}
\centering
\includegraphics[width=0.5\textwidth]{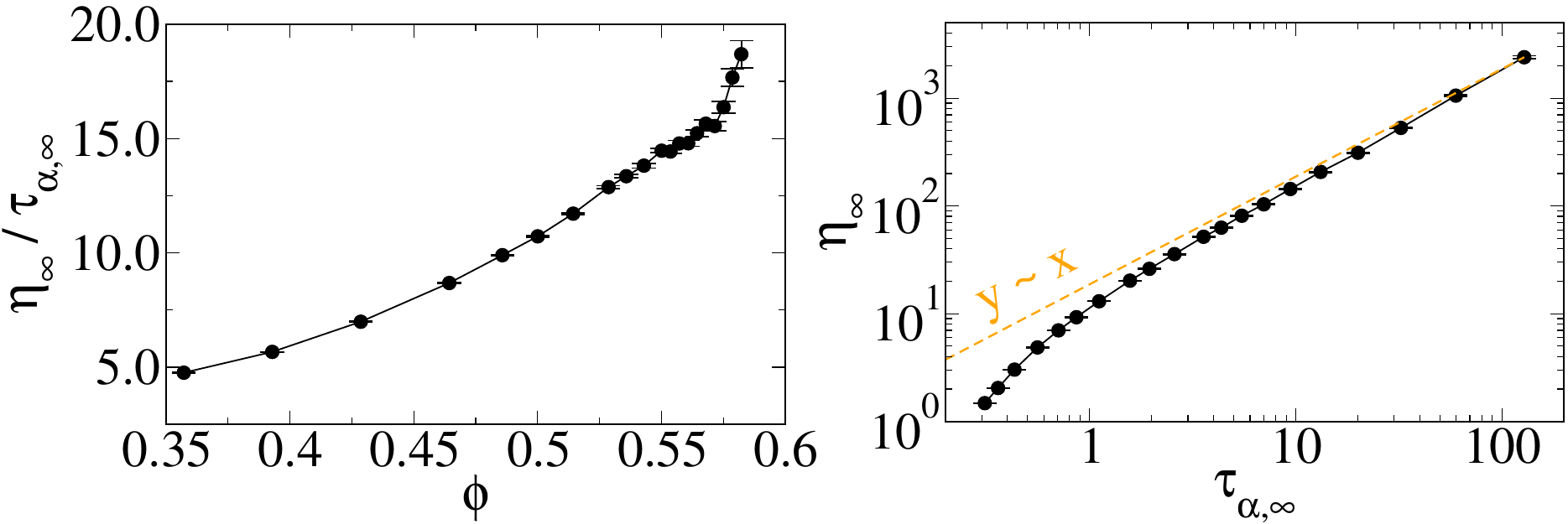}
\caption{Relationship between the structural relaxation time and the shear viscosity (top) and the total self-diffusivity (bottom), all in the thermodynamic-limit. The dashed lines are guides to the eye representing the linear scaling. Since uncertainties in $\tau_{\alpha}$ are not estimated in this work (see text), the error bars in the left panels reflect only the uncertainties in the transport coefficients.}
\label{fig:DVtau}
\end{figure}

\subsubsection{Scaling of $\tau_\alpha$ with transport coefficients} 
We first examine the widely assumed proportionality between the shear viscosity and the structural relaxation time, $\eta/T \sim \tau_{\alpha}$.\cite{shi2013r} This assumption is computationally attractive because the viscosity is more difficult to compute than $\tau_{\alpha}$.\cite{kumar2007r} Consequently, $\tau_\alpha$ is frequently used as a proxy for $\eta/T$ in the Stokes--Einstein relation (Eq.~\ref{eq:se}). In Fig.~\ref{fig:DVtau}(a), we examine the validity of this scaling through the density dependence of the ratio $\eta_\infty/\tau_{\alpha,\infty}$. This ratio would be constant if the assumed scaling held; however, the data instead exhibit a nearly linear increase. As noted earlier, uncertainties in $\tau_\alpha$ are not estimated in this work; therefore, the reported error bars represent contributions from uncertainty in $\eta$. Consequently, the observed linear trend would be even more statistically consistent with the data if the full uncertainties were taken into account. This conclusion is further supported by Fig.~\ref{fig:DVtau}(b), where the direct comparison between $\eta_\infty$ and $\tau_{\alpha,\infty}$ clearly deviates from a simple proportional relationship over the density range studied. Although the breakdown of the proportionality $\eta/T \propto \tau_\alpha$ has previously been reported for continuous interaction models,\cite{shi2013r} to the best of our knowledge, this behavior has not previously been demonstrated for hard-sphere fluids.

Given this observation, together with the breakdown of the Stokes--Einstein relation (Sec.~\ref{sec:se_relation}), we expect the inverse proportionality between the self-diffusivity and the structural relaxation time, $D \propto \tau_\alpha^{-1}$, to also fail. Figure~\ref{fig:DVtau}(c) shows the density dependence of the product $D_\infty \tau_{\alpha,\infty}$, which exhibits a pronounced and nontrivial density dependence. Hence, this simple proportionality is not supported by the present data. For completeness, Fig.~\ref{fig:DVtau}(d) presents the direct relationship between $\tau_{\alpha,\infty}$ and $D_\infty$, which likewise deviates markedly from a simple inverse proportionality.

It is worth noting that, based on the Maxwell relation for viscoelastic fluids, the shear viscosity is often related to the \textit{stress} relaxation time. However, for discontinuous interaction models such as hard spheres, the stress autocorrelation function diverges at $\tau=0$.\cite{allen2017book,haile_book} Consequently, a stress relaxation time cannot be defined in this case.

In summary, while it may be intuitive to expect that the common slowing down of all dynamical processes leads to a proportional relationship between these quantities, our data do not support such a scaling. Although approximate proportionality may emerge over limited regimes or in specific systems, we do not expect this connection to be universal. In particular, $\tau_\alpha$ is an arbitrarily defined timescale whose value depends on the chosen relaxation criterion (here, $F_{\rm s}(\tau_\alpha)=1/e$), whereas transport coefficients are uniquely and rigorously defined.

\subsection{Four-point correlation functions}
\begin{figure*}
\centering
\includegraphics[width=\textwidth]{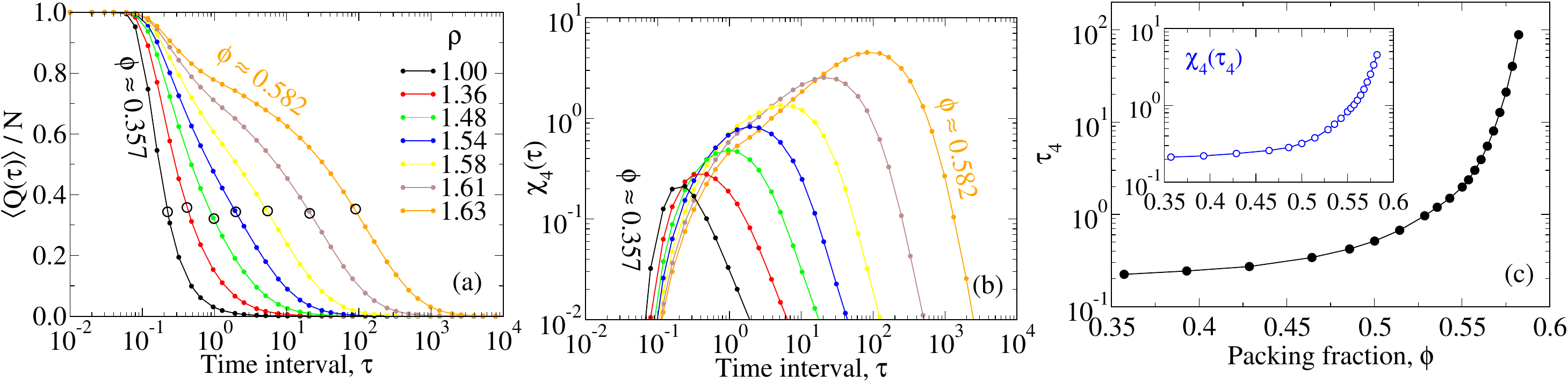}
\caption{Time dependence of (a) the average overlap function per particle and (b) the four-point dynamical susceptibility $\chi_4(\tau)$. Open symbols in Panel (a) indicate the peak times of $\chi_4(\tau)$ shown in Panel (b).  Panel (c) shows the density dependence of the peak time and height associated with $\chi_4(\tau)$. Lines connect the data points. Data are shown for the largest system size considered, $N=3200$. }
\label{fig:QChi4_phi}
\end{figure*}

Figure~\ref{fig:QChi4_phi}(a) shows the time dependence of the average overlap function (Eq.~\ref{eq:Qt}) at different densities. In a similar behavior to $F_{\rm s}(\tau)$, the decay of $Q(\tau)$ becomes progressively slower with the emergence of a plateau as the system gets denser. Panel (b) quantifies the resulting temporal dynamic heterogeneity through the four-point dynamic susceptibility, $\chi_4(\tau)$, computed using Eq.~\ref{eq:chi4}. At both short and long times $\chi_4(\tau)$ vanishes since nearly all particles are immobile at short times ($\langle Q(\tau)\rangle/N \approx 1$) and mobile at long times ($\langle Q(\tau)\rangle/N \approx 0$), respectively.

At intermediate times, however, $\chi_4(\tau)$ develops a density-dependent peak, whose maximum defines the characteristic timescale, $\tau_4$, at which dynamic heterogeneity is most pronounced. At this time scale, the coexistence of mobile and immobile regions gives rise to the largest temporal fluctuations in $Q(\tau)$. The corresponding average overlap values at the peak times, indicated by the open symbols in Panel (a), are nearly similar across all densities, with a value of approximately $0.35$.  

Panel (c) depicts the density dependence of the corresponding peak time, $\tau_4$, for a system size of $N=3200$ particles. Similar to the previously discussed characteristic times, $\tau_4$ exhibits a relatively weak density dependence at low densities, followed by a super-Arrhenius-like increase as the system enters the high-density regime. Over the density range investigated, $\tau_4$ spans nearly three orders of magnitude. Moreover, the density dependence of $\tau_4$ is comparable to $\tau_\alpha$, as shown in Sec. S2 of the supplementary material, suggesting that local cage rearrangements are the primary source of temporal dynamical fluctuations. The $\tau_4$ timescale exhibits finite-size effects; these are discussed in Sec. S1 C of the supplementary material, where we also report the corresponding thermodynamic-limit values, $\tau_{4,\infty}$. Nevertheless, the qualitative density dependence remains consistent with that shown here.

Given that $\chi_4(\tau)$ is directly proportional to the spatial integral of $g_4(r,\tau)$ (see Eq.~\ref{eq:chi4}), it provides a measure of the extent of spatially correlated dynamics and, hence, the size of dynamically correlated regions.\cite{flenner2010dynamic,flenner2011analysis,flenner2014u} The inset of Panel (c) shows the packing fraction dependence of $\chi_4(\tau_4)$ at the peak time, showing a growth by about an order of magnitude over the density range investigated. The increase signifies that an increasing number of particles participate in spatially correlated dynamics as increasingly glassy states are approached.

\section{CONCLUSIONS}
\label{sec:conclusions}
We have presented a comprehensive and robust event-driven MD study of the self-diffusivity, $D$, and shear viscosity, $\eta$, as well as the two- and four-point dynamical properties of an equimolar mixture of hard spheres over a wide range of packing fractions, $0.357\lesssim\phi\lesssim0.582$. Some key findings of this work include the following:
\begin{enumerate}[label=(\alph*)]
\item Transport coefficients in the thermodynamic-limit: The Einstein--Helfand method, combined with an efficient binary-based time-sampling scheme, provides reliable estimates of the self-diffusivity, $D$, and shear viscosity, $\eta$, together with their associated uncertainties. Significant finite-size effects are observed for $D$ at all densities and for $\eta$ only at high densities, with the magnitude of these effects increasing substantially with density. The finite-size effects of $D$ are well described by a power law with a density-dependent exponent, rather than the conventional hydrodynamic $N^{-1/3}$ scaling. In contrast, the finite-size dependence of $\eta$ is more complex, however statistical convergence is achieved with the largest system-size considered, $N=3200$. Extrapolation of both properties to the thermodynamic-limit, $1/N\to 0$, enables accurate estimation of their thermodynamic-limit values, $D_\infty$ and $\eta_\infty$.
\item  Density dependence of the transport coefficients: Both $D_{\infty}(\phi)$ and $\eta_{\infty}(\phi)$ exhibit an approximately logarithmic dependence at low densities, followed by a rapid, super-Arrhenius-like dependence in the high-density regime. The onset of this glassy behavior occurs at around $\phi\approx0.486$, coinciding with the emergence of particle caging, as evidenced by the inflection point in the logarithmic mean squared displacement plot. In the high-density regime, both transport coefficients are well-described by a standard four-parameter exponential fit. 
\item Critical packing fraction: Despite the good quality of the fits to $D_{\infty}(\phi)$ and $\eta_{\infty}(\phi)$, the resulting estimates of the critical packing fraction, $\phi_0$, are not reliable. Estimates from the two transport coefficients are statistically inconsistent and substantially below values reported in the literature, which extend to $\phi\approx0.65$. This discrepancy is due, in part, to the substantial extrapolation required beyond the highest equilibrium packing fraction studied here, $\phi=0.582$, as well as beyond the range explored in other equilibrium event-driven MD (or Monte Carlo) studies ($\phi_{\rm max}\approx 0.6$). This conclusion is further supported by the compressibility factor data, which show no evidence of divergence over the density range considered here. Because statistical thermodynamics dictates that this static property must diverge at the same $\phi_0$ as the dynamic properties, the current extrapolation fails to capture the underlying physics.
 \item Breakdown of the Stokes--Einstein relation: The Stokes--Einstein hydrodynamic relation, $D_{\infty}(\phi)\eta_{\infty}(\phi) \sim \text{const.}$, is approximately satisfied for large particles at low densities but progressively breaks down beyond the onset of glassy dynamics. For smaller particles, this breakdown persists across the entire range of densities considered. Furthermore, a simple fractional Stokes--Einstein relation, $D_{\infty}(\phi)\eta^{c}_{\infty}(\phi) \sim \text{const.}$, is not supported by our data either. This indicates that the decoupling between diffusive and viscous transport cannot be described by a simple relation.
 \item van Hove self-correlation function, $G_s(r,\tau)$: The spatial dependence of $G_s(r,\tau)$ for both species shows approximately Gaussian behavior at short and long times, with a strong deviation at intermediate times. This non-Gaussianity is quantified by the non-Gaussian parameter, $\alpha_2(\tau)$, which we show to peak at some intermediate time, $\tau^*$, associated with the particle cage-escape time. At packing fractions well above the onset of glassy dynamics ($\phi\gtrsim 0.570$), $G_s(r,\tau)$ develops a pronounced, long-displacement exponential tail at times around $\tau^*$, reflecting dynamic heterogeneity due to distinct mobile and immobile particle populations. This behavior is commonly associated with hopping events, consistent with its emergence around the cage-escape timescale. Moreover, the exponential decay length scales with time as $\lambda\approx\tau^{1/3}$, rather than the conventional square-root dependence. 
\item Scaling between shear viscosity and structural relaxation: The structural relaxation time, $\tau_\alpha$, grows by nearly three orders of magnitude across the density range, showing a qualitatively similar behavior to that of the viscosity. However, the current data do not support the commonly assumed scaling $\eta \propto\tau_{\alpha}$. Thus, the structural relaxation timescale cannot be regarded as a universal proxy for the shear viscosity, at least for hard-sphere fluids. Given the arbitrary definition of $\tau_\alpha$, we do not expect such a scaling to hold generally over arbitrary density or temperature ranges. This scaling analysis is performed using values extrapolated to the thermodynamic limit, as $\tau_\alpha$ exhibits substantial finite-size effects that are often neglected in the literature.
\item Dynamically correlated regions: The temporal heterogeneity is captured by the four-point susceptibility, $\chi_4(\tau)$. The function develops a pronounced peak at an intermediate timescale, $\tau_4$, closely tracks $\tau_\alpha$, suggesting a close connection to cage rearrangements. Simultaneously, the peak magnitude $\chi_4(\tau_4)$ increases with density, showing an increasingly large number of particles participate in this correlated dynamics.
\end{enumerate}

Overall, this study provides a comprehensive suite of event-driven MD data and a robust analytical framework for evaluating the fundamental dynamical properties of hard-sphere fluids. Given the high quality of the data reported here, these results can serve as a rigorous benchmark for both theoretical developments and experimental investigations. Furthermore, while this work focuses on a simple hard-sphere model, the analytical and computational frameworks established herein are readily generalizable to more complex molecular and ionic liquids.

\section*{SUPPLEMENTARY MATERIAL}
The supplementary material contains the complete dataset files for the dynamical properties presented in this study, along with additional results and discussion of finite-size effects. The provided datasets enable reproduction of the reported results and facilitate further analysis.

\section*{ACKNOWLEDGMENTS}
Computational resources were provided by the Center for Computational Research (CCR), University at Buffalo. We thank Jack Douglas, Patrick Charbonneau, and Grzegorz Szamel for helpful discussions and valuable insights.

\section*{AUTHOR DECLARATIONS}
\subsection*{Conflict of Interest}
The authors have no conflicts to disclose.

\subsection*{Author Contributions}
\textbf{Sabry G. Moustafa}: Conceptualization (equal); Data curation (lead); Formal analysis (lead); Methodology (lead); Validation (equal); Visualization (lead); Writing – original draft (lead); Writing – review \& editing (lead). \textbf{Andrew J. Schultz}: Conceptualization (equal); Methodology (equal); Software (equal); Writing – review \& editing (equal). 

\section*{DATA AVAILABILITY}
Data supporting the findings of this study are available from the corresponding author on a reasonable request.

\bibliography{glass} 

\begin{thebibliography}{76}%
\makeatletter
\providecommand \@ifxundefined [1]{%
 \@ifx{#1\undefined}
}%
\providecommand \@ifnum [1]{%
 \ifnum #1\expandafter \@firstoftwo
 \else \expandafter \@secondoftwo
 \fi
}%
\providecommand \@ifx [1]{%
 \ifx #1\expandafter \@firstoftwo
 \else \expandafter \@secondoftwo
 \fi
}%
\providecommand \natexlab [1]{#1}%
\providecommand \enquote  [1]{``#1''}%
\providecommand \bibnamefont  [1]{#1}%
\providecommand \bibfnamefont [1]{#1}%
\providecommand \citenamefont [1]{#1}%
\providecommand \href@noop [0]{\@secondoftwo}%
\providecommand \href [0]{\begingroup \@sanitize@url \@href}%
\providecommand \@href[1]{\@@startlink{#1}\@@href}%
\providecommand \@@href[1]{\endgroup#1\@@endlink}%
\providecommand \@sanitize@url [0]{\catcode `\\12\catcode `\$12\catcode `\&12\catcode `\#12\catcode `\^12\catcode `\_12\catcode `\%12\relax}%
\providecommand \@@startlink[1]{}%
\providecommand \@@endlink[0]{}%
\providecommand \url  [0]{\begingroup\@sanitize@url \@url }%
\providecommand \@url [1]{\endgroup\@href {#1}{\urlprefix }}%
\providecommand \urlprefix  [0]{URL }%
\providecommand \Eprint [0]{\href }%
\providecommand \doibase [0]{https://doi.org/}%
\providecommand \selectlanguage [0]{\@gobble}%
\providecommand \bibinfo  [0]{\@secondoftwo}%
\providecommand \bibfield  [0]{\@secondoftwo}%
\providecommand \translation [1]{[#1]}%
\providecommand \BibitemOpen [0]{}%
\providecommand \bibitemStop [0]{}%
\providecommand \bibitemNoStop [0]{.\EOS\space}%
\providecommand \EOS [0]{\spacefactor3000\relax}%
\providecommand \BibitemShut  [1]{\csname bibitem#1\endcsname}%
\let\auto@bib@innerbib\@empty
\bibitem [{\citenamefont {Angell}(1995)}]{angell1995}%
  \BibitemOpen
  \bibfield  {author} {\bibinfo {author} {\bibfnamefont {C.~A.}\ \bibnamefont {Angell}},\ }\bibfield  {title} {\enquote {\bibinfo {title} {Formation of glasses from liquids and biopolymers},}\ }\href@noop {} {\bibfield  {journal} {\bibinfo  {journal} {Science}\ }\textbf {\bibinfo {volume} {267}},\ \bibinfo {pages} {1924--1935} (\bibinfo {year} {1995})}\BibitemShut {NoStop}%
\bibitem [{\citenamefont {Debenedetti}\ and\ \citenamefont {Stillinger}(2001)}]{debenedetti2001rev}%
  \BibitemOpen
  \bibfield  {author} {\bibinfo {author} {\bibfnamefont {P.~G.}\ \bibnamefont {Debenedetti}}\ and\ \bibinfo {author} {\bibfnamefont {F.~H.}\ \bibnamefont {Stillinger}},\ }\bibfield  {title} {\enquote {\bibinfo {title} {Supercooled liquids and the glass transition},}\ }\href@noop {} {\bibfield  {journal} {\bibinfo  {journal} {Nature}\ }\textbf {\bibinfo {volume} {410}},\ \bibinfo {pages} {259--267} (\bibinfo {year} {2001})}\BibitemShut {NoStop}%
\bibitem [{\citenamefont {Hunter}\ and\ \citenamefont {Weeks}(2012)}]{hunter2012physics}%
  \BibitemOpen
  \bibfield  {author} {\bibinfo {author} {\bibfnamefont {G.~L.}\ \bibnamefont {Hunter}}\ and\ \bibinfo {author} {\bibfnamefont {E.~R.}\ \bibnamefont {Weeks}},\ }\bibfield  {title} {\enquote {\bibinfo {title} {The physics of the colloidal glass transition},}\ }\href@noop {} {\bibfield  {journal} {\bibinfo  {journal} {Reports on progress in physics}\ }\textbf {\bibinfo {volume} {75}},\ \bibinfo {pages} {066501} (\bibinfo {year} {2012})}\BibitemShut {NoStop}%
\bibitem [{\citenamefont {Weeks}(2017)}]{weeks2017intro}%
  \BibitemOpen
  \bibfield  {author} {\bibinfo {author} {\bibfnamefont {E.~R.}\ \bibnamefont {Weeks}},\ }\href@noop {} {\enquote {\bibinfo {title} {Introduction to the colloidal glass transition},}\ } (\bibinfo {year} {2017})\BibitemShut {NoStop}%
\bibitem [{\citenamefont {Berthier}\ and\ \citenamefont {Reichman}(2023)}]{berthier2023mod}%
  \BibitemOpen
  \bibfield  {author} {\bibinfo {author} {\bibfnamefont {L.}~\bibnamefont {Berthier}}\ and\ \bibinfo {author} {\bibfnamefont {D.~R.}\ \bibnamefont {Reichman}},\ }\bibfield  {title} {\enquote {\bibinfo {title} {Modern computational studies of the glass transition},}\ }\href@noop {} {\bibfield  {journal} {\bibinfo  {journal} {Nature Reviews Physics}\ }\textbf {\bibinfo {volume} {5}},\ \bibinfo {pages} {102--116} (\bibinfo {year} {2023})}\BibitemShut {NoStop}%
\bibitem [{\citenamefont {Allen}\ and\ \citenamefont {Tildesley}(2017)}]{allen2017book}%
  \BibitemOpen
  \bibfield  {author} {\bibinfo {author} {\bibfnamefont {M.}~\bibnamefont {Allen}}\ and\ \bibinfo {author} {\bibfnamefont {D.}~\bibnamefont {Tildesley}},\ }\href@noop {} {\emph {\bibinfo {title} {Computer Simulation of Liquids}}},\ Oxford science publications\ (\bibinfo  {publisher} {Oxford University Press},\ \bibinfo {year} {2017})\BibitemShut {NoStop}%
\bibitem [{\citenamefont {Zhang}\ \emph {et~al.}(2023)\citenamefont {Zhang}, \citenamefont {Luo}, \citenamefont {Zheng},\ and\ \citenamefont {Han}}]{zhang2023effects}%
  \BibitemOpen
  \bibfield  {author} {\bibinfo {author} {\bibfnamefont {H.}~\bibnamefont {Zhang}}, \bibinfo {author} {\bibfnamefont {C.}~\bibnamefont {Luo}}, \bibinfo {author} {\bibfnamefont {Z.}~\bibnamefont {Zheng}},\ and\ \bibinfo {author} {\bibfnamefont {Y.}~\bibnamefont {Han}},\ }\bibfield  {title} {\enquote {\bibinfo {title} {Effects of size ratio on particle packing in binary glasses},}\ }\href@noop {} {\bibfield  {journal} {\bibinfo  {journal} {Acta Materialia}\ }\textbf {\bibinfo {volume} {246}},\ \bibinfo {pages} {118700} (\bibinfo {year} {2023})}\BibitemShut {NoStop}%
\bibitem [{\citenamefont {Royall}\ \emph {et~al.}(2024)\citenamefont {Royall}, \citenamefont {Charbonneau}, \citenamefont {Dijkstra}, \citenamefont {Russo}, \citenamefont {Smallenburg}, \citenamefont {Speck},\ and\ \citenamefont {Valeriani}}]{royall2024colloidal}%
  \BibitemOpen
  \bibfield  {author} {\bibinfo {author} {\bibfnamefont {C.~P.}\ \bibnamefont {Royall}}, \bibinfo {author} {\bibfnamefont {P.}~\bibnamefont {Charbonneau}}, \bibinfo {author} {\bibfnamefont {M.}~\bibnamefont {Dijkstra}}, \bibinfo {author} {\bibfnamefont {J.}~\bibnamefont {Russo}}, \bibinfo {author} {\bibfnamefont {F.}~\bibnamefont {Smallenburg}}, \bibinfo {author} {\bibfnamefont {T.}~\bibnamefont {Speck}},\ and\ \bibinfo {author} {\bibfnamefont {C.}~\bibnamefont {Valeriani}},\ }\bibfield  {title} {\enquote {\bibinfo {title} {Colloidal hard spheres: Triumphs, challenges, and mysteries},}\ }\href@noop {} {\bibfield  {journal} {\bibinfo  {journal} {Reviews of Modern Physics}\ }\textbf {\bibinfo {volume} {96}},\ \bibinfo {pages} {045003} (\bibinfo {year} {2024})}\BibitemShut {NoStop}%
\bibitem [{\citenamefont {Royall}, \citenamefont {Poon},\ and\ \citenamefont {Weeks}(2013)}]{royall2013}%
  \BibitemOpen
  \bibfield  {author} {\bibinfo {author} {\bibfnamefont {C.~P.}\ \bibnamefont {Royall}}, \bibinfo {author} {\bibfnamefont {W.~C.}\ \bibnamefont {Poon}},\ and\ \bibinfo {author} {\bibfnamefont {E.~R.}\ \bibnamefont {Weeks}},\ }\bibfield  {title} {\enquote {\bibinfo {title} {In search of colloidal hard spheres},}\ }\href@noop {} {\bibfield  {journal} {\bibinfo  {journal} {Soft Matter}\ }\textbf {\bibinfo {volume} {9}},\ \bibinfo {pages} {17--27} (\bibinfo {year} {2013})}\BibitemShut {NoStop}%
\bibitem [{\citenamefont {Srivastava}\ \emph {et~al.}(2021)\citenamefont {Srivastava}, \citenamefont {Roberts}, \citenamefont {Clemmer}, \citenamefont {Silbert}, \citenamefont {Lechman},\ and\ \citenamefont {Grest}}]{srivastava2021}%
  \BibitemOpen
  \bibfield  {author} {\bibinfo {author} {\bibfnamefont {I.}~\bibnamefont {Srivastava}}, \bibinfo {author} {\bibfnamefont {S.~A.}\ \bibnamefont {Roberts}}, \bibinfo {author} {\bibfnamefont {J.~T.}\ \bibnamefont {Clemmer}}, \bibinfo {author} {\bibfnamefont {L.~E.}\ \bibnamefont {Silbert}}, \bibinfo {author} {\bibfnamefont {J.~B.}\ \bibnamefont {Lechman}},\ and\ \bibinfo {author} {\bibfnamefont {G.~S.}\ \bibnamefont {Grest}},\ }\bibfield  {title} {\enquote {\bibinfo {title} {Jamming of bidisperse frictional spheres},}\ }\href@noop {} {\bibfield  {journal} {\bibinfo  {journal} {Physical Review Research}\ }\textbf {\bibinfo {volume} {3}},\ \bibinfo {pages} {L032042} (\bibinfo {year} {2021})}\BibitemShut {NoStop}%
\bibitem [{\citenamefont {Angell}\ \emph {et~al.}(2000)\citenamefont {Angell}, \citenamefont {Ngai}, \citenamefont {McKenna}, \citenamefont {McMillan},\ and\ \citenamefont {Martin}}]{angell2000relaxation}%
  \BibitemOpen
  \bibfield  {author} {\bibinfo {author} {\bibfnamefont {C.~A.}\ \bibnamefont {Angell}}, \bibinfo {author} {\bibfnamefont {K.~L.}\ \bibnamefont {Ngai}}, \bibinfo {author} {\bibfnamefont {G.~B.}\ \bibnamefont {McKenna}}, \bibinfo {author} {\bibfnamefont {P.~F.}\ \bibnamefont {McMillan}},\ and\ \bibinfo {author} {\bibfnamefont {S.~W.}\ \bibnamefont {Martin}},\ }\bibfield  {title} {\enquote {\bibinfo {title} {Relaxation in glassforming liquids and amorphous solids},}\ }\href@noop {} {\bibfield  {journal} {\bibinfo  {journal} {Journal of applied physics}\ }\textbf {\bibinfo {volume} {88}},\ \bibinfo {pages} {3113--3157} (\bibinfo {year} {2000})}\BibitemShut {NoStop}%
\bibitem [{\citenamefont {Zhang}\ \emph {et~al.}(2014)\citenamefont {Zhang}, \citenamefont {Smith}, \citenamefont {Wang}, \citenamefont {Liu}, \citenamefont {Schroers}, \citenamefont {Shattuck},\ and\ \citenamefont {O'Hern}}]{zhang2014connection}%
  \BibitemOpen
  \bibfield  {author} {\bibinfo {author} {\bibfnamefont {K.}~\bibnamefont {Zhang}}, \bibinfo {author} {\bibfnamefont {W.~W.}\ \bibnamefont {Smith}}, \bibinfo {author} {\bibfnamefont {M.}~\bibnamefont {Wang}}, \bibinfo {author} {\bibfnamefont {Y.}~\bibnamefont {Liu}}, \bibinfo {author} {\bibfnamefont {J.}~\bibnamefont {Schroers}}, \bibinfo {author} {\bibfnamefont {M.~D.}\ \bibnamefont {Shattuck}},\ and\ \bibinfo {author} {\bibfnamefont {C.~S.}\ \bibnamefont {O'Hern}},\ }\bibfield  {title} {\enquote {\bibinfo {title} {Connection between the packing efficiency of binary hard spheres and the glass-forming ability of bulk metallic glasses},}\ }\href@noop {} {\bibfield  {journal} {\bibinfo  {journal} {Physical Review E}\ }\textbf {\bibinfo {volume} {90}},\ \bibinfo {pages} {032311} (\bibinfo {year} {2014})}\BibitemShut {NoStop}%
\bibitem [{\citenamefont {Narumi}\ \emph {et~al.}(2009)\citenamefont {Narumi}, \citenamefont {Franklin}, \citenamefont {Desmond}, \citenamefont {Tokuyama},\ and\ \citenamefont {Weeks}}]{weeks2011s}%
  \BibitemOpen
  \bibfield  {author} {\bibinfo {author} {\bibfnamefont {T.}~\bibnamefont {Narumi}}, \bibinfo {author} {\bibfnamefont {S.~V.}\ \bibnamefont {Franklin}}, \bibinfo {author} {\bibfnamefont {K.~W.}\ \bibnamefont {Desmond}}, \bibinfo {author} {\bibfnamefont {M.}~\bibnamefont {Tokuyama}},\ and\ \bibinfo {author} {\bibfnamefont {E.~R.}\ \bibnamefont {Weeks}},\ }\bibfield  {title} {\enquote {\bibinfo {title} {Spatial and temporal dynamical heterogeneities approaching the binary colloidal glass transition},}\ }\href@noop {} {\bibfield  {journal} {\bibinfo  {journal} {arXiv preprint arXiv:0911.0702}\ } (\bibinfo {year} {2009})}\BibitemShut {NoStop}%
\bibitem [{\citenamefont {Puertas}\ \emph {et~al.}(2007)\citenamefont {Puertas}, \citenamefont {De~Michele}, \citenamefont {Sciortino}, \citenamefont {Tartaglia},\ and\ \citenamefont {Zaccarelli}}]{puertas2007visco}%
  \BibitemOpen
  \bibfield  {author} {\bibinfo {author} {\bibfnamefont {A.~M.}\ \bibnamefont {Puertas}}, \bibinfo {author} {\bibfnamefont {C.}~\bibnamefont {De~Michele}}, \bibinfo {author} {\bibfnamefont {F.}~\bibnamefont {Sciortino}}, \bibinfo {author} {\bibfnamefont {P.}~\bibnamefont {Tartaglia}},\ and\ \bibinfo {author} {\bibfnamefont {E.}~\bibnamefont {Zaccarelli}},\ }\bibfield  {title} {\enquote {\bibinfo {title} {Viscoelasticity and stokes-einstein relation in repulsive and attractive colloidal glasses},}\ }\href@noop {} {\bibfield  {journal} {\bibinfo  {journal} {The Journal of chemical physics}\ }\textbf {\bibinfo {volume} {127}} (\bibinfo {year} {2007})}\BibitemShut {NoStop}%
\bibitem [{\citenamefont {Mittal}(2009)}]{mittal2009u}%
  \BibitemOpen
  \bibfield  {author} {\bibinfo {author} {\bibfnamefont {J.}~\bibnamefont {Mittal}},\ }\bibfield  {title} {\enquote {\bibinfo {title} {Using compressibility factor as a predictor of confined hard-sphere fluid dynamics},}\ }\href@noop {} {\bibfield  {journal} {\bibinfo  {journal} {The Journal of Physical Chemistry B}\ }\textbf {\bibinfo {volume} {113}},\ \bibinfo {pages} {13800--13804} (\bibinfo {year} {2009})}\BibitemShut {NoStop}%
\bibitem [{\citenamefont {Charbonneau}\ \emph {et~al.}(2013)\citenamefont {Charbonneau}, \citenamefont {Charbonneau}, \citenamefont {Jin}, \citenamefont {Parisi},\ and\ \citenamefont {Zamponi}}]{charbonneau2013dim}%
  \BibitemOpen
  \bibfield  {author} {\bibinfo {author} {\bibfnamefont {B.}~\bibnamefont {Charbonneau}}, \bibinfo {author} {\bibfnamefont {P.}~\bibnamefont {Charbonneau}}, \bibinfo {author} {\bibfnamefont {Y.}~\bibnamefont {Jin}}, \bibinfo {author} {\bibfnamefont {G.}~\bibnamefont {Parisi}},\ and\ \bibinfo {author} {\bibfnamefont {F.}~\bibnamefont {Zamponi}},\ }\bibfield  {title} {\enquote {\bibinfo {title} {Dimensional dependence of the stokes--einstein relation and its violation},}\ }\href@noop {} {\bibfield  {journal} {\bibinfo  {journal} {The Journal of chemical physics}\ }\textbf {\bibinfo {volume} {139}} (\bibinfo {year} {2013})}\BibitemShut {NoStop}%
\bibitem [{\citenamefont {Foffi}\ \emph {et~al.}(2003)\citenamefont {Foffi}, \citenamefont {G{\"o}tze}, \citenamefont {Sciortino}, \citenamefont {Tartaglia},\ and\ \citenamefont {Voigtmann}}]{foffi2003mixing}%
  \BibitemOpen
  \bibfield  {author} {\bibinfo {author} {\bibfnamefont {G.}~\bibnamefont {Foffi}}, \bibinfo {author} {\bibfnamefont {W.}~\bibnamefont {G{\"o}tze}}, \bibinfo {author} {\bibfnamefont {F.}~\bibnamefont {Sciortino}}, \bibinfo {author} {\bibfnamefont {P.}~\bibnamefont {Tartaglia}},\ and\ \bibinfo {author} {\bibfnamefont {T.}~\bibnamefont {Voigtmann}},\ }\bibfield  {title} {\enquote {\bibinfo {title} {Mixing effects for the structural relaxation in binary hard-sphere liquids},}\ }\href@noop {} {\bibfield  {journal} {\bibinfo  {journal} {Physical review letters}\ }\textbf {\bibinfo {volume} {91}},\ \bibinfo {pages} {085701} (\bibinfo {year} {2003})}\BibitemShut {NoStop}%
\bibitem [{\citenamefont {Flenner}, \citenamefont {Zhang},\ and\ \citenamefont {Szamel}(2011)}]{flenner2011analysis}%
  \BibitemOpen
  \bibfield  {author} {\bibinfo {author} {\bibfnamefont {E.}~\bibnamefont {Flenner}}, \bibinfo {author} {\bibfnamefont {M.}~\bibnamefont {Zhang}},\ and\ \bibinfo {author} {\bibfnamefont {G.}~\bibnamefont {Szamel}},\ }\bibfield  {title} {\enquote {\bibinfo {title} {Analysis of a growing dynamic length scale in a glass-forming binary hard-sphere mixture},}\ }\href@noop {} {\bibfield  {journal} {\bibinfo  {journal} {Physical Review E—Statistical, Nonlinear, and Soft Matter Physics}\ }\textbf {\bibinfo {volume} {83}},\ \bibinfo {pages} {051501} (\bibinfo {year} {2011})}\BibitemShut {NoStop}%
\bibitem [{\citenamefont {Charbonneau}\ and\ \citenamefont {Tarjus}(2013)}]{charbonneau2013decor}%
  \BibitemOpen
  \bibfield  {author} {\bibinfo {author} {\bibfnamefont {P.}~\bibnamefont {Charbonneau}}\ and\ \bibinfo {author} {\bibfnamefont {G.}~\bibnamefont {Tarjus}},\ }\bibfield  {title} {\enquote {\bibinfo {title} {Decorrelation of the static and dynamic length scales in hard-sphere glass formers},}\ }\href@noop {} {\bibfield  {journal} {\bibinfo  {journal} {Physical Review E—Statistical, Nonlinear, and Soft Matter Physics}\ }\textbf {\bibinfo {volume} {87}},\ \bibinfo {pages} {042305} (\bibinfo {year} {2013})}\BibitemShut {NoStop}%
\bibitem [{\citenamefont {Green}(1954)}]{green1954}%
  \BibitemOpen
  \bibfield  {author} {\bibinfo {author} {\bibfnamefont {M.~S.}\ \bibnamefont {Green}},\ }\bibfield  {title} {\enquote {\bibinfo {title} {Markoff random processes and the statistical mechanics of time-dependent phenomena. {II}. {I}rreversible processes in fluids},}\ }\href@noop {} {\bibfield  {journal} {\bibinfo  {journal} {The Journal of chemical physics}\ }\textbf {\bibinfo {volume} {22}},\ \bibinfo {pages} {398--413} (\bibinfo {year} {1954})}\BibitemShut {NoStop}%
\bibitem [{\citenamefont {Kubo}(1957)}]{kubo1957}%
  \BibitemOpen
  \bibfield  {author} {\bibinfo {author} {\bibfnamefont {R.}~\bibnamefont {Kubo}},\ }\bibfield  {title} {\enquote {\bibinfo {title} {Statistical-mechanical theory of irreversible processes. {I}. {G}eneral theory and simple applications to magnetic and conduction problems},}\ }\href@noop {} {\bibfield  {journal} {\bibinfo  {journal} {Journal of the Physical Society of Japan}\ }\textbf {\bibinfo {volume} {12}},\ \bibinfo {pages} {570--586} (\bibinfo {year} {1957})}\BibitemShut {NoStop}%
\bibitem [{\citenamefont {Helfand}(1960)}]{Helfand1960}%
  \BibitemOpen
  \bibfield  {author} {\bibinfo {author} {\bibfnamefont {E.}~\bibnamefont {Helfand}},\ }\bibfield  {title} {\enquote {\bibinfo {title} {Transport coefficients from dissipation in a canonical ensemble},}\ }\href {https://doi.org/10.1103/PhysRev.119.1} {\bibfield  {journal} {\bibinfo  {journal} {Phys. Rev.}\ }\textbf {\bibinfo {volume} {119}},\ \bibinfo {pages} {1--9} (\bibinfo {year} {1960})}\BibitemShut {NoStop}%
\bibitem [{\citenamefont {Hess}(2002)}]{hess2002d}%
  \BibitemOpen
  \bibfield  {author} {\bibinfo {author} {\bibfnamefont {B.}~\bibnamefont {Hess}},\ }\bibfield  {title} {\enquote {\bibinfo {title} {Determining the shear viscosity of model liquids from molecular dynamics simulations},}\ }\href@noop {} {\bibfield  {journal} {\bibinfo  {journal} {The Journal of chemical physics}\ }\textbf {\bibinfo {volume} {116}},\ \bibinfo {pages} {209--217} (\bibinfo {year} {2002})}\BibitemShut {NoStop}%
\bibitem [{\citenamefont {Nieszporek}, \citenamefont {Nieszporek},\ and\ \citenamefont {Trojak}(2016)}]{nieszporek2016c}%
  \BibitemOpen
  \bibfield  {author} {\bibinfo {author} {\bibfnamefont {K.}~\bibnamefont {Nieszporek}}, \bibinfo {author} {\bibfnamefont {J.}~\bibnamefont {Nieszporek}},\ and\ \bibinfo {author} {\bibfnamefont {M.}~\bibnamefont {Trojak}},\ }\bibfield  {title} {\enquote {\bibinfo {title} {Calculations of shear viscosity, electric conductivity and diffusion coefficients of aqueous sodium perchlorate solutions from molecular dynamics simulations},}\ }\href@noop {} {\bibfield  {journal} {\bibinfo  {journal} {Computational and Theoretical Chemistry}\ }\textbf {\bibinfo {volume} {1090}},\ \bibinfo {pages} {52--57} (\bibinfo {year} {2016})}\BibitemShut {NoStop}%
\bibitem [{\citenamefont {Jamali}\ \emph {et~al.}(2019)\citenamefont {Jamali}, \citenamefont {Wolff}, \citenamefont {Becker}, \citenamefont {De~Groen}, \citenamefont {Ramdin}, \citenamefont {Hartkamp}, \citenamefont {Bardow}, \citenamefont {Vlugt},\ and\ \citenamefont {Moultos}}]{jamali2019}%
  \BibitemOpen
  \bibfield  {author} {\bibinfo {author} {\bibfnamefont {S.~H.}\ \bibnamefont {Jamali}}, \bibinfo {author} {\bibfnamefont {L.}~\bibnamefont {Wolff}}, \bibinfo {author} {\bibfnamefont {T.~M.}\ \bibnamefont {Becker}}, \bibinfo {author} {\bibfnamefont {M.}~\bibnamefont {De~Groen}}, \bibinfo {author} {\bibfnamefont {M.}~\bibnamefont {Ramdin}}, \bibinfo {author} {\bibfnamefont {R.}~\bibnamefont {Hartkamp}}, \bibinfo {author} {\bibfnamefont {A.}~\bibnamefont {Bardow}}, \bibinfo {author} {\bibfnamefont {T.~J.}\ \bibnamefont {Vlugt}},\ and\ \bibinfo {author} {\bibfnamefont {O.~A.}\ \bibnamefont {Moultos}},\ }\bibfield  {title} {\enquote {\bibinfo {title} {{OCTP}: {A} tool for on-the-fly calculation of transport properties of fluids with the order-n algorithm in {LAMMPS}},}\ }\href@noop {} {\bibfield  {journal} {\bibinfo  {journal} {Journal of chemical information and modeling}\ }\textbf {\bibinfo {volume} {59}},\ \bibinfo {pages} {1290--1294} (\bibinfo {year} {2019})}\BibitemShut {NoStop}%
\bibitem [{\citenamefont {Mangaud}\ and\ \citenamefont {Rotenberg}(2020)}]{mangaud2020sampling}%
  \BibitemOpen
  \bibfield  {author} {\bibinfo {author} {\bibfnamefont {E.}~\bibnamefont {Mangaud}}\ and\ \bibinfo {author} {\bibfnamefont {B.}~\bibnamefont {Rotenberg}},\ }\bibfield  {title} {\enquote {\bibinfo {title} {Sampling mobility profiles of confined fluids with equilibrium molecular dynamics simulations},}\ }\href@noop {} {\bibfield  {journal} {\bibinfo  {journal} {The Journal of Chemical Physics}\ }\textbf {\bibinfo {volume} {153}} (\bibinfo {year} {2020})}\BibitemShut {NoStop}%
\bibitem [{\citenamefont {Celebi}\ \emph {et~al.}(2021)\citenamefont {Celebi}, \citenamefont {Jamali}, \citenamefont {Bardow}, \citenamefont {Vlugt},\ and\ \citenamefont {Moultos}}]{celebi2021finite}%
  \BibitemOpen
  \bibfield  {author} {\bibinfo {author} {\bibfnamefont {A.~T.}\ \bibnamefont {Celebi}}, \bibinfo {author} {\bibfnamefont {S.~H.}\ \bibnamefont {Jamali}}, \bibinfo {author} {\bibfnamefont {A.}~\bibnamefont {Bardow}}, \bibinfo {author} {\bibfnamefont {T.~J.}\ \bibnamefont {Vlugt}},\ and\ \bibinfo {author} {\bibfnamefont {O.~A.}\ \bibnamefont {Moultos}},\ }\bibfield  {title} {\enquote {\bibinfo {title} {Finite-size effects of diffusion coefficients computed from molecular dynamics: a review of what we have learned so far},}\ }\href@noop {} {\bibfield  {journal} {\bibinfo  {journal} {Molecular Simulation}\ }\textbf {\bibinfo {volume} {47}},\ \bibinfo {pages} {831--845} (\bibinfo {year} {2021})}\BibitemShut {NoStop}%
\bibitem [{\citenamefont {Li}\ and\ \citenamefont {Ni}(2023)}]{li2023md2d}%
  \BibitemOpen
  \bibfield  {author} {\bibinfo {author} {\bibfnamefont {Y.}~\bibnamefont {Li}}\ and\ \bibinfo {author} {\bibfnamefont {H.}~\bibnamefont {Ni}},\ }\bibfield  {title} {\enquote {\bibinfo {title} {{MD2D}: {A} {P}ython module for accurate determination of diffusion coefficient from molecular dynamics},}\ }\href@noop {} {\bibfield  {journal} {\bibinfo  {journal} {Computer Physics Communications}\ }\textbf {\bibinfo {volume} {284}},\ \bibinfo {pages} {108599} (\bibinfo {year} {2023})}\BibitemShut {NoStop}%
\bibitem [{\citenamefont {Malaspina}\ \emph {et~al.}(2023)\citenamefont {Malaspina}, \citenamefont {L{\'\i}sal}, \citenamefont {Larentzos}, \citenamefont {Brennan}, \citenamefont {Mackie},\ and\ \citenamefont {Avalos}}]{malaspina2023}%
  \BibitemOpen
  \bibfield  {author} {\bibinfo {author} {\bibfnamefont {D.}~\bibnamefont {Malaspina}}, \bibinfo {author} {\bibfnamefont {M.}~\bibnamefont {L{\'\i}sal}}, \bibinfo {author} {\bibfnamefont {J.}~\bibnamefont {Larentzos}}, \bibinfo {author} {\bibfnamefont {J.}~\bibnamefont {Brennan}}, \bibinfo {author} {\bibfnamefont {A.}~\bibnamefont {Mackie}},\ and\ \bibinfo {author} {\bibfnamefont {J.~B.}\ \bibnamefont {Avalos}},\ }\bibfield  {title} {\enquote {\bibinfo {title} {Transport coefficients from {E}instein--{H}elfand relations using standard and energy-conserving dissipative particle dynamics methods},}\ }\href@noop {} {\bibfield  {journal} {\bibinfo  {journal} {Physical Chemistry Chemical Physics}\ }\textbf {\bibinfo {volume} {25}},\ \bibinfo {pages} {12025--12040} (\bibinfo {year} {2023})}\BibitemShut {NoStop}%
\bibitem [{\citenamefont {Mabillard}\ and\ \citenamefont {Gaspard}(2023)}]{mabillard2023poles}%
  \BibitemOpen
  \bibfield  {author} {\bibinfo {author} {\bibfnamefont {J.}~\bibnamefont {Mabillard}}\ and\ \bibinfo {author} {\bibfnamefont {P.}~\bibnamefont {Gaspard}},\ }\bibfield  {title} {\enquote {\bibinfo {title} {Poles of hydrodynamic spectral functions and einstein--helfand formulas for transport coefficients},}\ }\href@noop {} {\bibfield  {journal} {\bibinfo  {journal} {Journal of Statistical Mechanics: Theory and Experiment}\ }\textbf {\bibinfo {volume} {2023}},\ \bibinfo {pages} {073206} (\bibinfo {year} {2023})}\BibitemShut {NoStop}%
\bibitem [{\citenamefont {Garc{\'\i}a-Rojo}, \citenamefont {Luding},\ and\ \citenamefont {Brey}(2006)}]{garcia2006t}%
  \BibitemOpen
  \bibfield  {author} {\bibinfo {author} {\bibfnamefont {R.}~\bibnamefont {Garc{\'\i}a-Rojo}}, \bibinfo {author} {\bibfnamefont {S.}~\bibnamefont {Luding}},\ and\ \bibinfo {author} {\bibfnamefont {J.~J.}\ \bibnamefont {Brey}},\ }\bibfield  {title} {\enquote {\bibinfo {title} {Transport coefficients for dense hard-disk systems},}\ }\href@noop {} {\bibfield  {journal} {\bibinfo  {journal} {Physical Review E—Statistical, Nonlinear, and Soft Matter Physics}\ }\textbf {\bibinfo {volume} {74}},\ \bibinfo {pages} {061305} (\bibinfo {year} {2006})}\BibitemShut {NoStop}%
\bibitem [{\citenamefont {Bannerman}\ and\ \citenamefont {Lue}(2009)}]{bannerman2009t}%
  \BibitemOpen
  \bibfield  {author} {\bibinfo {author} {\bibfnamefont {M.~N.}\ \bibnamefont {Bannerman}}\ and\ \bibinfo {author} {\bibfnamefont {L.}~\bibnamefont {Lue}},\ }\bibfield  {title} {\enquote {\bibinfo {title} {Transport properties of highly asymmetric hard-sphere mixtures},}\ }\href@noop {} {\bibfield  {journal} {\bibinfo  {journal} {The Journal of chemical physics}\ }\textbf {\bibinfo {volume} {130}} (\bibinfo {year} {2009})}\BibitemShut {NoStop}%
\bibitem [{\citenamefont {Heyes}, \citenamefont {Pieprzyk},\ and\ \citenamefont {Bra{\'n}ka}(2022)}]{heyes2022bulk}%
  \BibitemOpen
  \bibfield  {author} {\bibinfo {author} {\bibfnamefont {D.}~\bibnamefont {Heyes}}, \bibinfo {author} {\bibfnamefont {S.}~\bibnamefont {Pieprzyk}},\ and\ \bibinfo {author} {\bibfnamefont {A.}~\bibnamefont {Bra{\'n}ka}},\ }\bibfield  {title} {\enquote {\bibinfo {title} {Bulk viscosity of hard sphere fluids by equilibrium and nonequilibrium molecular dynamics simulations},}\ }\href@noop {} {\bibfield  {journal} {\bibinfo  {journal} {The Journal of Chemical Physics}\ }\textbf {\bibinfo {volume} {157}} (\bibinfo {year} {2022})}\BibitemShut {NoStop}%
\bibitem [{\citenamefont {Pieprzyk}\ \emph {et~al.}(2024)\citenamefont {Pieprzyk}, \citenamefont {Bra{\'n}ka}, \citenamefont {Heyes},\ and\ \citenamefont {Bannerman}}]{pieprzyk2024r}%
  \BibitemOpen
  \bibfield  {author} {\bibinfo {author} {\bibfnamefont {S.}~\bibnamefont {Pieprzyk}}, \bibinfo {author} {\bibfnamefont {A.~C.}\ \bibnamefont {Bra{\'n}ka}}, \bibinfo {author} {\bibfnamefont {D.~M.}\ \bibnamefont {Heyes}},\ and\ \bibinfo {author} {\bibfnamefont {M.~N.}\ \bibnamefont {Bannerman}},\ }\bibfield  {title} {\enquote {\bibinfo {title} {Revised enskog theory and molecular dynamics simulations of the viscosities and thermal conductivity of the hard-sphere fluid and crystal},}\ }\href@noop {} {\bibfield  {journal} {\bibinfo  {journal} {Physical Review E}\ }\textbf {\bibinfo {volume} {109}},\ \bibinfo {pages} {054119} (\bibinfo {year} {2024})}\BibitemShut {NoStop}%
\bibitem [{\citenamefont {Pieprzyk}\ \emph {et~al.}(2019)\citenamefont {Pieprzyk}, \citenamefont {Bannerman}, \citenamefont {Bra{\'n}ka}, \citenamefont {Chudak},\ and\ \citenamefont {Heyes}}]{pieprzyk2019t}%
  \BibitemOpen
  \bibfield  {author} {\bibinfo {author} {\bibfnamefont {S.}~\bibnamefont {Pieprzyk}}, \bibinfo {author} {\bibfnamefont {M.~N.}\ \bibnamefont {Bannerman}}, \bibinfo {author} {\bibfnamefont {A.~C.}\ \bibnamefont {Bra{\'n}ka}}, \bibinfo {author} {\bibfnamefont {M.}~\bibnamefont {Chudak}},\ and\ \bibinfo {author} {\bibfnamefont {D.~M.}\ \bibnamefont {Heyes}},\ }\bibfield  {title} {\enquote {\bibinfo {title} {Thermodynamic and dynamical properties of the hard sphere system revisited by molecular dynamics simulation},}\ }\href@noop {} {\bibfield  {journal} {\bibinfo  {journal} {Physical Chemistry Chemical Physics}\ }\textbf {\bibinfo {volume} {21}},\ \bibinfo {pages} {6886--6899} (\bibinfo {year} {2019})}\BibitemShut {NoStop}%
\bibitem [{\citenamefont {Moustafa}, \citenamefont {Schultz},\ and\ \citenamefont {Douglas}(2024)}]{moustafa2024}%
  \BibitemOpen
  \bibfield  {author} {\bibinfo {author} {\bibfnamefont {S.~G.}\ \bibnamefont {Moustafa}}, \bibinfo {author} {\bibfnamefont {A.~J.}\ \bibnamefont {Schultz}},\ and\ \bibinfo {author} {\bibfnamefont {J.~F.}\ \bibnamefont {Douglas}},\ }\bibfield  {title} {\enquote {\bibinfo {title} {Efficient single-run implementation of generalized einstein relation to compute transport coefficients: A binary-based time sampling},}\ }\href@noop {} {\bibfield  {journal} {\bibinfo  {journal} {The Journal of Chemical Physics}\ }\textbf {\bibinfo {volume} {160}} (\bibinfo {year} {2024})}\BibitemShut {NoStop}%
\bibitem [{\citenamefont {Yeh}\ and\ \citenamefont {Hummer}(2004)}]{yeh2004system}%
  \BibitemOpen
  \bibfield  {author} {\bibinfo {author} {\bibfnamefont {I.-C.}\ \bibnamefont {Yeh}}\ and\ \bibinfo {author} {\bibfnamefont {G.}~\bibnamefont {Hummer}},\ }\bibfield  {title} {\enquote {\bibinfo {title} {System-size dependence of diffusion coefficients and viscosities from molecular dynamics simulations with periodic boundary conditions},}\ }\href@noop {} {\bibfield  {journal} {\bibinfo  {journal} {The Journal of Physical Chemistry B}\ }\textbf {\bibinfo {volume} {108}},\ \bibinfo {pages} {15873--15879} (\bibinfo {year} {2004})}\BibitemShut {NoStop}%
\bibitem [{\citenamefont {Krekelberg}\ \emph {et~al.}(2007)\citenamefont {Krekelberg}, \citenamefont {Mittal}, \citenamefont {Ganesan},\ and\ \citenamefont {Truskett}}]{krekelberg2007short}%
  \BibitemOpen
  \bibfield  {author} {\bibinfo {author} {\bibfnamefont {W.~P.}\ \bibnamefont {Krekelberg}}, \bibinfo {author} {\bibfnamefont {J.}~\bibnamefont {Mittal}}, \bibinfo {author} {\bibfnamefont {V.}~\bibnamefont {Ganesan}},\ and\ \bibinfo {author} {\bibfnamefont {T.~M.}\ \bibnamefont {Truskett}},\ }\bibfield  {title} {\enquote {\bibinfo {title} {How short-range attractions impact the structural order, self-diffusivity, and viscosity of a fluid},}\ }\href@noop {} {\bibfield  {journal} {\bibinfo  {journal} {The Journal of chemical physics}\ }\textbf {\bibinfo {volume} {127}} (\bibinfo {year} {2007})}\BibitemShut {NoStop}%
\bibitem [{\citenamefont {Flenner}\ and\ \citenamefont {Szamel}(2010)}]{flenner2010dynamic}%
  \BibitemOpen
  \bibfield  {author} {\bibinfo {author} {\bibfnamefont {E.}~\bibnamefont {Flenner}}\ and\ \bibinfo {author} {\bibfnamefont {G.}~\bibnamefont {Szamel}},\ }\bibfield  {title} {\enquote {\bibinfo {title} {Dynamic heterogeneity in a glass forming fluid: Susceptibility,structure factor, and correlation length},}\ }\href@noop {} {\bibfield  {journal} {\bibinfo  {journal} {Physical review letters}\ }\textbf {\bibinfo {volume} {105}},\ \bibinfo {pages} {217801} (\bibinfo {year} {2010})}\BibitemShut {NoStop}%
\bibitem [{\citenamefont {Schober}\ and\ \citenamefont {Peng}(2016)}]{schober2016h}%
  \BibitemOpen
  \bibfield  {author} {\bibinfo {author} {\bibfnamefont {H.}~\bibnamefont {Schober}}\ and\ \bibinfo {author} {\bibfnamefont {H.}~\bibnamefont {Peng}},\ }\bibfield  {title} {\enquote {\bibinfo {title} {Heterogeneous diffusion, viscosity, and the stokes-einstein relation in binary liquids},}\ }\href@noop {} {\bibfield  {journal} {\bibinfo  {journal} {Physical Review E}\ }\textbf {\bibinfo {volume} {93}},\ \bibinfo {pages} {052607} (\bibinfo {year} {2016})}\BibitemShut {NoStop}%
\bibitem [{\citenamefont {Yamamoto}\ and\ \citenamefont {Onuki}(1998)}]{yamamoto1998}%
  \BibitemOpen
  \bibfield  {author} {\bibinfo {author} {\bibfnamefont {R.}~\bibnamefont {Yamamoto}}\ and\ \bibinfo {author} {\bibfnamefont {A.}~\bibnamefont {Onuki}},\ }\bibfield  {title} {\enquote {\bibinfo {title} {Heterogeneous diffusion in highly supercooled liquids},}\ }\href@noop {} {\bibfield  {journal} {\bibinfo  {journal} {Physical review letters}\ }\textbf {\bibinfo {volume} {81}},\ \bibinfo {pages} {4915} (\bibinfo {year} {1998})}\BibitemShut {NoStop}%
\bibitem [{\citenamefont {Sengupta}\ \emph {et~al.}(2013)\citenamefont {Sengupta}, \citenamefont {Karmakar}, \citenamefont {Dasgupta},\ and\ \citenamefont {Sastry}}]{sengupta2013}%
  \BibitemOpen
  \bibfield  {author} {\bibinfo {author} {\bibfnamefont {S.}~\bibnamefont {Sengupta}}, \bibinfo {author} {\bibfnamefont {S.}~\bibnamefont {Karmakar}}, \bibinfo {author} {\bibfnamefont {C.}~\bibnamefont {Dasgupta}},\ and\ \bibinfo {author} {\bibfnamefont {S.}~\bibnamefont {Sastry}},\ }\bibfield  {title} {\enquote {\bibinfo {title} {Breakdown of the stokes-einstein relation in two, three, and four dimensions},}\ }\href@noop {} {\bibfield  {journal} {\bibinfo  {journal} {The Journal of chemical physics}\ }\textbf {\bibinfo {volume} {138}} (\bibinfo {year} {2013})}\BibitemShut {NoStop}%
\bibitem [{\citenamefont {Kawasaki}\ and\ \citenamefont {Kim}(2017)}]{kawasaki2017}%
  \BibitemOpen
  \bibfield  {author} {\bibinfo {author} {\bibfnamefont {T.}~\bibnamefont {Kawasaki}}\ and\ \bibinfo {author} {\bibfnamefont {K.}~\bibnamefont {Kim}},\ }\bibfield  {title} {\enquote {\bibinfo {title} {Identifying time scales for violation/preservation of stokes-einstein relation in supercooled water},}\ }\href@noop {} {\bibfield  {journal} {\bibinfo  {journal} {Science Advances}\ }\textbf {\bibinfo {volume} {3}},\ \bibinfo {pages} {e1700399} (\bibinfo {year} {2017})}\BibitemShut {NoStop}%
\bibitem [{\citenamefont {Kawasaki}, \citenamefont {Kim},\ and\ \citenamefont {Onuki}(2014)}]{kawasaki2014d}%
  \BibitemOpen
  \bibfield  {author} {\bibinfo {author} {\bibfnamefont {T.}~\bibnamefont {Kawasaki}}, \bibinfo {author} {\bibfnamefont {K.}~\bibnamefont {Kim}},\ and\ \bibinfo {author} {\bibfnamefont {A.}~\bibnamefont {Onuki}},\ }\bibfield  {title} {\enquote {\bibinfo {title} {Dynamics in a tetrahedral network glassformer: Vibrations, network rearrangements, and diffusion},}\ }\href@noop {} {\bibfield  {journal} {\bibinfo  {journal} {The Journal of Chemical Physics}\ }\textbf {\bibinfo {volume} {140}} (\bibinfo {year} {2014})}\BibitemShut {NoStop}%
\bibitem [{\citenamefont {Shi}, \citenamefont {Debenedetti},\ and\ \citenamefont {Stillinger}(2013)}]{shi2013r}%
  \BibitemOpen
  \bibfield  {author} {\bibinfo {author} {\bibfnamefont {Z.}~\bibnamefont {Shi}}, \bibinfo {author} {\bibfnamefont {P.~G.}\ \bibnamefont {Debenedetti}},\ and\ \bibinfo {author} {\bibfnamefont {F.~H.}\ \bibnamefont {Stillinger}},\ }\bibfield  {title} {\enquote {\bibinfo {title} {Relaxation processes in liquids: Variations on a theme by stokes and einstein},}\ }\href@noop {} {\bibfield  {journal} {\bibinfo  {journal} {The Journal of chemical physics}\ }\textbf {\bibinfo {volume} {138}} (\bibinfo {year} {2013})}\BibitemShut {NoStop}%
\bibitem [{\citenamefont {Flenner}\ and\ \citenamefont {Szamel}(2019)}]{flenner2019v}%
  \BibitemOpen
  \bibfield  {author} {\bibinfo {author} {\bibfnamefont {E.}~\bibnamefont {Flenner}}\ and\ \bibinfo {author} {\bibfnamefont {G.}~\bibnamefont {Szamel}},\ }\bibfield  {title} {\enquote {\bibinfo {title} {Viscoelastic shear stress relaxation in two-dimensional glass-forming liquids},}\ }\href@noop {} {\bibfield  {journal} {\bibinfo  {journal} {Proceedings of the National Academy of Sciences}\ }\textbf {\bibinfo {volume} {116}},\ \bibinfo {pages} {2015--2020} (\bibinfo {year} {2019})}\BibitemShut {NoStop}%
\bibitem [{\citenamefont {Chaudhuri}, \citenamefont {Berthier},\ and\ \citenamefont {Kob}(2007)}]{chaudhuri2007u}%
  \BibitemOpen
  \bibfield  {author} {\bibinfo {author} {\bibfnamefont {P.}~\bibnamefont {Chaudhuri}}, \bibinfo {author} {\bibfnamefont {L.}~\bibnamefont {Berthier}},\ and\ \bibinfo {author} {\bibfnamefont {W.}~\bibnamefont {Kob}},\ }\bibfield  {title} {\enquote {\bibinfo {title} {Universal nature of particle displacements close to glass and jamming transitions},}\ }\href@noop {} {\bibfield  {journal} {\bibinfo  {journal} {Physical review letters}\ }\textbf {\bibinfo {volume} {99}},\ \bibinfo {pages} {060604} (\bibinfo {year} {2007})}\BibitemShut {NoStop}%
\bibitem [{\citenamefont {Kob}\ \emph {et~al.}(1997)\citenamefont {Kob}, \citenamefont {Donati}, \citenamefont {Plimpton}, \citenamefont {Poole},\ and\ \citenamefont {Glotzer}}]{kob1997d}%
  \BibitemOpen
  \bibfield  {author} {\bibinfo {author} {\bibfnamefont {W.}~\bibnamefont {Kob}}, \bibinfo {author} {\bibfnamefont {C.}~\bibnamefont {Donati}}, \bibinfo {author} {\bibfnamefont {S.~J.}\ \bibnamefont {Plimpton}}, \bibinfo {author} {\bibfnamefont {P.~H.}\ \bibnamefont {Poole}},\ and\ \bibinfo {author} {\bibfnamefont {S.~C.}\ \bibnamefont {Glotzer}},\ }\bibfield  {title} {\enquote {\bibinfo {title} {Dynamical heterogeneities in a supercooled lennard-jones liquid},}\ }\href@noop {} {\bibfield  {journal} {\bibinfo  {journal} {Physical review letters}\ }\textbf {\bibinfo {volume} {79}},\ \bibinfo {pages} {2827} (\bibinfo {year} {1997})}\BibitemShut {NoStop}%
\bibitem [{\citenamefont {Glotzer}, \citenamefont {Novikov},\ and\ \citenamefont {Schr{\o}der}(2000)}]{glotzer2000time}%
  \BibitemOpen
  \bibfield  {author} {\bibinfo {author} {\bibfnamefont {S.~C.}\ \bibnamefont {Glotzer}}, \bibinfo {author} {\bibfnamefont {V.~N.}\ \bibnamefont {Novikov}},\ and\ \bibinfo {author} {\bibfnamefont {T.~B.}\ \bibnamefont {Schr{\o}der}},\ }\bibfield  {title} {\enquote {\bibinfo {title} {Time-dependent, four-point density correlation function description of dynamical heterogeneity and decoupling in supercooled liquids},}\ }\href@noop {} {\bibfield  {journal} {\bibinfo  {journal} {The Journal of Chemical Physics}\ }\textbf {\bibinfo {volume} {112}},\ \bibinfo {pages} {509--512} (\bibinfo {year} {2000})}\BibitemShut {NoStop}%
\bibitem [{\citenamefont {Alder}, \citenamefont {Gass},\ and\ \citenamefont {Wainwright}(1970)}]{Alder1970}%
  \BibitemOpen
  \bibfield  {author} {\bibinfo {author} {\bibfnamefont {B.~J.}\ \bibnamefont {Alder}}, \bibinfo {author} {\bibfnamefont {D.~M.}\ \bibnamefont {Gass}},\ and\ \bibinfo {author} {\bibfnamefont {T.~E.}\ \bibnamefont {Wainwright}},\ }\bibfield  {title} {\enquote {\bibinfo {title} {Studies in molecular dynamics. viii. the transport coefficients for a hard‐sphere fluid},}\ }\href {https://doi.org/10.1063/1.1673845} {\bibfield  {journal} {\bibinfo  {journal} {The Journal of Chemical Physics}\ }\textbf {\bibinfo {volume} {53}},\ \bibinfo {pages} {3813--3826} (\bibinfo {year} {1970})}\BibitemShut {NoStop}%
\bibitem [{\citenamefont {Haile}(1992)}]{haile_book}%
  \BibitemOpen
  \bibfield  {author} {\bibinfo {author} {\bibfnamefont {J.~M.}\ \bibnamefont {Haile}},\ }\href@noop {} {\emph {\bibinfo {title} {Molecular dynamics simulation: {E}lementary methods}}}\ (\bibinfo  {publisher} {John Wiley \& Sons, Inc.},\ \bibinfo {year} {1992})\BibitemShut {NoStop}%
\bibitem [{\citenamefont {Alder}\ and\ \citenamefont {Wainwright}(1960)}]{alder1960s}%
  \BibitemOpen
  \bibfield  {author} {\bibinfo {author} {\bibfnamefont {B.~J.}\ \bibnamefont {Alder}}\ and\ \bibinfo {author} {\bibfnamefont {T.~E.}\ \bibnamefont {Wainwright}},\ }\bibfield  {title} {\enquote {\bibinfo {title} {Studies in molecular dynamics. ii. behavior of a small number of elastic spheres},}\ }\href@noop {} {\bibfield  {journal} {\bibinfo  {journal} {The Journal of Chemical Physics}\ }\textbf {\bibinfo {volume} {33}},\ \bibinfo {pages} {1439--1451} (\bibinfo {year} {1960})}\BibitemShut {NoStop}%
\bibitem [{\citenamefont {Verma}, \citenamefont {Thorat},\ and\ \citenamefont {Shah}(2026)}]{verma2026md}%
  \BibitemOpen
  \bibfield  {author} {\bibinfo {author} {\bibfnamefont {A.~K.}\ \bibnamefont {Verma}}, \bibinfo {author} {\bibfnamefont {A.~S.}\ \bibnamefont {Thorat}},\ and\ \bibinfo {author} {\bibfnamefont {J.~K.}\ \bibnamefont {Shah}},\ }\bibfield  {title} {\enquote {\bibinfo {title} {Mdtransport: A modular, open-source, extensible python tool for computing transport properties from gromacs and lammps simulations},}\ }\href@noop {} {\bibfield  {journal} {\bibinfo  {journal} {Journal of Chemical Information and Modeling}\ } (\bibinfo {year} {2026})}\BibitemShut {NoStop}%
\bibitem [{Han(2013)}]{HansenBook2013}%
  \BibitemOpen
  \bibfield  {title} {\enquote {\bibinfo {title} {Theory of simple liquids},}\ }in\ \href {https://doi.org/https://doi.org/10.1016/B978-0-12-387032-2.00013-1} {\emph {\bibinfo {booktitle} {Theory of Simple Liquids (Fourth Edition)}}},\ \bibinfo {editor} {edited by\ \bibinfo {editor} {\bibfnamefont {J.-P.}\ \bibnamefont {Hansen}}\ and\ \bibinfo {editor} {\bibfnamefont {I.~R.}\ \bibnamefont {McDonald}}}\ (\bibinfo  {publisher} {Academic Press},\ \bibinfo {address} {Oxford},\ \bibinfo {year} {2013})\ \bibinfo {edition} {fourth edition}\ ed.,\ p.~\bibinfo {pages} {i}\BibitemShut {NoStop}%
\bibitem [{\citenamefont {Dandekar}, \citenamefont {Bose},\ and\ \citenamefont {Dutta}(2020)}]{dandekar2020n}%
  \BibitemOpen
  \bibfield  {author} {\bibinfo {author} {\bibfnamefont {R.}~\bibnamefont {Dandekar}}, \bibinfo {author} {\bibfnamefont {S.}~\bibnamefont {Bose}},\ and\ \bibinfo {author} {\bibfnamefont {S.}~\bibnamefont {Dutta}},\ }\bibfield  {title} {\enquote {\bibinfo {title} {Non-gaussian information of heterogeneity in soft matter},}\ }\href@noop {} {\bibfield  {journal} {\bibinfo  {journal} {Europhysics Letters}\ }\textbf {\bibinfo {volume} {131}},\ \bibinfo {pages} {18002} (\bibinfo {year} {2020})}\BibitemShut {NoStop}%
\bibitem [{\citenamefont {O'Hern}\ \emph {et~al.}(2002)\citenamefont {O'Hern}, \citenamefont {Langer}, \citenamefont {Liu},\ and\ \citenamefont {Nagel}}]{hern2002random}%
  \BibitemOpen
  \bibfield  {author} {\bibinfo {author} {\bibfnamefont {C.~S.}\ \bibnamefont {O'Hern}}, \bibinfo {author} {\bibfnamefont {S.~A.}\ \bibnamefont {Langer}}, \bibinfo {author} {\bibfnamefont {A.~J.}\ \bibnamefont {Liu}},\ and\ \bibinfo {author} {\bibfnamefont {S.~R.}\ \bibnamefont {Nagel}},\ }\bibfield  {title} {\enquote {\bibinfo {title} {Random packings of frictionless particles},}\ }\href@noop {} {\bibfield  {journal} {\bibinfo  {journal} {Physical Review Letters}\ }\textbf {\bibinfo {volume} {88}},\ \bibinfo {pages} {075507} (\bibinfo {year} {2002})}\BibitemShut {NoStop}%
\bibitem [{\citenamefont {Brambilla}\ \emph {et~al.}(2009)\citenamefont {Brambilla}, \citenamefont {El~Masri}, \citenamefont {Pierno}, \citenamefont {Berthier}, \citenamefont {Cipelletti}, \citenamefont {Petekidis},\ and\ \citenamefont {Schofield}}]{brambilla2009probing}%
  \BibitemOpen
  \bibfield  {author} {\bibinfo {author} {\bibfnamefont {G.}~\bibnamefont {Brambilla}}, \bibinfo {author} {\bibfnamefont {D.}~\bibnamefont {El~Masri}}, \bibinfo {author} {\bibfnamefont {M.}~\bibnamefont {Pierno}}, \bibinfo {author} {\bibfnamefont {L.}~\bibnamefont {Berthier}}, \bibinfo {author} {\bibfnamefont {L.}~\bibnamefont {Cipelletti}}, \bibinfo {author} {\bibfnamefont {G.}~\bibnamefont {Petekidis}},\ and\ \bibinfo {author} {\bibfnamefont {A.~B.}\ \bibnamefont {Schofield}},\ }\bibfield  {title} {\enquote {\bibinfo {title} {Probing the equilibrium dynamics of colloidal hard spheres above the mode-coupling glass transition},}\ }\href@noop {} {\bibfield  {journal} {\bibinfo  {journal} {Physical review letters}\ }\textbf {\bibinfo {volume} {102}},\ \bibinfo {pages} {085703} (\bibinfo {year} {2009})}\BibitemShut {NoStop}%
\bibitem [{\citenamefont {Flenner}, \citenamefont {Staley},\ and\ \citenamefont {Szamel}(2014)}]{flenner2014u}%
  \BibitemOpen
  \bibfield  {author} {\bibinfo {author} {\bibfnamefont {E.}~\bibnamefont {Flenner}}, \bibinfo {author} {\bibfnamefont {H.}~\bibnamefont {Staley}},\ and\ \bibinfo {author} {\bibfnamefont {G.}~\bibnamefont {Szamel}},\ }\bibfield  {title} {\enquote {\bibinfo {title} {Universal features of dynamic heterogeneity in supercooled liquids},}\ }\href@noop {} {\bibfield  {journal} {\bibinfo  {journal} {Physical review letters}\ }\textbf {\bibinfo {volume} {112}},\ \bibinfo {pages} {097801} (\bibinfo {year} {2014})}\BibitemShut {NoStop}%
\bibitem [{\citenamefont {Callaham}\ and\ \citenamefont {Machta}(2017)}]{callaham2017p}%
  \BibitemOpen
  \bibfield  {author} {\bibinfo {author} {\bibfnamefont {J.}~\bibnamefont {Callaham}}\ and\ \bibinfo {author} {\bibfnamefont {J.}~\bibnamefont {Machta}},\ }\bibfield  {title} {\enquote {\bibinfo {title} {Population annealing simulations of a binary hard-sphere mixture},}\ }\href@noop {} {\bibfield  {journal} {\bibinfo  {journal} {Physical Review E}\ }\textbf {\bibinfo {volume} {95}},\ \bibinfo {pages} {063315} (\bibinfo {year} {2017})}\BibitemShut {NoStop}%
\bibitem [{\citenamefont {Schultz}\ and\ \citenamefont {Kofke}(2015)}]{schultz2015etomica}%
  \BibitemOpen
  \bibfield  {author} {\bibinfo {author} {\bibfnamefont {A.~J.}\ \bibnamefont {Schultz}}\ and\ \bibinfo {author} {\bibfnamefont {D.~A.}\ \bibnamefont {Kofke}},\ }\href@noop {} {\enquote {\bibinfo {title} {Etomica: An object-oriented framework for molecular simulation},}\ } (\bibinfo {year} {2015})\BibitemShut {NoStop}%
\bibitem [{\citenamefont {van Swol}\ and\ \citenamefont {Petsev}(2014)}]{Swol2014m}%
  \BibitemOpen
  \bibfield  {author} {\bibinfo {author} {\bibfnamefont {F.}~\bibnamefont {van Swol}}\ and\ \bibinfo {author} {\bibfnamefont {D.~N.}\ \bibnamefont {Petsev}},\ }\bibfield  {title} {\enquote {\bibinfo {title} {Molecular dynamics simulation of binary hard sphere colloids near the glass transition},}\ }\href@noop {} {\bibfield  {journal} {\bibinfo  {journal} {RSC Advances}\ }\textbf {\bibinfo {volume} {4}},\ \bibinfo {pages} {21631--21637} (\bibinfo {year} {2014})}\BibitemShut {NoStop}%
\bibitem [{\citenamefont {Heyes}\ \emph {et~al.}(2007)\citenamefont {Heyes}, \citenamefont {Cass}, \citenamefont {Powles},\ and\ \citenamefont {Evans}}]{heyes2007self}%
  \BibitemOpen
  \bibfield  {author} {\bibinfo {author} {\bibfnamefont {D.~M.}\ \bibnamefont {Heyes}}, \bibinfo {author} {\bibfnamefont {M.}~\bibnamefont {Cass}}, \bibinfo {author} {\bibfnamefont {J.~G.}\ \bibnamefont {Powles}},\ and\ \bibinfo {author} {\bibfnamefont {W.}~\bibnamefont {Evans}},\ }\bibfield  {title} {\enquote {\bibinfo {title} {Self-diffusion coefficient of the hard-sphere fluid: System size dependence and empirical correlations},}\ }\href@noop {} {\bibfield  {journal} {\bibinfo  {journal} {The Journal of Physical Chemistry B}\ }\textbf {\bibinfo {volume} {111}},\ \bibinfo {pages} {1455--1464} (\bibinfo {year} {2007})}\BibitemShut {NoStop}%
\bibitem [{\citenamefont {Daivis}\ and\ \citenamefont {Evans}(1995)}]{daivis1995transport}%
  \BibitemOpen
  \bibfield  {author} {\bibinfo {author} {\bibfnamefont {P.~J.}\ \bibnamefont {Daivis}}\ and\ \bibinfo {author} {\bibfnamefont {D.~J.}\ \bibnamefont {Evans}},\ }\bibfield  {title} {\enquote {\bibinfo {title} {Transport coefficients of liquid butane near the boiling point by equilibrium molecular dynamics},}\ }\href@noop {} {\bibfield  {journal} {\bibinfo  {journal} {The Journal of chemical physics}\ }\textbf {\bibinfo {volume} {103}},\ \bibinfo {pages} {4261--4265} (\bibinfo {year} {1995})}\BibitemShut {NoStop}%
\bibitem [{\citenamefont {Meier}, \citenamefont {Laesecke},\ and\ \citenamefont {Kabelac}(2004)}]{meier2004transport}%
  \BibitemOpen
  \bibfield  {author} {\bibinfo {author} {\bibfnamefont {K.}~\bibnamefont {Meier}}, \bibinfo {author} {\bibfnamefont {A.}~\bibnamefont {Laesecke}},\ and\ \bibinfo {author} {\bibfnamefont {S.}~\bibnamefont {Kabelac}},\ }\bibfield  {title} {\enquote {\bibinfo {title} {Transport coefficients of the lennard-jones model fluid. i. viscosity},}\ }\href@noop {} {\bibfield  {journal} {\bibinfo  {journal} {The Journal of chemical physics}\ }\textbf {\bibinfo {volume} {121}},\ \bibinfo {pages} {3671--3687} (\bibinfo {year} {2004})}\BibitemShut {NoStop}%
\bibitem [{\citenamefont {Viscardy}, \citenamefont {Servantie},\ and\ \citenamefont {Gaspard}(2007)}]{viscardy2007t}%
  \BibitemOpen
  \bibfield  {author} {\bibinfo {author} {\bibfnamefont {S.}~\bibnamefont {Viscardy}}, \bibinfo {author} {\bibfnamefont {J.}~\bibnamefont {Servantie}},\ and\ \bibinfo {author} {\bibfnamefont {P.}~\bibnamefont {Gaspard}},\ }\bibfield  {title} {\enquote {\bibinfo {title} {Transport and helfand moments in the lennard-jones fluid. i. shear viscosity},}\ }\href@noop {} {\bibfield  {journal} {\bibinfo  {journal} {The Journal of chemical physics}\ }\textbf {\bibinfo {volume} {126}} (\bibinfo {year} {2007})}\BibitemShut {NoStop}%
\bibitem [{\citenamefont {Moultos}\ \emph {et~al.}(2016)\citenamefont {Moultos}, \citenamefont {Zhang}, \citenamefont {Tsimpanogiannis}, \citenamefont {Economou},\ and\ \citenamefont {Maginn}}]{moultos2016system}%
  \BibitemOpen
  \bibfield  {author} {\bibinfo {author} {\bibfnamefont {O.~A.}\ \bibnamefont {Moultos}}, \bibinfo {author} {\bibfnamefont {Y.}~\bibnamefont {Zhang}}, \bibinfo {author} {\bibfnamefont {I.~N.}\ \bibnamefont {Tsimpanogiannis}}, \bibinfo {author} {\bibfnamefont {I.~G.}\ \bibnamefont {Economou}},\ and\ \bibinfo {author} {\bibfnamefont {E.~J.}\ \bibnamefont {Maginn}},\ }\bibfield  {title} {\enquote {\bibinfo {title} {System-size corrections for self-diffusion coefficients calculated from molecular dynamics simulations: The case of co2, n-alkanes, and poly (ethylene glycol) dimethyl ethers},}\ }\href@noop {} {\bibfield  {journal} {\bibinfo  {journal} {The Journal of Chemical Physics}\ }\textbf {\bibinfo {volume} {145}} (\bibinfo {year} {2016})}\BibitemShut {NoStop}%
\bibitem [{\citenamefont {Jamali}\ \emph {et~al.}(2018)\citenamefont {Jamali}, \citenamefont {Wolff}, \citenamefont {Becker}, \citenamefont {Bardow}, \citenamefont {Vlugt},\ and\ \citenamefont {Moultos}}]{jamali2018finite}%
  \BibitemOpen
  \bibfield  {author} {\bibinfo {author} {\bibfnamefont {S.~H.}\ \bibnamefont {Jamali}}, \bibinfo {author} {\bibfnamefont {L.}~\bibnamefont {Wolff}}, \bibinfo {author} {\bibfnamefont {T.~M.}\ \bibnamefont {Becker}}, \bibinfo {author} {\bibfnamefont {A.}~\bibnamefont {Bardow}}, \bibinfo {author} {\bibfnamefont {T.~J.}\ \bibnamefont {Vlugt}},\ and\ \bibinfo {author} {\bibfnamefont {O.~A.}\ \bibnamefont {Moultos}},\ }\bibfield  {title} {\enquote {\bibinfo {title} {Finite-size effects of binary mutual diffusion coefficients from molecular dynamics},}\ }\href@noop {} {\bibfield  {journal} {\bibinfo  {journal} {Journal of chemical theory and computation}\ }\textbf {\bibinfo {volume} {14}},\ \bibinfo {pages} {2667--2677} (\bibinfo {year} {2018})}\BibitemShut {NoStop}%
\bibitem [{\citenamefont {Charbonneau}, \citenamefont {Charbonneau},\ and\ \citenamefont {Tarjus}(2012)}]{charbonneau2012g}%
  \BibitemOpen
  \bibfield  {author} {\bibinfo {author} {\bibfnamefont {B.}~\bibnamefont {Charbonneau}}, \bibinfo {author} {\bibfnamefont {P.}~\bibnamefont {Charbonneau}},\ and\ \bibinfo {author} {\bibfnamefont {G.}~\bibnamefont {Tarjus}},\ }\bibfield  {title} {\enquote {\bibinfo {title} {Geometrical frustration and static correlations in a simple glass former},}\ }\href@noop {} {\bibfield  {journal} {\bibinfo  {journal} {Physical review letters}\ }\textbf {\bibinfo {volume} {108}},\ \bibinfo {pages} {035701} (\bibinfo {year} {2012})}\BibitemShut {NoStop}%
\bibitem [{\citenamefont {Berthier}\ and\ \citenamefont {Witten}(2009)}]{berthier2009glass}%
  \BibitemOpen
  \bibfield  {author} {\bibinfo {author} {\bibfnamefont {L.}~\bibnamefont {Berthier}}\ and\ \bibinfo {author} {\bibfnamefont {T.~A.}\ \bibnamefont {Witten}},\ }\bibfield  {title} {\enquote {\bibinfo {title} {Glass transition of dense fluids of hard and compressible spheres},}\ }\href@noop {} {\bibfield  {journal} {\bibinfo  {journal} {Physical Review E—Statistical, Nonlinear, and Soft Matter Physics}\ }\textbf {\bibinfo {volume} {80}},\ \bibinfo {pages} {021502} (\bibinfo {year} {2009})}\BibitemShut {NoStop}%
\bibitem [{\citenamefont {Odriozola}\ and\ \citenamefont {Berthier}(2011)}]{odriozola2011}%
  \BibitemOpen
  \bibfield  {author} {\bibinfo {author} {\bibfnamefont {G.}~\bibnamefont {Odriozola}}\ and\ \bibinfo {author} {\bibfnamefont {L.}~\bibnamefont {Berthier}},\ }\bibfield  {title} {\enquote {\bibinfo {title} {Equilibrium equation of state of a hard sphere binary mixture at very large densities using replica exchange monte carlo simulations},}\ }\href@noop {} {\bibfield  {journal} {\bibinfo  {journal} {The Journal of chemical physics}\ }\textbf {\bibinfo {volume} {134}} (\bibinfo {year} {2011})}\BibitemShut {NoStop}%
\bibitem [{\citenamefont {Ozawa}\ \emph {et~al.}(2012)\citenamefont {Ozawa}, \citenamefont {Kuroiwa}, \citenamefont {Ikeda},\ and\ \citenamefont {Miyazaki}}]{ozawa2012j}%
  \BibitemOpen
  \bibfield  {author} {\bibinfo {author} {\bibfnamefont {M.}~\bibnamefont {Ozawa}}, \bibinfo {author} {\bibfnamefont {T.}~\bibnamefont {Kuroiwa}}, \bibinfo {author} {\bibfnamefont {A.}~\bibnamefont {Ikeda}},\ and\ \bibinfo {author} {\bibfnamefont {K.}~\bibnamefont {Miyazaki}},\ }\bibfield  {title} {\enquote {\bibinfo {title} {Jamming transition and inherent structures of hard spheres and disks},}\ }\href@noop {} {\bibfield  {journal} {\bibinfo  {journal} {Physical review letters}\ }\textbf {\bibinfo {volume} {109}},\ \bibinfo {pages} {205701} (\bibinfo {year} {2012})}\BibitemShut {NoStop}%
\bibitem [{\citenamefont {Chaudhuri}, \citenamefont {Berthier},\ and\ \citenamefont {Sastry}(2010)}]{chaudhuri2010j}%
  \BibitemOpen
  \bibfield  {author} {\bibinfo {author} {\bibfnamefont {P.}~\bibnamefont {Chaudhuri}}, \bibinfo {author} {\bibfnamefont {L.}~\bibnamefont {Berthier}},\ and\ \bibinfo {author} {\bibfnamefont {S.}~\bibnamefont {Sastry}},\ }\bibfield  {title} {\enquote {\bibinfo {title} {Jamming transitions in amorphous packings of frictionless spheres occur over a continuous range of volume fractions},}\ }\href@noop {} {\bibfield  {journal} {\bibinfo  {journal} {Physical review letters}\ }\textbf {\bibinfo {volume} {104}},\ \bibinfo {pages} {165701} (\bibinfo {year} {2010})}\BibitemShut {NoStop}%
\bibitem [{\citenamefont {Heyes}(2007)}]{heyes2007sys}%
  \BibitemOpen
  \bibfield  {author} {\bibinfo {author} {\bibfnamefont {D.}~\bibnamefont {Heyes}},\ }\bibfield  {title} {\enquote {\bibinfo {title} {System size dependence of the transport coefficients and stokes--einstein relationship of hard sphere and weeks--chandler--andersen fluids},}\ }\href@noop {} {\bibfield  {journal} {\bibinfo  {journal} {Journal of Physics: Condensed Matter}\ }\textbf {\bibinfo {volume} {19}},\ \bibinfo {pages} {376106} (\bibinfo {year} {2007})}\BibitemShut {NoStop}%
\bibitem [{\citenamefont {Kumar}, \citenamefont {Szamel},\ and\ \citenamefont {Douglas}(2006)}]{kumar2006nature}%
  \BibitemOpen
  \bibfield  {author} {\bibinfo {author} {\bibfnamefont {S.~K.}\ \bibnamefont {Kumar}}, \bibinfo {author} {\bibfnamefont {G.}~\bibnamefont {Szamel}},\ and\ \bibinfo {author} {\bibfnamefont {J.~F.}\ \bibnamefont {Douglas}},\ }\bibfield  {title} {\enquote {\bibinfo {title} {Nature of the breakdown in the stokes-einstein relationship in a hard sphere fluid},}\ }\href@noop {} {\bibfield  {journal} {\bibinfo  {journal} {The Journal of chemical physics}\ }\textbf {\bibinfo {volume} {124}} (\bibinfo {year} {2006})}\BibitemShut {NoStop}%
\bibitem [{\citenamefont {Janssen}(2018)}]{janssen2018m}%
  \BibitemOpen
  \bibfield  {author} {\bibinfo {author} {\bibfnamefont {L.~M.}\ \bibnamefont {Janssen}},\ }\bibfield  {title} {\enquote {\bibinfo {title} {Mode-coupling theory of the glass transition: A primer},}\ }\href@noop {} {\bibfield  {journal} {\bibinfo  {journal} {Frontiers in Physics}\ }\textbf {\bibinfo {volume} {6}},\ \bibinfo {pages} {97} (\bibinfo {year} {2018})}\BibitemShut {NoStop}%
\bibitem [{\citenamefont {Kumar}\ \emph {et~al.}(2007)\citenamefont {Kumar}, \citenamefont {Buldyrev}, \citenamefont {Becker}, \citenamefont {Poole}, \citenamefont {Starr},\ and\ \citenamefont {Stanley}}]{kumar2007r}%
  \BibitemOpen
  \bibfield  {author} {\bibinfo {author} {\bibfnamefont {P.}~\bibnamefont {Kumar}}, \bibinfo {author} {\bibfnamefont {S.}~\bibnamefont {Buldyrev}}, \bibinfo {author} {\bibfnamefont {S.}~\bibnamefont {Becker}}, \bibinfo {author} {\bibfnamefont {P.}~\bibnamefont {Poole}}, \bibinfo {author} {\bibfnamefont {F.}~\bibnamefont {Starr}},\ and\ \bibinfo {author} {\bibfnamefont {H.}~\bibnamefont {Stanley}},\ }\bibfield  {title} {\enquote {\bibinfo {title} {Relation between the widom line and the breakdown of the stokes--einstein relation in supercooled water},}\ }\href@noop {} {\bibfield  {journal} {\bibinfo  {journal} {Proceedings of the National Academy of Sciences}\ }\textbf {\bibinfo {volume} {104}},\ \bibinfo {pages} {9575--9579} (\bibinfo {year} {2007})}\BibitemShut {NoStop}%
\end{thebibliography}%


\begin{thebibliography}{0}%
\makeatletter
\providecommand \@ifxundefined [1]{%
 \@ifx{#1\undefined}
}%
\providecommand \@ifnum [1]{%
 \ifnum #1\expandafter \@firstoftwo
 \else \expandafter \@secondoftwo
 \fi
}%
\providecommand \@ifx [1]{%
 \ifx #1\expandafter \@firstoftwo
 \else \expandafter \@secondoftwo
 \fi
}%
\providecommand \natexlab [1]{#1}%
\providecommand \enquote  [1]{``#1''}%
\providecommand \bibnamefont  [1]{#1}%
\providecommand \bibfnamefont [1]{#1}%
\providecommand \citenamefont [1]{#1}%
\providecommand \href@noop [0]{\@secondoftwo}%
\providecommand \href [0]{\begingroup \@sanitize@url \@href}%
\providecommand \@href[1]{\@@startlink{#1}\@@href}%
\providecommand \@@href[1]{\endgroup#1\@@endlink}%
\providecommand \@sanitize@url [0]{\catcode `\\12\catcode `\$12\catcode `\&12\catcode `\#12\catcode `\^12\catcode `\_12\catcode `\%12\relax}%
\providecommand \@@startlink[1]{}%
\providecommand \@@endlink[0]{}%
\providecommand \url  [0]{\begingroup\@sanitize@url \@url }%
\providecommand \@url [1]{\endgroup\@href {#1}{\urlprefix }}%
\providecommand \urlprefix  [0]{URL }%
\providecommand \Eprint [0]{\href }%
\providecommand \doibase [0]{https://doi.org/}%
\providecommand \selectlanguage [0]{\@gobble}%
\providecommand \bibinfo  [0]{\@secondoftwo}%
\providecommand \bibfield  [0]{\@secondoftwo}%
\providecommand \translation [1]{[#1]}%
\providecommand \BibitemOpen [0]{}%
\providecommand \bibitemStop [0]{}%
\providecommand \bibitemNoStop [0]{.\EOS\space}%
\providecommand \EOS [0]{\spacefactor3000\relax}%
\providecommand \BibitemShut  [1]{\csname bibitem#1\endcsname}%
\let\auto@bib@innerbib\@empty
\end{thebibliography}%
\end{document}


\title{\textit{Supplementary Material}:\\
Comprehensive molecular dynamics study of the dynamical properties of a dense binary hard-sphere mixture}

\author{Sabry G. Moustafa}
\email[Electronic mail: ]{smoustafa@una.edu}
\affiliation{Department of Engineering and Industrial Professions, University of North Alabama, Florence, Alabama 35632, USA}
\author{Andrew J. Schultz}
\affiliation{Department of Chemical and Biological Engineering, University at Buffalo, The State University of New York, Buffalo, New York 14260, USA}

\maketitle

\section{Finite-Size Effects}
In the following, we examine finite-size effects (FSE) in the transport, dynamical, structural, and thermodynamic properties discussed in the main text. Specifically, we consider the time-dependent self-diffusivity, $D(\tau)$, shear viscosity, $\eta(\tau)$, non-Gaussian parameter, $\alpha_2(\tau)$, self-intermediate scattering function, $F_s(\tau)$, four-point dynamic susceptibility, $\chi_4(\tau)$, and the spatial dependence of the self-part of the van Hove correlation function, $G_s(r,\tau)$. We also examine the corresponding FSE in the characteristic relaxation times and the compressibility factor, $Z$, including its thermodynamic-limit dependence on packing fraction, $Z_\infty(\phi)$.

\subsection{Diffusion and viscosity}
\label{sec:DV_t}
\begin{figure}[!htb]
\centering
\includegraphics[width=\textwidth]{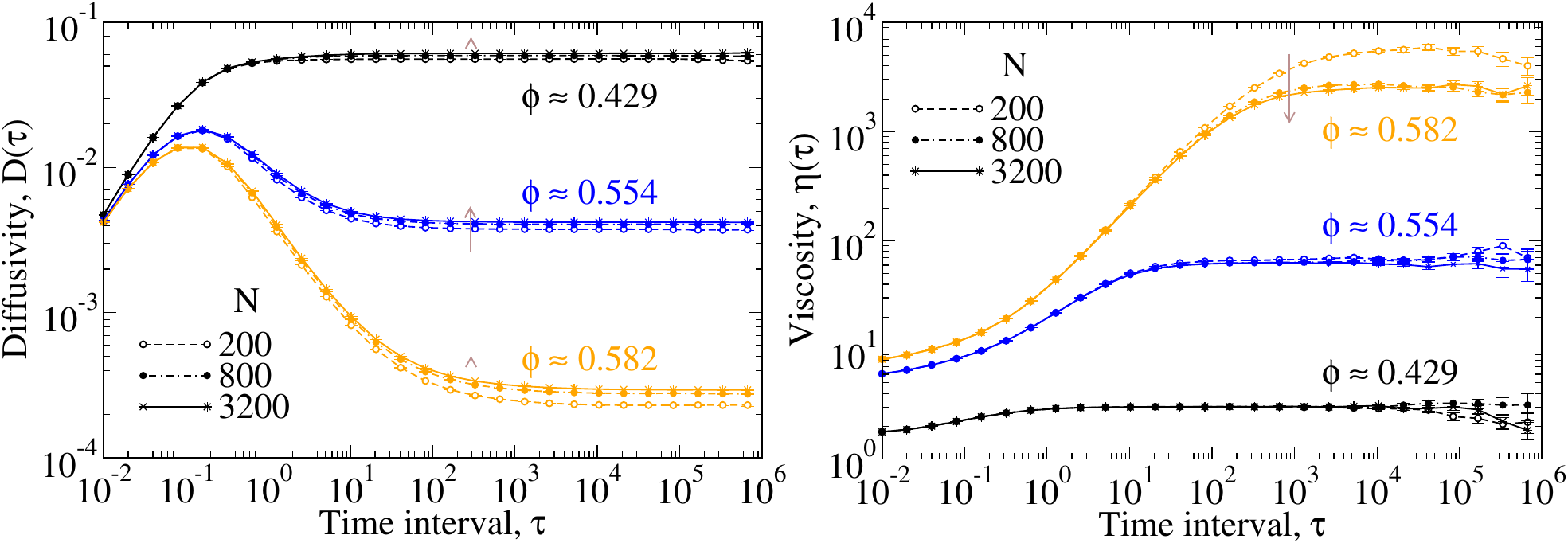}
\caption{Time-dependent total self-diffusivity (left) and shear viscosity (right) at three densities ($\rho = 1.20,\ 1.55,\ 1.63$). Arrows indicate increasing system size, $N=200, 800, 3200$. Lines are guides to the eye.}
\label{fig:fse_DV_t}
\end{figure}
Figure~\ref{fig:fse_DV_t} illustrates the finite-size dependence of the time-dependent transport coefficients, $D(\tau)$ and $\eta(\tau)$, at three representative packing fractions ($\phi \approx 0.429, 0.554$, and $0.582$). For packing fractions above $\phi \approx 0.464$, the time-dependent self-diffusivity exhibits a distinct peak at short time intervals before settling into a stable long-time plateau across all system sizes. This long-time plateau value increases systematically with expanding system size $N$, a behavior discussed in detail in the main text. Conversely, the time-dependent shear viscosity $\eta(\tau)$ grows monotonically toward its long-time limit. While $\eta(\tau)$ exhibits no noticeable system-size dependence at the lowest density, it decreases with increasing $N$ at higher densities, ultimately demonstrating statistical convergence for the larger system sizes ($N \ge 1600$).

\subsection{Non-Gaussian parameter}
\label{sec:alpha2_fse}
\begin{figure}[!htb]
\centering
\includegraphics[width=0.7\textwidth]{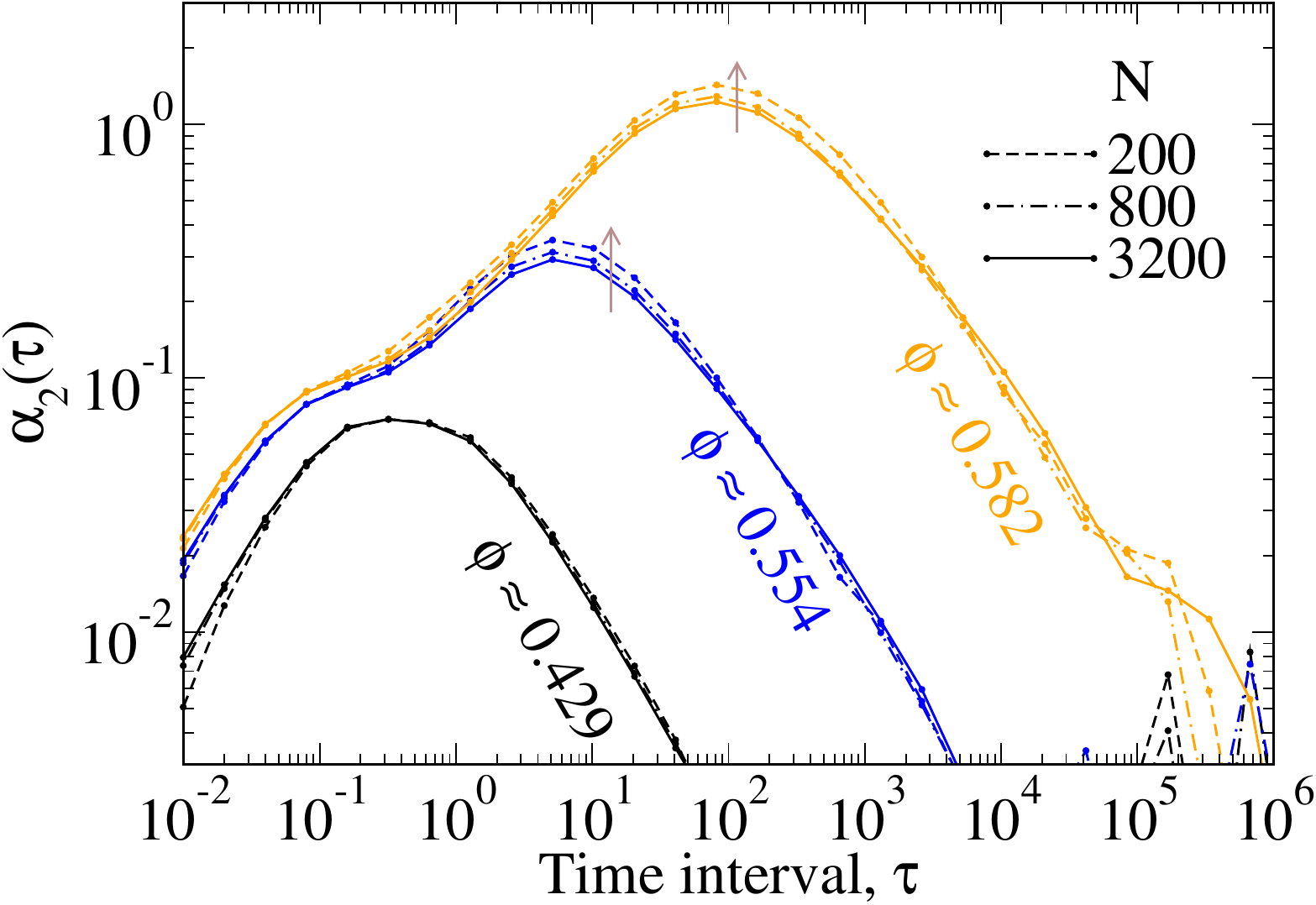}
\caption{Finite-size effects on the non-Gaussian parameter $\alpha_2(\tau)$ for the large particles (A) at three representative densities ($\rho=1.20, 1.55,$ and $1.63$) and for three system sizes ($N = 200, 800,$ and $3200$). Lines are guides to the eye.}
\label{fig:fse_alpha2_t}
\end{figure}
Figure~\ref{fig:fse_alpha2_t} presents the finite-size effects on the non-Gaussian parameter of the large particles (A). As observed for other dynamical properties, $\alpha_2(\tau)$ exhibits more pronounced FSE at higher densities. However, the corresponding peak time, $\tau^*$, does not appear to be sensitive to variations in system size. A qualitatively similar behavior is observed for the smaller particles (B), though the corresponding data are not shown here for brevity.

\subsection{Structural and four-point relaxation times}
\label{sec:fs_chi4_fse}
\begin{figure}[!htb]
\centering
\includegraphics[width=\textwidth]{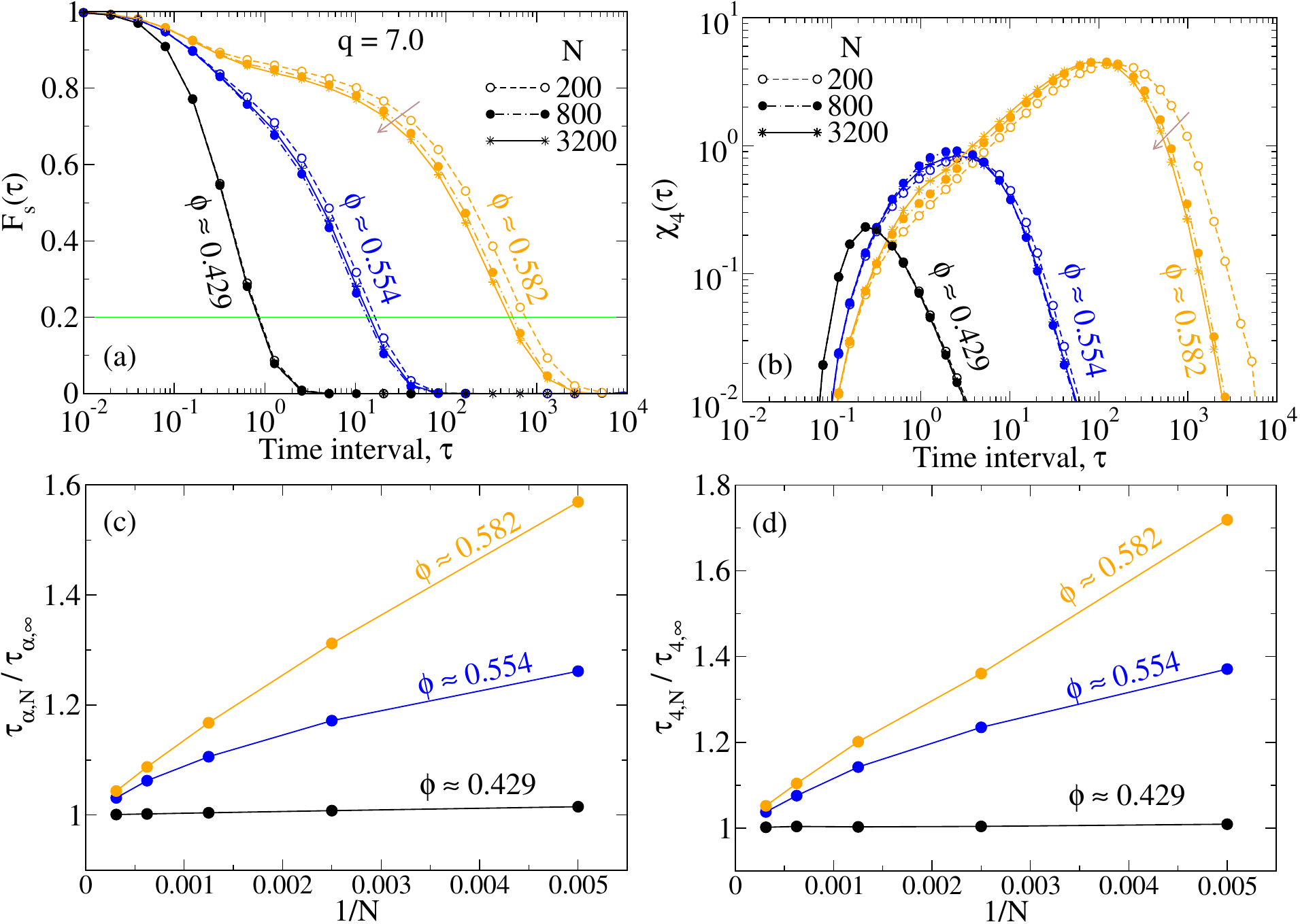}
\caption{Top: Finite-size effects on the self-intermediate scattering function (left) and four-point dynamic susceptibility (right) at three densities ($\rho=1.20$, $1.55$, and $1.63$). Arrows indicate increasing $N$. Bottom: System-size dependence of the corresponding relaxation times, normalized by their thermodynamic-limit values. Lines are guides to the eye.}
\label{fig:fse_fs}
\end{figure}
In Fig.~\ref{fig:fse_fs}, we examine the finite-size effects on the self-intermediate scattering function and the four-point dynamic susceptibility, along with their associated characteristic relaxation times. As the system size increases, both $F_s(\tau)$ and $\chi_4(\tau)$ relax faster, which manifests as a systematic decrease in their characteristic relaxation times. This behavior reflects the reduced influence of spatial confinement and enhanced particle mobility in larger periodic domains (see main text). Furthermore, these finite-size effects become increasingly pronounced with increasing density, exhibiting a qualitative trend similar to that observed for the shear viscosity.

\subsection{Self-part of the van Hove correlation function}
\label{sec:gs_fse}
\begin{figure}[!htb]
\centering
\includegraphics[width=\textwidth]{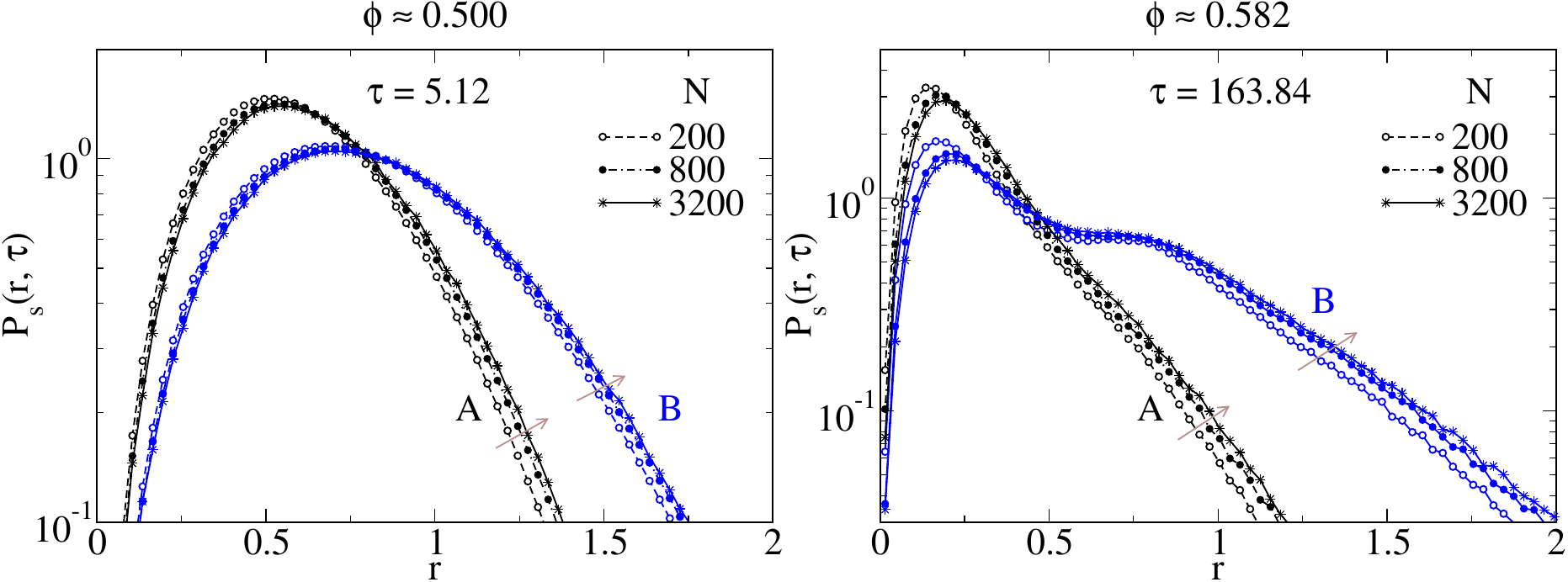}
\caption{System-size effects on the radial probability density function for the large (A) and small (B) particles at low (left; $\rho=1.40$, $\tau=5.12$) and high (right; $\rho=1.63$, $\tau=163.84$) densities. Arrows indicate increasing $N$. Lines are guides to the eye.}
\label{fig:fse_gs}
\end{figure}

Figure~\ref{fig:fse_gs} illustrates the effect of system size on the radial probability density function, $P_s(r,\tau)=4\pi r^2G_s(r,\tau)$, where $G_s(r,\tau)$ is the van Hove self-correlation function. The data are shown for both particles A and B, at two representative low and high densities and evaluated at intermediate times. As the system size increases, $P_s\left(r, \tau\right)$ broadens, in a manner that preserves its integral to unity. This behavior is consistent with the results in Fig.~\ref{fig:fse_DV_t} and follows directly from the relation $D(\tau)=\frac{\langle r^2(\tau)\rangle}{6\tau}
=\frac{1}{6\tau}\int_0^\infty r^2P_s(r,\tau) dr$. 

\subsection{Compressibility factor}
\label{sec:z_fse}
\begin{figure}[!htb]
\centering
\includegraphics[width=0.75\textwidth]{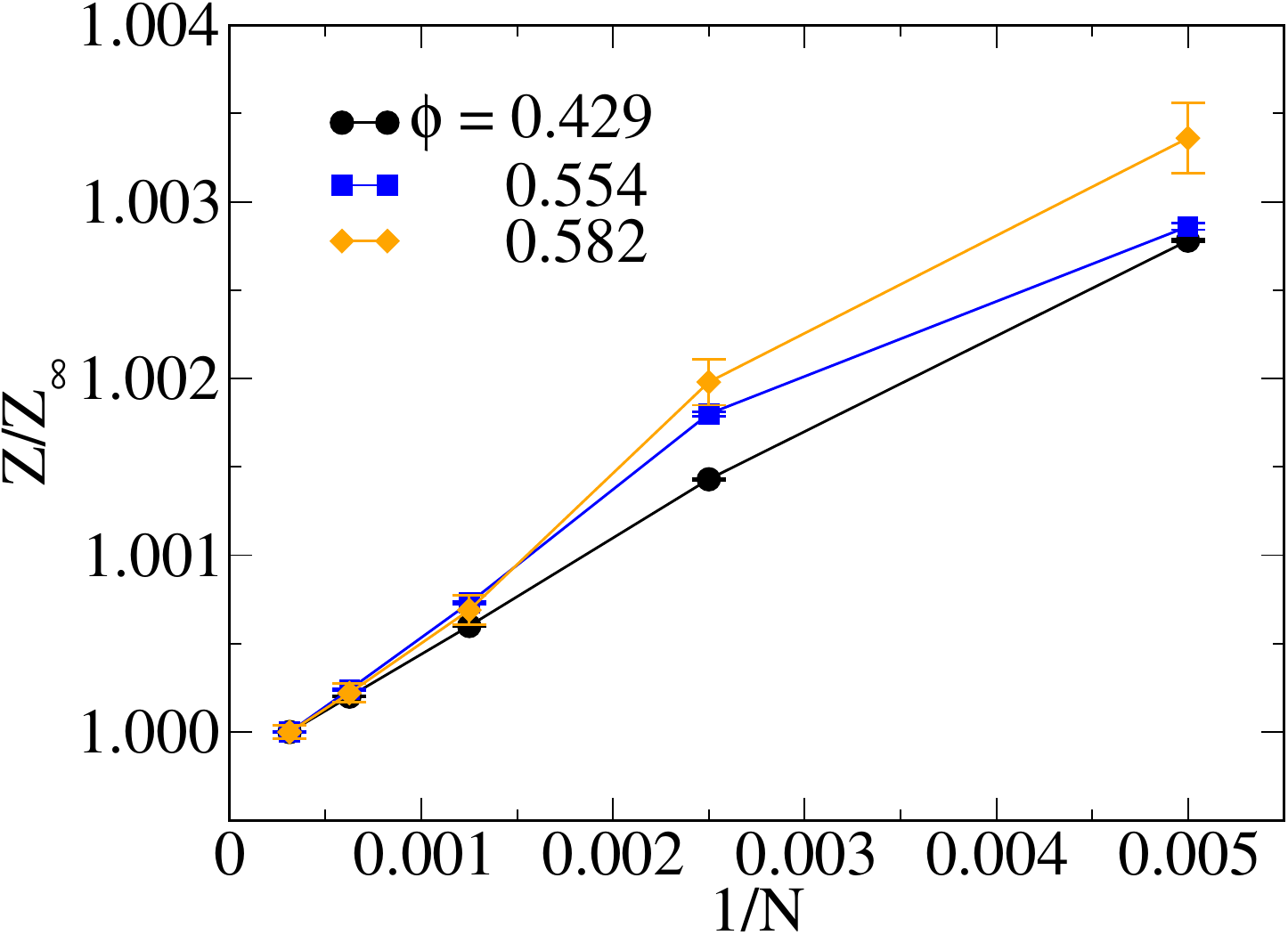}
\caption{Finite-size effects on the compressibility factor, normalized by its thermodynamic-limit value. Data are shown for three densities, $\rho=1.20$, $1.55$, and $1.63$. Lines are guides to the eye.}
\label{fig:fse_Z}
\end{figure}

Figure~\ref{fig:fse_Z} shows the finite-size scaling of the compressibility factor normalized by its thermodynamic-limit value, $Z/Z_\infty$. The data are presented at three representative low-, intermediate-, and high-density state points. We note that the thermodynamic properties are less sensitive to system size than the dynamic properties, particularly for larger system sizes. For example, at a system size of $N=800$, a deviation of less than 0.1\% from the thermodynamic limit is observed. Moreover, the variations appear to scale linearly with $1/N$ for system sizes $N \ge 1600$. Consequently, the thermodynamic-limit estimates reported here are obtained by fitting the three data points corresponding to $N \ge 1600$ to a straight line in $1/N$.

\begin{figure}[!htb]
\centering
\includegraphics[width=0.7\textwidth]{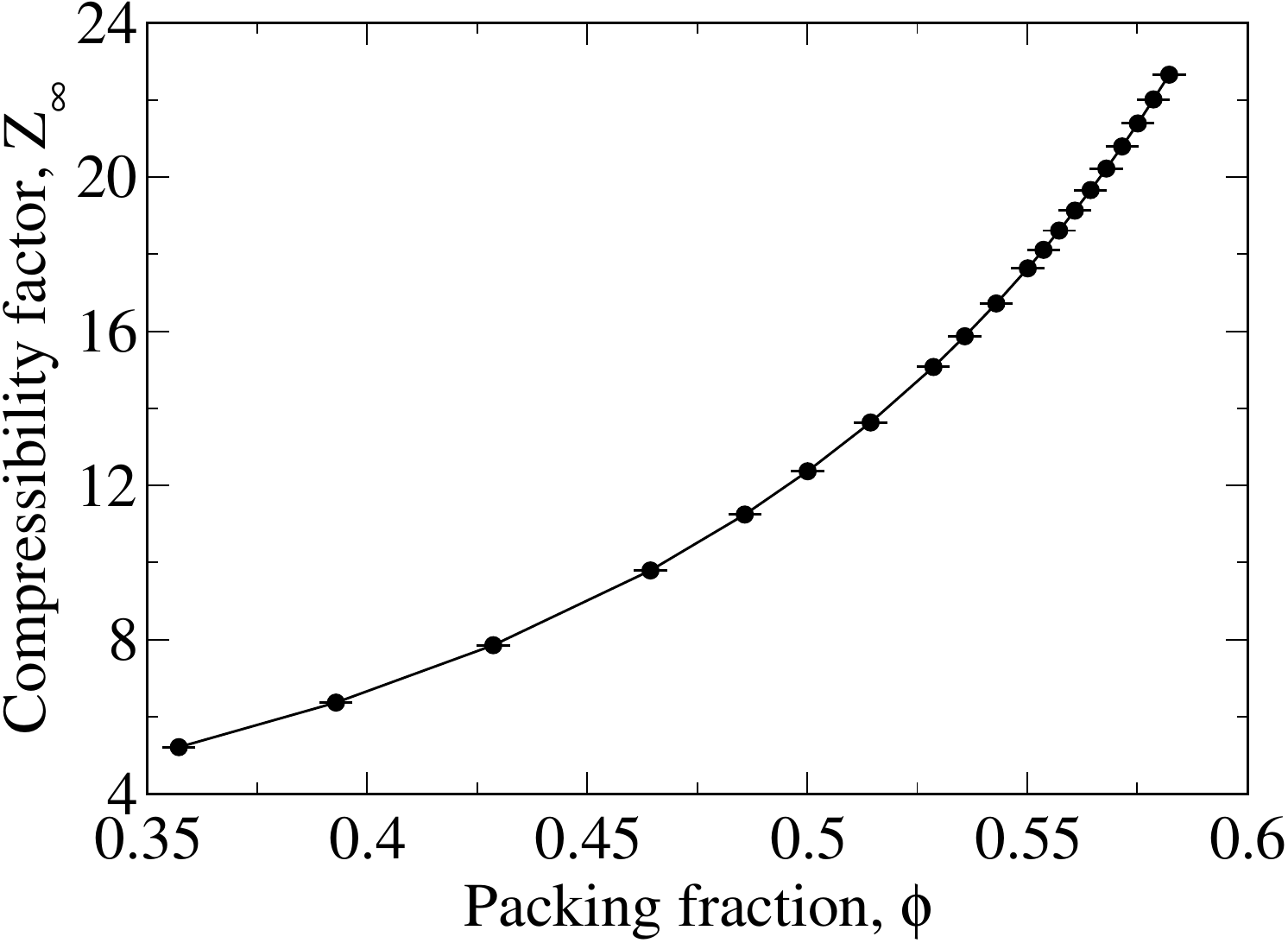}
\caption{Compressibility factor in the thermodynamic-limit, $Z_\infty$, as a function of packing fraction, $\phi$. Statistical uncertainties (error bars) are smaller than the symbol sizes. Lines are guides to the eye.}
\label{fig:Z_phi}
\end{figure}

Figure~\ref{fig:Z_phi} shows the dependence of the thermodynamic-limit compressibility factor, $Z_\infty(\phi)$, on packing fraction $\phi$ on a linear scale, covering the entire range of studied fluid densities. The compressibility factor increases monotonically with packing fraction and exhibits an increasingly pronounced upward curvature as the mixture becomes denser. This increase reflects the rapidly growing collision frequency at higher packing fractions and the resulting increase in pressure.

\section{Relaxation times}
\label{sec:relax_sm}
\begin{figure}[!h]
\centering
\includegraphics[width=\textwidth]{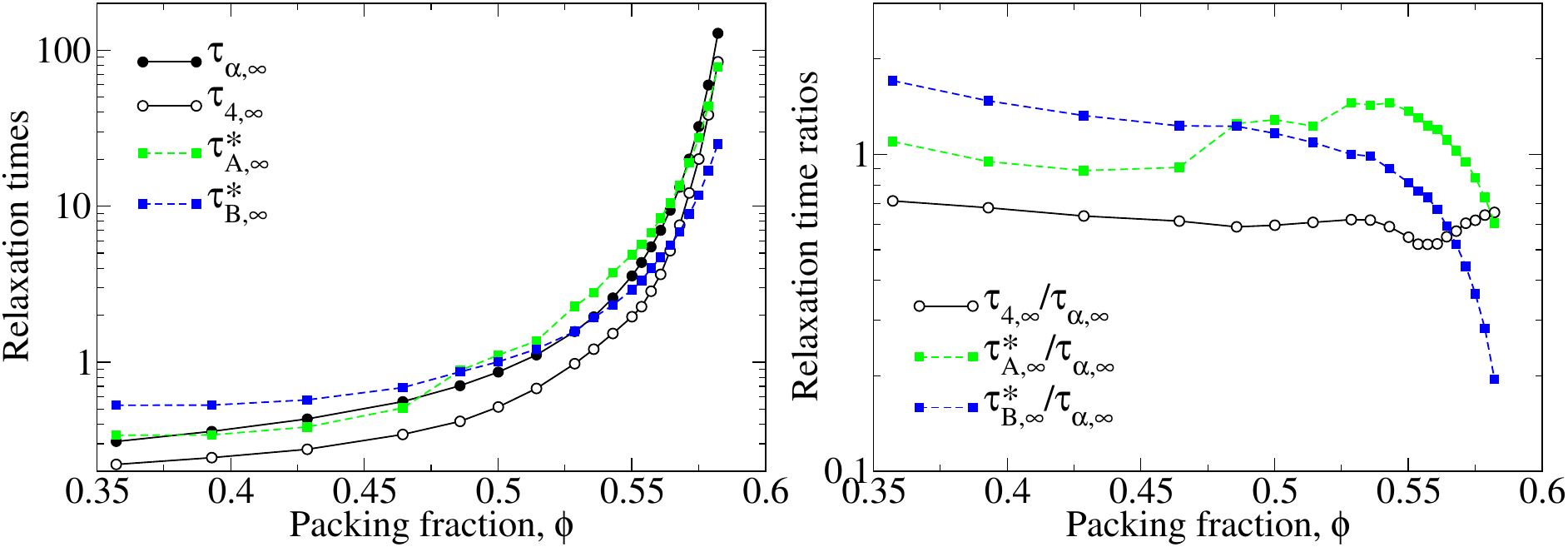}
\caption{Panel (a): Packing fraction dependence of the relaxation times. Panel (b): Ratios of the different relaxation times to the structural relaxation time. All data are evaluated in the thermodynamic limit. Lines are guides to the eye.}
\label{fig:taus_phi}
\end{figure}

Figure~\ref{fig:taus_phi}(left) summarizes the packing-fraction dependence of the characteristic relaxation times, $\tau_{\alpha}$, $\tau_4$, and $\tau_i^{*}$, introduced in the main text, but evaluated in the thermodynamic limit. Since the relaxation times are shown on a logarithmic scale, the vertical separation between the curves directly reflects their relative magnitudes and, hence, the ratios between the different timescales. Overall, the behavior is qualitatively similar to that observed for the finite system size, $N=3200$, presented in the main text.

The ratios of the characteristic relaxation times to the structural relaxation time, $\tau_{\alpha}$, are shown explicitly in Fig.~\ref{fig:taus_phi}(right). The ratio $\tau_4/\tau_{\alpha}$ remains nearly constant over the entire range of packing fractions, indicating that these two timescales exhibit an approximately invariant relative scaling. In contrast, the ratio $\tau_i^{*}/\tau_{\alpha}$ displays a pronounced non-monotonic dependence on packing fraction, with a particularly sharp variation at high packing fractions.

